\documentclass[10pt,a4paper]{article}
\pdfoutput=1
\usepackage{amssymb,amsmath,enumerate,mathrsfs,enumerate}
\usepackage{graphicx,rotate,multicol}
\usepackage{multirow}
\usepackage{array}
\usepackage{jheppub}

\usepackage[]{changes} 

\usepackage{dcolumn}
\usepackage{bm}
\usepackage{dcolumn}
\usepackage{textcomp}
\usepackage{float}
\usepackage{subfig}
\usepackage{hypcap}
\usepackage[]{hyperref}
\usepackage{makecell}

\usepackage{xcolor}
\usepackage{color}
\usepackage{pifont}
\usepackage{appendix}
\usepackage{url}
\usepackage{cancel}
\allowdisplaybreaks[1]
\usepackage{import}
\usepackage{subfiles}
\usepackage{filecontents}
\usepackage{slashed}
\usepackage{lineno}
\usepackage{placeins}

\let\Oldsection\section
\renewcommand{\section}{\FloatBarrier\Oldsection}

\let\Oldsubsection\subsection
\renewcommand{\subsection}{\FloatBarrier\Oldsubsection}

\let\Oldsubsubsection\subsubsection
\renewcommand{\subsubsection}{\FloatBarrier\Oldsubsubsection}
\hypersetup{
  bookmarks=true,         
  unicode=false,          
  pdftoolbar=true,        
 pdfmenubar=true,        
 pdffitwindow=true,     
 pdfstartview={FitH},    
 pdfsubject={GUT and Topological defects},   
 pdfnewwindow=true,      
 pdfcreator={RevTeX},
 colorlinks=true,       
 linkcolor=red,          
 citecolor=blue,        
 filecolor=black,      
 urlcolor=blue,           
  }

\usepackage{amsmath}
\usepackage{slashed}

\usepackage{rotating}
\usepackage{float}

\usepackage{scalerel}
\def\msquare{\square}
\usepackage{color}
\long\def\rpl#1!!#2!!{\textcolor{red}{#1} \textcolor{blue}{#2}}
\long\def\commagda#1!!{\textcolor{blue}{#1}}

\def \order(#1){{\cal O} \left(#1 \right)}
\DeclareMathOperator{\Tr}{Tr}

\allowdisplaybreaks

\title
{Discriminating the \texttt{HTM} and \texttt{MLRSM} models in collider studies via doubly charged Higgs boson pair production
and the subsequent leptonic decays}

\author[a,b]{Janusz Gluza,}  
\author[a]{Magdalena Kordiaczy\'{n}ska,}
\author[c]{Tripurari Srivastava}

  \affiliation[a]{Institute of Physics, University of Silesia, Katowice, Poland}
  \affiliation[b]{Faculty of Science, University of Hradec Kr\'alov\'e, Czech Republic} 
   \affiliation[c]{Theoretical Physics Division, Physical Research Laboratory, Ahmedabad-380009, India}  

\emailAdd{janusz.gluza@us.edu.pl}
\emailAdd{mkordiaczynska@us.edu.pl}
\emailAdd{tripurari@prl.res.in}

\abstract{
We present a case study for the doubly charged Higgs bosons $H^{\pm\pm}$ pair production in $e^+e^-$ and $pp$ colliders with their subsequent decays to four charged leptons. 
We consider the Higgs Triplet Model  (\texttt{HTM}) not restricted by the custodial symmetry  and  the Minimal Left-Right Symmetric Model (\texttt{MLRSM}). The models include scalar triplets with different complexity of scalar potentials and, due to experimental restrictions, completely different scales of non-standard triplet vacuum expectation values. 
 In both models, a doubly charged Higgs boson  $H^{\pm\pm}$ can acquire a mass of hundreds of gigaelectronvolts, which can be probed at HL-LHC, future $e^+e^-$, and hadron colliders.  
We take into account  a comprehensive set of  constraints on the parameters of both models coming from neutrino oscillations, LHC, $e^+e^-$ and low-energy lepton flavour violating data and assume the same mass of $H^{\pm\pm}$. Our finding is that the $H^{\pm\pm}$ pair production in lepton and hadron colliders is comparable in both models, though more pronounced in \texttt{MLRSM}. We show that  the decay branching ratios can be different within both models, leading to distinguishable four lepton signals and that the strongest are $4\mu$ events yielded by \texttt{MLRSM}. Typically we find that \texttt{MLRSM} signals are  one order of magnitude larger that in \texttt{HTM}. For example, the $pp \to 4\mu$ \texttt{MLRSM} signal for 1 TeV  $H^{\pm \pm}$ mass  results in a clearly detectable significance of $S \simeq 11$ for HL-LHC and $S \simeq 290$ for FCC-hh.
Finally we provide quantitative predictions for the dilepton invariant mass distributions and lepton separations which help to identify non-standard signals.

}

\begin{document}
\maketitle
\flushbottom
\newpage

\section{Introduction}
\label{s:intro}
\subsection{Discovery of the first Higgs scalar boson at LHC and the Standard Model}
The spectacular discovery of the chargeless Higgs particle $(H^0)$ at the LHC~\cite{Aad:2012tfa,Chatrchyan:2012xdj,1576} is consistent with the prediction of the Standard Model (SM), confirming the basic concept of the spontaneous symmetry breaking mechanism and elementary particle mass generation. Observed $H^0$ decays into gauge boson particles $W^+W^-$ and $ZZ$~\cite{Chatrchyan:2012jja,Aad:2013xqa,Chatrchyan:2011tz} fits beautifully into this picture. Similarly, determination of $t \bar t H^0$ couplings in gluon fusion~\cite{Khachatryan:2014qaa,Aad:2014lma,Khachatryan:2016vau} and $t\bar t H^0$ production~\cite{Aad:2015iha} confirm Higgs boson role in fermion mass generation. With gathered statistics, we know more and more about this particle, namely its decay rate to  $\gamma \gamma$~\cite{Djouadi:1993ji,Aad:2015gba}, spin-parity which is dominantly $J^P=0^+$~\cite{Chatrchyan:2012jja,Aad:2013xqa,Aad:2015mxa,Aad:2015rwa,DiMarco:2016uwi}.
Also mass suppressed decay rate to muon pairs when comparing to top pairs is evident~\cite{Aad:2015gba,Sirunyan:2017khh}.
Yet another spectacular success of the LHC physics is a clear discovery that the Higgs boson decays to the third generation of fermions, namely to the pairs of $\tau$ leptons and b-quarks. Especially determination of the Yukawa Higgs boson coupling to b-quarks is tricky, as though this channel amounts to about 60\% of Higgs boson decays, the QCD b-quark background is overwhelming \cite{plb201859}. The story of the Higgs boson studies continues. Very recently the measurements of the Higgs boson’s properties have reached a new stage in precision through detection of a rare decay mode where the Higgs boson decays into two muons \cite{CMS-PAS-HIG-19-006,Aad:2020xfq}.    
Aiming at sub-percent precision for Higgs boson decays, quantitative tests of the SM for Higgs boson couplings need further scrutinization in studies at HL-LHC and future Higgs factories. These also include the investigation of the Higgs boson self-coupling \cite{Blondel:2018aan}. 

\subsection{Searching for new scalar bosons at future colliders and a choice of tested models}

Detection of the Standard Model scalar particle does not preclude validity of more elaborated physical scenarios with extended scalar sectors. The simplest extensions beyond the SM doublet scalar multiplet include their copies, like the two Higgs doublet model~\cite{Branco:2011iw}, supersymmetric extensions of the SM~\cite{Gunion:1984yn,Gunion:1989we} or, stepping up in this construction,  scenarios with triplet scalar representations either in their  supersymmetric~\cite{ESPINOSA1992113,DiChiara:2008rg} or non-supersymmetric versions~\cite{KONETSCHNY1977433,GELMINI1981411,PhysRevLett.44.912}.
Here we will consider the latter. 
There are many possibilities for triplet representations, depending on the hypercharge \mbox{$Y~\!\!\equiv\!\!~2(Q-T_3)$} \cite{PhysRevD.40.1546,VEGA1990533,PhysRevD.42.1673,Huitu:1996su}. We will explore the simplest one which involves doubly charged Higgs fields in the triplet representation with hypercharge $Y =2$, the Higgs Triplet Model (\texttt{HTM})~\cite{Chun:2003ej}. For that, we will not assume any special symmetries or constructions \cite{Georgi:1985nv,Chiang:2012cn}, so that $v_\Delta$, the triplet vacuum expectation value (VEV) will be extremely tiny, at the scale of electronvolts which makes experiments more challenging. We will also consider a much more complex model where the Standard Model $SU(2)\times U(1)$ gauge symmetry is extended by an additional $SU(2)$ group, the so-called minimal left-right symmetric model
(\texttt{MLRSM})~\cite{Mohapatra:1974gc,Senjanovic:1975rk,PhysRevLett.44.912,PhysRevD.44.837,Duka:1999uc,Barenboim:2001vu}.
Thus we consider a setting where both the \texttt{HTM} and the \texttt{MLRSM} models include doubly charged Higgs bosons.

\texttt{HTM} received a considerable amount of attention recently \cite{PhysRevD.84.095005,PhysRevD.85.055018,PhysRevD.85.055007,PhysRevD.85.115009,PhysRevD.86.035015,Blum2015,Blunier:2016peh,Dev:2017ouk,Biswas:2017tnw,Du:2018eaw,deMelo:2019asm,Primulando:2019evb,Dev:2019hev,Fuks:2019clu}. This model when confronted with experimental data, features a strong restriction in which  $v_\Delta$ is very small, 
$\order(1)$ ~(GeV), or below. Here, in particular, we concentrate on the cases where $v_\Delta$ is of the order of neutrino masses. Then the triplet Yukawa couplings will be of the $\order(1)$ order and $H^{\pm\pm}$ decays dominantly into the same-sign
dilepton channel. In this case, the LHC direct search bound on the doubly charged scalar mass, $m_{H^{\pm \pm}}\gtrsim 850$~GeV~\cite{Aaboud:2017qph}
applies. At the same time, the constraints from different lepton flavor violating (LFV) processes and non-universality of leptonic couplings start to weigh in. There is thus a direct relationship among the triplet VEV $v_\Delta$, neutrino masses, their mixing and doubly charged Higgs couplings. That is why the production and decays of $H^{\pm \pm}$ scalars at high energies depend substantially on the oscillation data and limits on LFV processes in \texttt{HTM}. 

On the other hand, in \texttt{MLRSM} the dominating non-standard effects in phenomenological studies are connected with the right-handed breaking scale $v_R$ which affects the couplings and masses of a wide set of non-standard heavy particles of spin 0,1,1/2 present in the model. Low-energy precision SM and rare processes, as well as  high-energy collider studies, limit the possible values of $v_R$ from below. The scale of relevance for $v_R$ starts from  the $\mathcal{O}(1)$ TeV level up \cite{Chakrabortty:2016wkl}.
\\
Consequently in both models we have two completely different VEV scales, $v_\Delta$ and $v_R$. 
\\
How to discriminate such two distinct models experimentally? Indeed it is not easy as any non-standard effect considered or thought of so far in phenomenological studies in search for BSM models failed to show unambiguous excess rates (reported excesses were vanishing with higher statistic). 

One of the most appealing rare process capable of exposing  BSM signals involving doubly charged Higgs bosons at high energy colliders would be the $H^{\pm \pm}$ pair production and the subsequent decays to four charged leptons. Here the same charge sign dileptons appear from parent's $H^{\pm \pm}$, $H^{\pm \pm} \to l^\pm l^\pm$, which is distinguishable from the SM background. On top of that, we compare the $H^{\pm \pm}$ production and decay signals in the two considered BSM models taking into account all relevant experimental limits. 
In this work we investigate this scenario in detail. 
  
The \texttt{HTM} model which we discuss is the simplest theoretical scenario with the triplet scalar representation, without ad-hoc symmetries put in.
On the opposite side of the theoretical complexity stands the \texttt{MLRSM} model. This model poses a broad spectrum of non-standard features: additional gauge group, so other gauge bosons and right-handed currents, heavy neutral leptons, a plethora of Higgs scalars, including two doubly charged Higgs bosons $H_{(1,2)}^{\pm \pm}$. Details on scalar potentials and fields are given in the Appendix. On top of that, as already mentioned, the process of the $H_{}^{\pm \pm}$ pair production at colliders is peculiar because it exhibits a small background. 
We assume a scenario that the excess signal of four charged leptons
$e^+ e^- (pp) \to H_{}^{++} H_{}^{--} \to  4l$  over the background is identified. In case when no other non-standard signals appear (e.g. connected with right-handed currents), the question is how to find to which non-standard model does the signal belong? 
In practice, such identification will be not trivial. 
In our opinion the problem of distinguishing two models based on rare processes where particles with the same masses play a crucial role is an important topic as needless to say such statements are essential for future post-LHC studies. 
Usually, in phenomenological analysis any specific models are considered.   
The exception is the effective field theory approach where non-standard interactions and energy scales are probed. However, to say anything about specific models, if positive signals and deviations from the SM signals will be found, comparative studies as given here based on particular models will be crucial. Apart from these general statements, other exciting subtleties can be probed in these studies connected with the neutrino sector. We will come to this topic in a moment.

To get reliable predictions for BSM processes, essential restrictions on the BSM model parameters coming from rare and so far not observed LFV processes must be considered. 
As even a single unambiguous LFV event detection  would be a signal of beyond SM physics, there are many efforts to upgrade or create new experimental setups for that, see e.g.~\cite{Lindner:2016bgg,Calibbi:2017uvl}.
Present bounds for low energy LFV signals, such as nuclear $\mu$ to $e$ conversion will become more stringent through the so-called intensity frontier experiments~\cite{Kuno:2013mha, Brown:2015cka}. 
The same is true for  $(\beta \beta)_{0 \nu}$ experiments, see e.g.~\cite{Abgrall:2013rze,Bolton:2019pcu}.
In this work, we consider these processes to predict reliable BSM $H^{\pm \pm}$ collider signals.

Concerning high energy colliders, there are presently several  options considered internationally for future electron colliders \cite{Strategy:2019vxc}, namely, FCC (Future Circular Collider) \cite{Abada:2019zxq,fcccdrweb}, CLIC (Compact Linear Collider)~\cite{Linssen:2012hp,clic} -- both at CERN, the ILC (International Linear Collider\footnote{Recently ILC and CLIC unite to advance the global development work for the next-generation linear \mbox{collider~\cite{ilcwww}.}})
\cite{Djouadi:2007ik,Moortgat-Picka:2015yla}.
The CEPC (Chinese Electron Positron Collider) \cite{CEPCStudyGroup:2018rmc, cepc} in China is of the circular type and similarly to FCC
is expected to collide electrons with positrons at 90-365 GeV center of mass energies. 
The ILC collider could potentially be positioned in Japan, and its centre of mass collision energies would reach 1 TeV while CLIC would cover the energies between 380 GeV and 3 TeV. In the future, extreme energies may become possible in	Plasma Wakefield Linear Colliders \cite{Adli:2019hnt}. In case of FCC-ee, four running stages are considered \cite{Abada:2019zxq,Blondel:2018mad,Blondel:2019qlh}, with focus on Z,W,H and top quark production. This means that the maximal energy will be not enough to search for direct $H^{\pm \pm}$ pair production signals. What remains is the high luminosity LHC (HL-LHC) \cite{hllhc,Strategy:2019vxc}, and the FCC-hh proton-proton option with center of mass energies of collided protons reaching 100 TeV \cite{Contino:2016spe,Golling:2016gvc}.

A significant part of  calculations done in this paper was performed using the MadGraph \cite{MadGraph:2014} and Pythia \cite{Sjostrand:2007gs,Sjostrand:2006za} programs. The UFO files were generated using FeynRules \cite{FeynRules:2013} and built on our model file, based on the default Standard Model implementation.\\

\section{
Doubly charged Higgs bosons and neutrinos in the considered triplet BSM scenarios \label{tripltesneutr}}

Regarding the scalar particle masses, we have constructed a mass spectrum in which $M_{H^{\pm\pm}}=700$ GeV. Corresponding parameters of scalar potentials in both models are given in Table \ref{HTM_LRSM_parameters}.

\begin{table}[H]
\centering
\begin{tabular}{cc}
\hline
 \parbox[t]{8mm}{\multirow{5}{*}{\rotatebox[origin=c]{90}{{\large \texttt{HTM}}}}} \\
& $\mu = 1.7 \times 10^{-7}, \quad \lambda = 0.519, \quad \lambda_1 = 0.519, \quad \lambda_2 = 0, \quad \lambda_3 = -1, \quad \lambda_4 = 0.$ \\ 
&\\
& \begin{tabular}{llll}
$M_{h}$ = 125.3 GeV,& $M_{H}$ = 700 GeV,& $M_{H^\pm}$ = 700 GeV,& $M_{H^{\pm\pm}}$ = 700 GeV.\\
& & & \\
\end{tabular} \\
&\\
\hline
\parbox[t]{8mm}{\multirow{7}{*}{\rotatebox[origin=c]{90}{{\large \texttt{MLRSM}}}}} \\
& $\lambda_1 = 0.129, \; \rho_1 = 0.0037, \; \rho_2 = 0.0037, \; \rho_3-2\rho_1 = 0.015, \; \alpha_3 = 4.0816, \; 2\lambda_2 - \lambda_3 = 0$. \\ &\\
& \begin{tabular}{llll}
$M_{H_0^0}$ = 125.3 GeV,& $M_{H_1^0}$ = 10 TeV,& $M_{H_2^0}$ = 600 GeV,& $M_{H_3^0}$ = 605.4 GeV,\\
& & & \\
$M_{H_1^{\pm\pm}}$ = 700 GeV,& $M_{H_2^{\pm\pm}}$ = 700 GeV,& $M_{H_1^\pm}$ = 654.4 GeV,& $M_{H_2^\pm}$ = 10 003.1 GeV.
\end{tabular}\\
&\\
\hline
&\\
\end{tabular}
\caption{\label{HTM_LRSM_parameters}Benchmark points and corresponding potential parameters for \texttt{HTM} and \texttt{LRSM} with $M_{H^{\pm\pm}}=M_{H^{\pm\pm}_{1,2}}$ = 700 GeV. The scalar potential parameters, fields and relations for masses are defined in the Appendix, Eq.~\ref{e:potential} and Eq.~\ref{VLRSM}. {{We identify $h$ and $H_0^0$ as the SM Higgs boson~($H^0$)}}. }
\end{table}

The mass benchmark points are constructed in order to satisfy several theoretical conditions like potential stability, unitarity and the T-parameter restriction and bounds from $h \to \gamma \gamma$ \cite{Chun:2012jw,Akeroyd:2012ms,Shen_2015_EPL_2,Das:2016,Chakrabortty:2016wkl,Gluza:2020icp}, see also sections \ref{HTMconstr} and \ref{MLRSMconstr}. \\ 

As we can see in   Tab.~\ref{HTM_LRSM_parameters}, there are more scalar fields in \texttt{MLRSM} than in \texttt{HTM}. Then any detectable signal connected with a neutral, singly or doubly charged Higgs bosons which are present in \texttt{MLRSM} but are not present in \texttt{HTM} would be in favour of \texttt{MLRSM}.  
However, we should note that though \texttt{MLRSM} is very rich in particle content and non-standard interactions, despite enormous theoretical and experimental efforts over last several decades, what we get so far are exclusion limits on the parameters of this model. 
All experimental data considered so far gives no indication for neutral, singly or doubly charged scalars, extra neutral heavy leptons, extra gauge bosons. So, the starting point is actually the same: we do not know if and which BSM model is realized in nature and we are still looking for a first experimental indication towards any non-standard signals in one or another model.  As the models' parameters are already severely constrained, we have to consider very rare processes and hence faint signals. 
We focus on the cleanest BSM colliders signal connected with doubly charged scalars: their pair production and subsequent decays (correlations between the same-sign leptons in the final state originated from $H^{\pm \pm}$ decays). To leave no stone unturned, we will focus especially on the case in which two doubly charged Higgs bosons in \texttt{MLRSM} have the same masses as otherwise the second scalar $H_2^{\pm \pm}$ connected with right-handed triplet in \texttt{MLRSM} (see Appendix for fields definitions) would help to  discriminate between both models in favor of \texttt{MLRSM}.  A case with different $H_1^{\pm \pm}$ and $H_2^{\pm \pm}$ masses  in $e^+e^-$ CLIC center of mass energies will be shortly discussed in Section \ref{comparison}, non-degenerate mass cases for hadron colliders have been discussed already in \cite{Bambhaniya:2013wza}.
For the same masses of $H_1^{\pm \pm}$ and $H_2^{\pm \pm}$, the production rates are higher in \texttt{MLRSM} than in \texttt{HTM}. We will see what is the contribution of $H_2^{\pm \pm}$ against $H_1^{\pm \pm}$ in production processes at lepton and hadron colliders, in case of the same doubly charged boson masses and how these contributions  change with center of mass energy. As we will see, there are scenarios with model parameters where the difference of signals in both models can be further enhanced by studying leptonic branching ratios of doubly charged Higgs bosons and kinematic cuts.
 
\begin{figure}[h!]
\begin{center}
\includegraphics[width=0.8\textwidth]{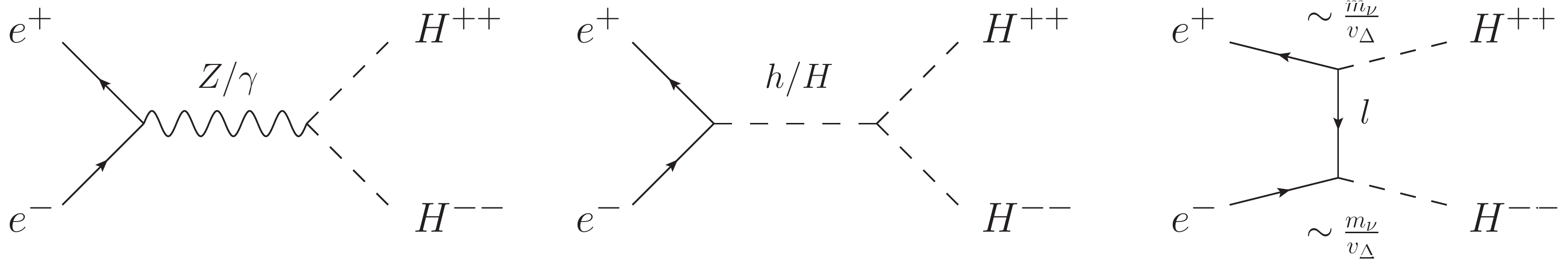} 
\caption{Pair production of doubly charged Higgs bosons $e^+ e^- \to H^{++} H^{--}$ in the \texttt{HTM} model. For the t-channel the couplings depend on neutrino parameters and $v_\Delta$. The exact form of the coupling is given by Eq.~(\ref{e:yukawa4}).
\label{ee_hcchcc_production}}
\end{center}
\end{figure}

\begin{figure}[h!]
\begin{center}
\includegraphics[width=0.8\textwidth]{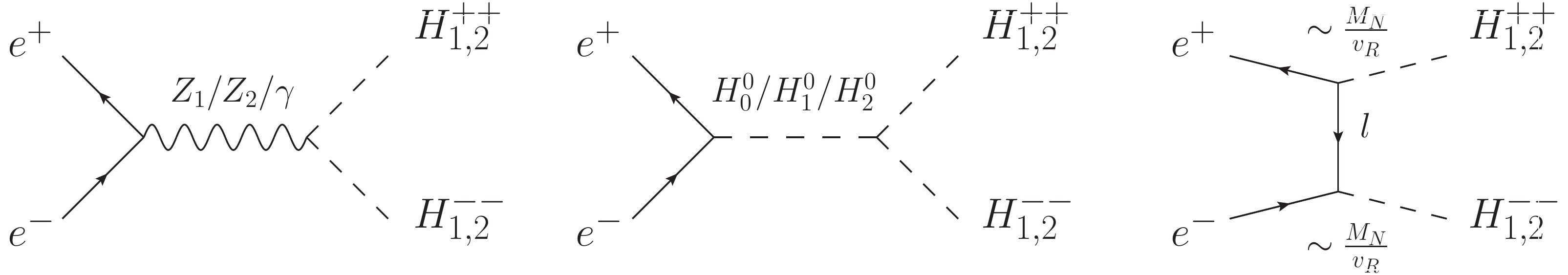} 
\caption{Feynman diagrams at the tree level for pair production of doubly charged Higgs bosons $e^+ e^- \to H^{++} H^{--}$ in the \texttt{MLRSM} model. For the t-channel the couplings depend on the heavy neutrino masses and $v_R$. The exact form of the coupling is given by Eq.~(\ref{hmvertex}). \label{ee_hcchcc_production_diag}}
\end{center}
\end{figure}

Fixing the scalar mass spectrum lets us take a first look at the production processes   $e^+ e^- \to H_{(1,2)}^{++} H_{(1,2)}^{--}$ and $pp \to H_{(1,2)}^{++} H_{(1,2)}^{--}$  for \texttt{HTM} and \texttt{MLRSM}. This will bring us to the discussion of the importance of the neutrino sector.
Fig.~\ref{ee_hcchcc_production} and Fig.~\ref{ee_hcchcc_production_diag} show classes of Feynman diagrams for the $H^{\pm\pm}$ pair production in $e^+e^-$ collisions in both models. There are s-channel diagrams mediated by neutral gauge bosons $Z$ and $\gamma$ and Higgs bosons, and the t-channel diagram. Due to experimental restrictions discussed in the next two sections, the contributions coming from the s-channel diagrams are comparable to the off-resonance regions, and the resonance regions for the considered center of mass energies and masses lie away from the allowed region of parameters (see Figs.~\ref{epemhtm}, \ref{epemLR}). 
It gives possibility to discuss how the t-channel diagrams in Fig.~\ref{ee_hcchcc_production} and Fig.~\ref{ee_hcchcc_production_diag} affects the process. 

As schematically depicted in the figures the relevant $H^{\pm\pm}-l^{\mp}-l'^{\mp}$ vertices come from Yukawa couplings.  

In \texttt{HTM} the Yukawa term (\ref{e:tripY}) with neutrino fields generates Majorana masses
\begin{equation}
\label{e:yukawa1}
{\mathscr L}_Y^{\rm \Delta} \rightarrow {\mathscr L}_\nu^{\rm \Delta} 
= \frac{1}{2} \: \bar{\nu}_{\ell} \: \frac{v_{\Delta}}{\sqrt{2}} \: \mathcal{Y}_{\ell\ell'} \: \nu_{\ell'}
\equiv \frac{1}{2} \: \bar{\nu}_{\ell} \: (M_\nu)_{_{\ell\ell'}} \: \nu_{\ell'}   \mbox{.}
\end{equation}
This term also contains the $H^{\pm\pm}-l^{\mp}-l'^{\mp}$ vertex which leads to lepton flavor violation. We can diagonalize the neutrino mass matrix by $U$ as follows \cite{Xing:2010}
\begin{equation}
\label{e:yukawa2}
U^\dagger \: M_\nu \: U^* = \frac{1}{2} D_\nu = \frac{1}{2} \: \mbox{diag}\{m_1, m_2, m_3\}.
\end{equation}
The matrix $U$ relates the mass eigenstates $| \nu_i \rangle $ through a superposition of the flavor states $| \nu_\ell \rangle$: $|\nu_i \rangle=~U^T |\nu_\ell \rangle$, so it is directly connected with the \texttt{PMNS} matrix (\ref{e:vupms})
and the exact relation between them is $U^* = V_{P\!M\!N\!S}$. 
Now we can write the Yukawa couplings as a function of the \texttt{PMNS} matrix and the masses of neutrinos.
From Eq.~(\ref{e:yukawa2}), $\mathcal{Y}_{\ell\ell'}$ can be written in the following form 
\begin{equation}
\label{e:yukawa4}
\mathcal{Y}_{\ell\ell'} = \frac{1}{\sqrt{2} v_{\Delta}} \ V_{P\!M\!N\!S}^* \: D_\nu \: V_{P\!M\!N\!S}^\dagger .
\end{equation}
We discuss the parametrization of $V_{\texttt{PMNS}}$ and the employed range of the oscillation parameters in Section \ref{s:hcc_and_neutrino_oscillation}.\\
The $\mathcal{Y}_{\ell\ell'}$ coupling depends on $v_{\Delta}$, neutrino masses and oscillation parameters. From perturbativity, $\mathcal{Y}_{\ell\ell'}^{\phantom{ll'}2} \le 4\pi$. Apart from this restriction, there are stringent limits on $\mathcal{Y}_{\ell\ell'}$ coming from various experimental data discussed in the next section.

In \texttt{MLRSM} the t-channel with the $H^{\pm\pm}-l_i-l_j$ vertex is  inversely proportional to $v_R$. We assume, see section \ref{vertex_LRSM}, vanishing off-diagonal couplings. In this case the vertex is 
\begin{equation}
H^{\pm\pm}-l_i-l_i = \frac{\sqrt{2}}{v_R} M_{N_i}.
\label{hmvertex}
\end{equation}

As we can see, the coupling Eq.~(\ref{e:yukawa4}) in \texttt{HTM} can be enhanced in the case of small values of $v_\Delta \to 0$. However, it is at the same time proportional to the light neutrino masses. The analogous coupling $H^{\pm\pm}-l_i-l_j$ in \texttt{MLRSM} is related to the heavy neutrino masses and $v_R$, which are limited by, e.g. bounds on heavy gauge boson masses, see section \ref{MLRSMconstr}. In the next two sections, we will consider details of the considered models to find the allowed space of the models' physical parameters, including the neutrino sector, which as seen enters the considered processes with very different light and heavy masses.  

In general, it would be tempting to find a way to show when the processes of doubly charged Higgs boson pair production decouple from the neutrino masses. Though such relations are a feature of considered models, if the signals which we predict in both models would not fit experimental data, this would be a sign for another mechanism that takes place. For \texttt{MLRSM} and \texttt{HTM} the basic neutrino mass mechanisms are the seesaw type-I and type-II, respectively.  

\begin{figure}[h!]
\begin{center}
\includegraphics[ width=0.8\textwidth]{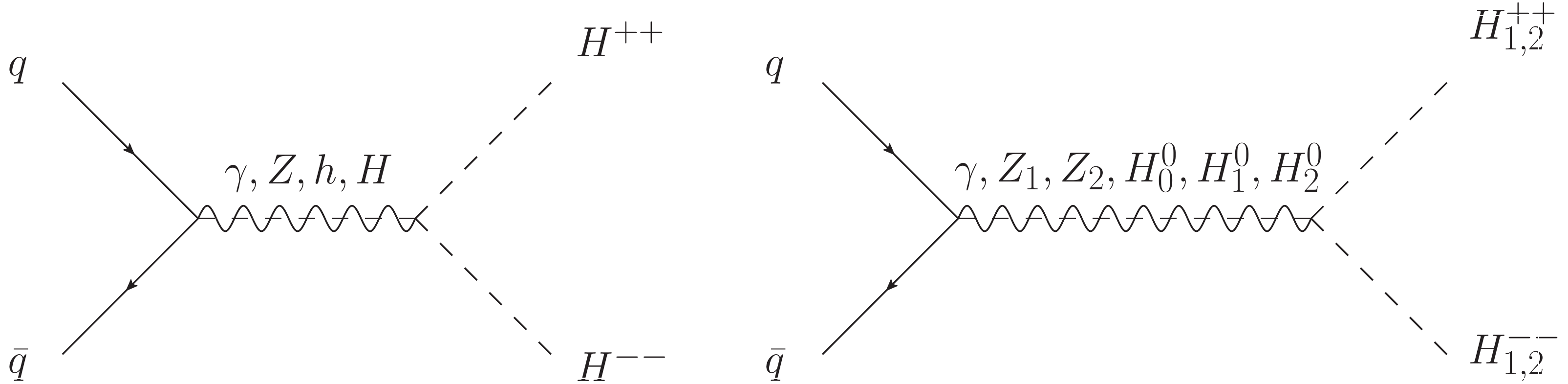}\\
a) \texttt{HTM} \hspace{130pt} b) \texttt{MLRSM} \hspace{10pt}
\caption{\label{fig_pp_hcchcc} Feynman diagrams for the doubly charged scalar particles' pair production in proton-proton colliders within (a) \texttt{HTM}  and (b) \texttt{MLRSM} models.}
\end{center}
\end{figure}

The $H^{\pm\pm}$ pair can be produced in the proton-proton collider via photon, $Z$ boson and neutral scalar particles in the s-channel, see Fig.~\ref{fig_pp_hcchcc}. As will be discussed in section \ref{subseepp}, due to existing experimental constraints, also here the production process is very similar in both models. 
What will bring the difference are doubly charged Higgs boson decays which lead to the four charged lepton final signals.   
To discuss it properly, in the next two sections we will present relevant experimental constraints on the models' parameters.

\section{The \texttt{HTM} model and the relevant experimental parameter constraints}
\label{HTMconstr}

\begin{figure}[h!]
\begin{center}
\includegraphics[width=0.42\textwidth]{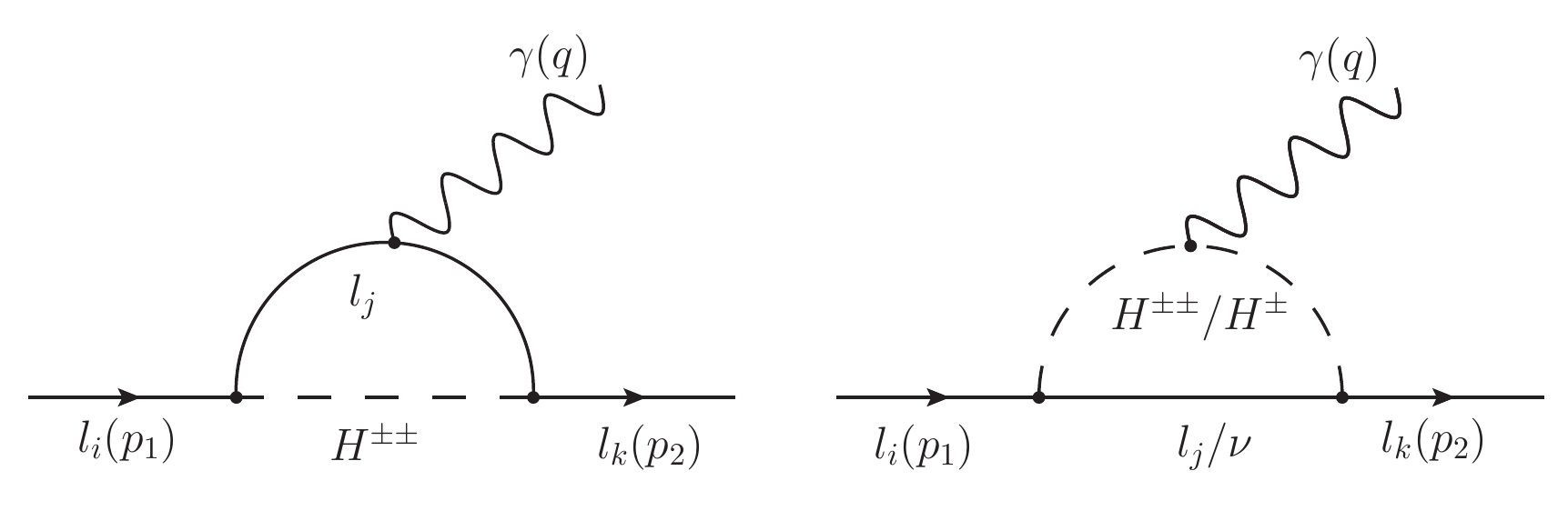}
\includegraphics[width=0.42\textwidth]{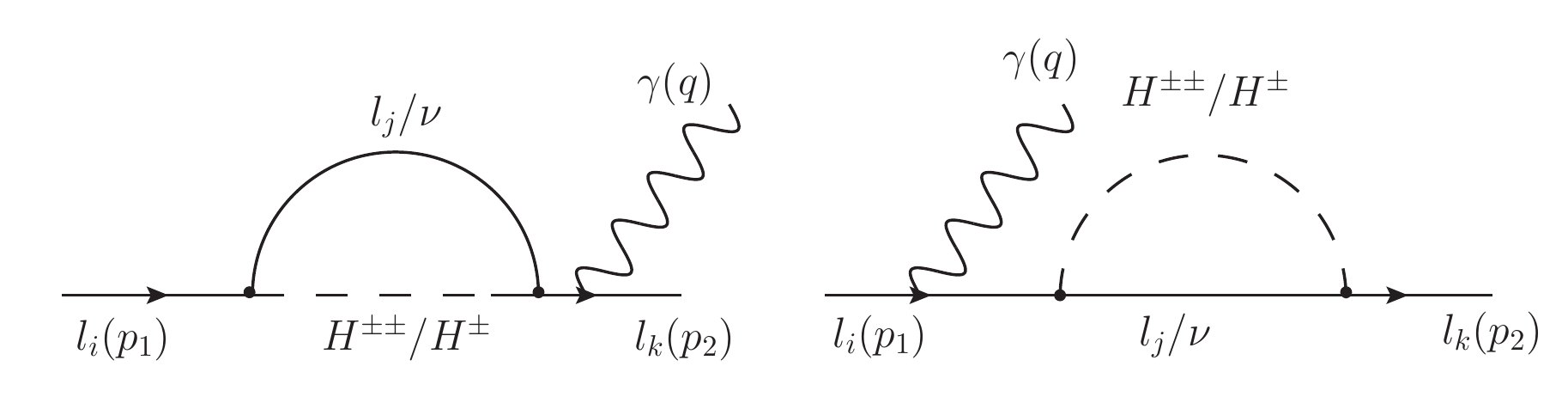}\\
(a) \hspace{180pt}(b)\\
\includegraphics[width=6.8cm,height=2cm]{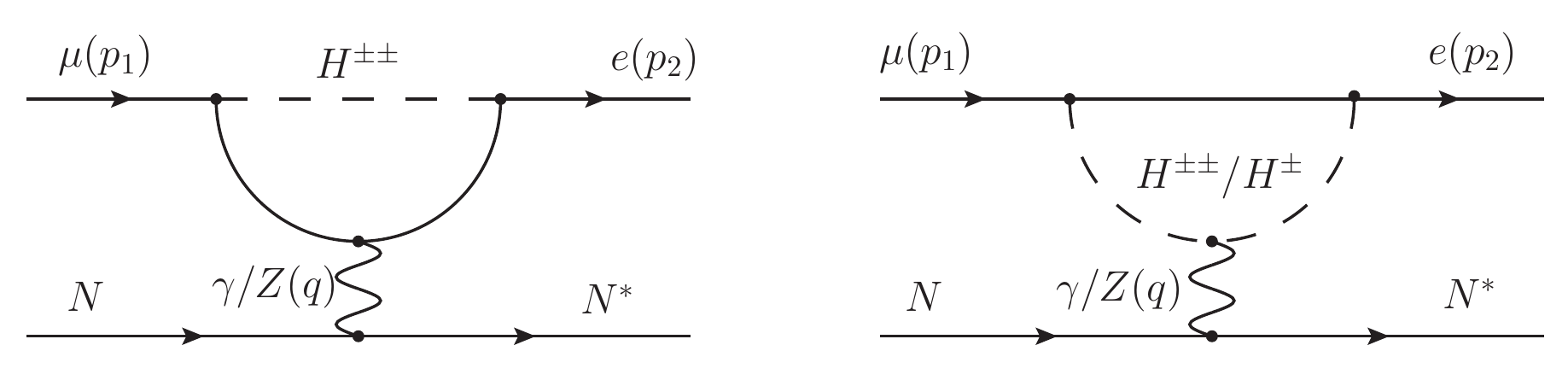}
\includegraphics[width=6.8cm,height=2cm]{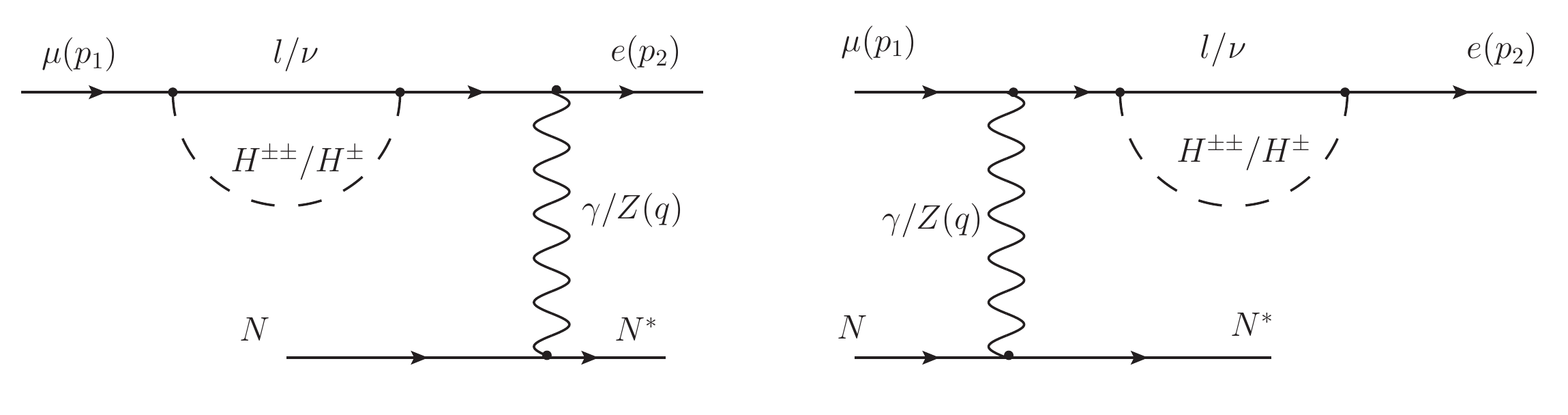}\\
(c)\hspace{180pt}(d)\\
\vspace{15pt}
\includegraphics[width=0.22\textwidth]{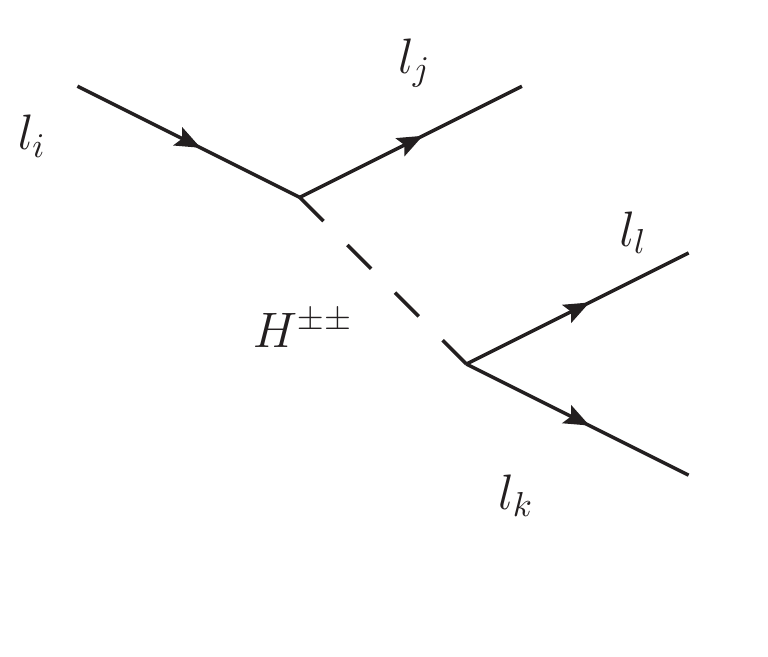}\hspace{70pt}
\includegraphics[width=0.23\textwidth]{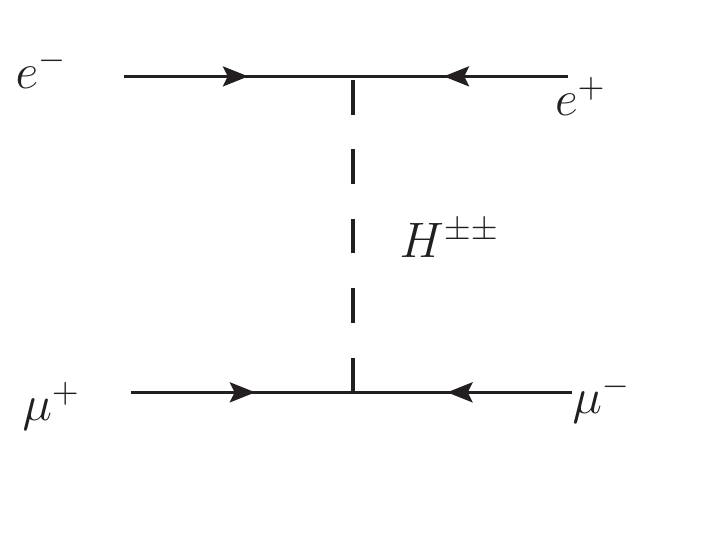}\\
(e)\hspace{180pt}
(f)
\caption{Feynman diagrams representing the contributions to various Lepton flavour violating processes mediated by charged scalar in the \texttt{HTM}. \label{LFV-rad} \label{LFV-tri}\label{LFV-Mu} (a) and (b) are representing the radiative decay $l_i\to l_j\gamma$, (c) and (d) corresponds to $\mu$ to $e$ conversion. Three body decay of lepton contribution is shown by diagram (e) and diagram (f) represents muonium-antimuonium conversion. }
\end{center}
\end{figure}

The Yukawa term (\ref{e:tripY}) includes the  $H^\pm-l-\nu$ and considered in the last section $H^{\pm\pm}-l-l$ vertices. 
They can contribute to several LFV processes like radiative decay of charged leptons
$l_i \rightarrow l_j \gamma$, three body decay of charged leptons $l_i \rightarrow l_j l_k l_l$, $\mu$-to-$e$ 
conversion in nuclei $\mu N \rightarrow e N^*$. We show the contributing diagrams for \texttt{HTM} in Fig.~\ref{LFV-rad}(a)-(e). 
In Fig.~\ref{LFV-rad}(f) we include the muonium-antimuonium conversion $\mu^- e^+ \to \mu^+ e^-$.
Corresponding limits on the $H^{\pm \pm}$ parameters are gathered in Tab.~\ref{limits_table1}.
Tab.~\ref{limits_table2} gathers relevant SM processes: M\o{}ller scattering $e^- e^- \to e^- e^-$, Bhabha scattering $e^+ e^- \to e^+ e^-$, the anomalous magnetic moment of the muon $(g-2)_\mu$ from which also useful limits on  the $H^{\pm \pm}$ parameters are derived. These processes have been discussed in the context of \texttt{HTM} in many works \cite{Ma:2000xh,Chun:2003ej,Fukuyama:2009xk,Akeroyd:2009nu,Dinh:2012bp,Chakrabortty:2012vp, Crivellin:2018ahj,Chakrabarty:2018qtt}). 
The branching ratios (BR) in Table~\ref{limits_table1} depend on the charged scalar masses and the Yukawa couplings $\mathcal{Y}_{\ell\ell'}$, defining the allowed space of mass and coupling parameters for charged scalars. 

The radiative decays $l_i\to l_j \gamma$ and the $\mu$-to-$e$ conversion process   mediated by doubly and singly charged scalar bosons originate from the following Lagrangian~\cite{Dinh:2013tvc}
\begin{eqnarray}
\mathcal{L}\subset -\frac{4 e G_F}{\sqrt{2}} m_l A_R(q^2)~ \bar{l}^{'} \sigma^{\nu\mu} P_R~ l~ F_{\mu\nu}-\frac{e^2 G_F}{\sqrt{2}} A_L(q^2)~\bar{l}^{'}\gamma^\nu P_L l\sum_{q=u,d}q_Q \bar{Q}\gamma_\nu Q +h.c.
\label{eq:lagran-formfac}
\end{eqnarray}
Branching ratios depend on the form factors $A_L$ and $A_R$, which actual form depends on Higgs scalar contributions to the considered processes.
 For the doubly charged scalar there are four relevant diagrams as shown in Fig.~\ref{LFV-rad} (a) and (b).
 The amplitude for $H^{\pm \pm}$ for the first two diagrams Fig.~\ref{LFV-rad}~(a), at the leading order of the doubly charged scalar mass is
\begin{eqnarray}
\mathcal{M}_{M_{H^{\pm\pm}}}^I &\subset & -\frac{(\mathcal{Y}^*)_{ei} (\mathcal{Y})_{\mu i}\gamma_\mu P_L}{128\pi^2}\Big(\frac{2}{\epsilon}+\log\frac{4\pi\mu^2}{M_{H^{\pm\pm}}^2}\Big)\nonumber\\&&
+\frac{(\mathcal{Y}^*)_{ei} (\mathcal{Y})_{\mu i}\gamma_\mu P_L}{64\pi^2}\Big(-\frac{1}{4}-\frac{r}{36}+\frac{s_i}{2}+\frac{r}{6}f(r,s_i)\Big)\nonumber\\&& 
+\frac{(\mathcal{Y}^*)_{ei} (\mathcal{Y})_{\mu i} P_L}{384\pi^2 M_{H^{\pm\pm}}^2}\Big[\Big(-\frac{5}{6}+{f(r,s_i)}\Big)(\slashed{p}_1\gamma_\mu \slashed{p_1}+\slashed{p_2}\gamma_\mu \slashed{p}_2)   \nonumber \\&& 
+\Big(\frac{1}{6}+{f(r,s_i)}\Big)(\slashed{p}_1\gamma_\mu \slashed{p}_2)+\Big(\frac{17}{6}-{f(r,s_i)}\Big)(\slashed{p}_2\gamma_\mu \slashed{p}_1)\Big]\nonumber \\&&-\frac{(\mathcal{Y}^*)_{ei} (\mathcal{Y})_{\mu i} P_R}{1152\pi^2M_{H^{\pm\pm}}^2}\Big({\slashed{p}_1 p_{1\mu}}+{5\slashed{p}_1 p_{2\mu}}+5{\slashed{p}_2 p_{1\mu}}+{\slashed{p}_2 p_{2\mu}}\Big),
\label{eq:mutoedc1}
\end{eqnarray}
where
$f(r,s_i)=\frac{4 s_i}{r}+\log(s_i)+\Big(1-\frac{2 s_i}{r}\Big) \sqrt{1+\frac{4s_i}{r}}\log \Big(\frac{\sqrt{r}+\sqrt{r+4s_i}}{\sqrt{r}-\sqrt{r+4s_i}}\Big)$,  $r=\frac{-q^2}{m_{H^{\pm\pm}}^2}\mbox{,}   s_i= \frac{m_i^2}{m_{H^{\pm\pm}}^2}$. \\ \noindent $\mu$ is a mass parameter introduced in dimensional regularization, $\epsilon=4-D$ and $D$ is dimension. \\

\noindent The contribution from other two diagrams Fig.~\ref{LFV-rad}(b) mediated by the doubly charged scalar boson is
\begin{eqnarray}
\mathcal{M}_{M_{H^{\pm\pm}}}^{II} &\subset& \frac{(\mathcal{Y}^*)_{ei} (\mathcal{Y})_{\mu i}\gamma_\mu P_L}{128\pi^2}\Big(\frac{2}{\epsilon}+\log\frac{4\pi\mu^2}{M_{H^{\pm\pm}}^2}\Big)\nonumber\\&&+\frac{(\mathcal{Y}^*)_{ei} (\mathcal{Y})_{\mu i}\gamma_\mu P_L}{128\pi^2}\Big(\frac{1}{2}+\frac{s_e+s_\mu}{3}-{s_i}\Big)\nonumber\\&&+\frac{(\mathcal{Y}^*)_{ei} (\mathcal{Y})_{\mu i} P_R}{384\pi^2M_{H^{\pm\pm}}^2}\Big({\slashed{p}_2\gamma_\mu \slashed{p}_1}\Big).\label{eq:mutoedc2}
\end{eqnarray}
By adding \eqref{eq:mutoedc1} and \eqref{eq:mutoedc2} we can see that the final contribution is finite and after doing some algebra the contribution of the doubly charged scalar form factors can be written in a compact way
\begin{eqnarray}
\mathcal{M}_{M_{H^{\pm\pm}}} &\subset& -\frac{(\mathcal{Y}^*)_{ei} (\mathcal{Y})_{\mu i} }{192\pi^2 M_{H^{\pm\pm}}^2}f(r,s_i)(q^2 \gamma_\mu-q_\mu q_\nu \gamma^\nu)P_L-
\nonumber \\&&\frac{(\mathcal{Y}^*)_{ei} (\mathcal{Y})_{\mu i}}{192\pi^2 M_{H^{\pm\pm}}^2}(m_e P_L i\sigma_{\mu\nu}q^\nu+m_\mu P_R i\sigma_{\mu\nu}q^\nu).
\end{eqnarray}
In a similar way one can compute the contributions from diagrams mediated by singly charged scalar bosons and the total amplitude in  \texttt{HTM} can be written as 
\begin{eqnarray}
\mathcal{M}_{HTM} &\subset &-\frac{(\mathcal{Y}^*)_{ei} (\mathcal{Y})_{\mu i} }{192\pi^2 }\Big(\frac{1}{12 M_{H^{\pm}}^2 }+\frac{f(r,s_i)}{M_{H^{\pm\pm}}^2}\Big)(q^2 \gamma_\mu-q_\mu q_\nu \gamma^\nu)P_L-
\nonumber \\&&\frac{(\mathcal{Y}^*)_{ei} (\mathcal{Y})_{\mu i}}{192\pi^2 }\Big(\frac{1}{8 M_{H^{\pm}}^2}+\frac{1}{M_{H^{\pm\pm}}^2}\Big)(m_e P_L i\sigma_{\mu\nu}q^\nu+m_\mu P_R i\sigma_{\mu\nu}q^\nu).
\label{eq:htm_formfac}
\end{eqnarray}
Matching Eq.~\eqref{eq:htm_formfac} to Eq.~\eqref{eq:lagran-formfac},
we can extract the form of the $A_L(A_R)$ form factors and compute the analytic formulas for the radiative decays and $\mu$-to-$e$ conversion processes.  The final analytic formula for considered LFV processes are gathered in the~Appendix.

If the massive neutrinos couple to leptons and are of Majorana type, the lepton number can be violated by two units, $\Delta L =2$. This leads to the neutrinoless double beta decay $\beta\beta 0 \nu$  process   \cite{Schechter:1981bd,Wolfenstein:1982bf} and as it has not been observed so far, it puts a constraint on the model parameters. This process has been analyzed within \texttt{HTM} in \cite{Petcov:2009zr}, the non-standard contribution is negligibly small.   

Above we have discussed LFV processes which have not been observed so far, leading to stringent bounds on BSM physics and parameters. Useful information about limits on the BSM physics can be also obtained by exploring observed SM processes and analysing experimental results and SM predictions. One of suchfinite processes is the Bhabha scattering present in electron-positron collisions. It serves as a calibration method for colliders as it is a QED dominated t-channel process, see Section~II and Figs.~1-2 in \cite{Actis:2008br}. The LEP data \cite{Achard:2003mv} sets a lower limit on $H^{\pm\pm}$, namely $\mathcal{Y}_{\ell\ell'} \ge 10^{-7}$ (to ensure $H^{\pm\pm}$ decay before entering the detector). 

Another SM experiment which seems to provide a promising signature of the BSM physics is the observed excess in the anomalous magnetic moment of the muon $(g-2)_\mu$. There is a lasting discrepancy of more than $3\sigma$ in the measurement of $(g-2)_\mu$  with the corresponding SM value \cite{Campanario:2019mjh}. At present the deviation, as given by PDG, is  \cite{PhysRevD.98.030001}:
\begin{equation}
\Delta a_{(g-2)_\mu} \equiv a_\mu^{\rm exp} - a_\mu^{\rm SM} =  268(63)(43)  \times 10^{-11}.
\end{equation}
The experimental limits for Bhabha, M\o{}ller and $(g-2)_\mu$ SM processes are collected in Tab.~\ref{limits_table2}. 
Charged scalars can contribute to the $(g-2)_\mu$ at the one-loop level. There are many studies of BSM contribution to $(g-2)_\mu$ in the literature.
The contribution from a doubly charged Higgs boson to $(g-2)_\mu$ is discussed in \cite{Chakrabortty:2015zpm} and in the context of \texttt{HTM} in \cite{Fukuyama:2009xk}. 
The diagrams mediated by singly and doubly charged scalars contributions to $(g-2)_\mu$ are given by Fig. \ref{LFV-rad} (a) and (b) where both $l_i, l_j$ are $\mu$ (muons). Contributions of singly and doubly charged scalar bosons  to $(g-2)_\mu$ 
amounts a negative number  \cite{Lindner:2016bgg} and 
$(g-2)_\mu$ anomaly is hard to explain by $H^{\pm \pm}$. However, it is worth mentioning that the observed anomaly is an open problem as still some discrepancies among different low-energy experiments exist \cite{Campanario:2019mjh}.
\begin{table}[h!]
\begin{center}
\begin{tabular}{c@{\hskip 0.5in}c@{\hskip 0.5in}r@{}c@{}lcr@{}c@{}lc}
\hline 
\hline
 & Process & \multicolumn{4}{c}{Present limits} & \multicolumn{4}{c}{Future limits}\\
\hline
 & & & & & & & & & \\
\parbox[t]{2mm}{\multirow{17}{*}{\rotatebox[origin=c]{90}{{\large LFV processes }}}}
& BR($\mu \rightarrow e \gamma$) 		&{\hskip 0.5in} 4.2 & $\times$ & $10^{-13}$ & \cite{TheMEG:2016wtm} 
						& 6.0 & $\times$ & $10^{-14}$  & \cite{Baldini:2013ke} \\
& BR($\tau \rightarrow e \gamma$) 	& 3.3 & $\times$ & $10^{-8}$  & \cite{Aubert:2009ag} 
						& 1.0 & $\times$ & $10^{-8}$   & \cite{Wang:2015hdf} \\

& BR($\tau \rightarrow \mu \gamma$) 	& 4.4 & $\times$ & $10^{-8}$  & \cite{Aubert:2009ag} 
						& 3.0 & $\times$ & $10^{-9}$   & \cite{Aushev:2010bq} \\

& & & & & & & & & \\
& BR($\mu \rightarrow e e e$)		& 1.0 & $\times$ & $10^{-12}$ & \cite{Bellgardt:1987du} 
						& \multicolumn{3}{c}{$\sim 10^{-16}$} & \cite{Blondel:2013ia}\\
& BR($\tau \rightarrow e e e$) 		& 2.7 & $\times$ & $10^{-8}$  & \cite{Hayasaka:2010np} 
						& 5.0 & $\times$ & $10^{-10}$   & \cite{Wang:2015hdf} \\
& BR($\tau \rightarrow \mu \mu \mu$) 		& 2.1 & $\times$ & $10^{-8}$  & \cite{Hayasaka:2010np} 
						& 4.0 & $\times$ & $10^{-10}$   & \cite{Wang:2015hdf} \\
& BR($\tau^- \rightarrow \mu^+ e^- \mu^-$) & 2.7 & $\times$ & $10^{-8}$ & \cite{Hayasaka:2010np} 
						& 5.0 & $\times$ & $10^{-10}$ & \cite{Wang:2015hdf} \\
& BR($\tau^- \rightarrow \mu^+ e^- e^- $)  & 1.5 & $\times$ & $10^{-8}$ & \cite{Hayasaka:2010np}  
						& 3.0 & $\times$ & $10^{-10}$ & \cite{Wang:2015hdf} \\
& BR($\tau^- \rightarrow e^+ \mu^- \mu^-$) & 1.7 & $\times$ & $10^{-8}$ & \cite{Hayasaka:2010np}  
						& 3.0 & $\times$ & $10^{-10}$ & \cite{Wang:2015hdf} \\
& BR($\tau^- \rightarrow e^+ e^- \mu^- $)  & 1.8 & $\times$ & $10^{-8}$ & \cite{Hayasaka:2010np}  
						& 3.0 & $\times$ & $10^{-10}$ & \cite{Wang:2015hdf} \\
& & & & & & & & & \\
& R($\mu N \rightarrow e N^*$)
				& 
		\multicolumn{3}{@{}c@{}}{
			\begin{tabular}{@{}c@{}c@{}c@{}}
				7.0 & $\times$ & $10^{-13}$\\
				\multicolumn{3}{@{}l@{}}{ {\small (for Au)} }\\
			\end{tabular}
			}  & \cite{SINDRUMII:2006} 
						& 
		\multicolumn{3}{@{}c@{}}{
			\begin{tabular}{@{}c@{}c@{}c@{}}
				2.87 & $\times$ & $10^{-17}$\\
				\multicolumn{3}{l}{ {\small (for Al)} }\\
			\end{tabular}
			} & \cite{Bartoszek:2014mya} \\
& & & & & & & & & \\
 & $\mu^+ e^- \: \to \: \mu^- e^+$ &
		\multicolumn{3}{@{}c@{}}{ $ \sqrt{\mathcal{Y}_{ee}\cdot\mathcal{Y}_{\mu\mu}} < \frac{0.44 \: \cdot \: M_{H^{\pm\pm}}}{10^3 \mbox{{\tiny GeV}}}  $ }  & \cite{Maalampi:2002vx} & & \\
& & & & & & & & & \\
\hline
& & & & & & & & & \\
\end{tabular}
\end{center}
\caption{Current and future limits on the processes with doubly charged scalar contributions, LFV processes (90\% CL). \label{limits_table1}}
\end{table}
\begin{table}[h!]
\begin{center}
\begin{tabular}{c@{\hskip 0.5in}c@{\hskip 0.5in}r@{}c@{}lcr@{}c@{}lc}
\hline  
\hline
& & & & & & & & & \\
 & Process & \multicolumn{4}{c}{Present limits} & \multicolumn{4}{c}{}\\
 \parbox[t]{6mm}{\multirow{17}{*}{\rotatebox[origin=c]{90}{{\large SM processes}}}}
\\
\hline
 & & & & & & & & & \\
& 			& \multicolumn{3}{l}{$|\mathcal{Y}_{ee}| \le 
						\frac{\sqrt{4\pi} \: M_{H^{\pm\pm}}}{8.7 \times 10^3 \: \mbox{{\tiny GeV}}}$} &
						\cite{Nomura:2017abh}
						&     &		 &	       & \\
& & & & & & & & & \\
& $e^+ e^- \to l^+ l^-$ & \multicolumn{3}{l}{$|\mathcal{Y}_{e\mu}| \le 
						\frac{1}{\sqrt{2}} \frac{\sqrt{4\pi} \: M_{H^{\pm\pm}}}{12.2 \times 10^3 \: \mbox{{\tiny GeV}}}$} &
						\cite{Nomura:2017abh}
						&     &		 &	       & \\
& {\footnotesize (LEP)} & & & & & & & & \\
& 			& \multicolumn{3}{l}{$|\mathcal{Y}_{e\tau}| \le 
						\frac{1}{\sqrt{2}} \frac{\sqrt{4\pi} \: M_{H^{\pm\pm}}}{9.1 \times 10^3 \: \mbox{{\tiny GeV}}}$} &
						\cite{Nomura:2017abh}
						&     &		 &	       & \\
& & & & & & & & & \\
& & & & & & & & & \\
& $e^- e^- \to e^- e^-$	& \multicolumn{3}{l}{$|\mathcal{Y}_{ee}| \le 
						\frac{M_{H^{\pm\pm}}}{3.7 \times 10^3 \: \mbox{{\tiny GeV}}}$} &
						\cite{Dev:2018sel}
						&     &		 &	       & \\ 
& {\footnotesize (M\O{}LLER)} & & & & & & & & \\	
& & & & & & & & & \\
 & $(g-2)_\mu$ & \multicolumn{3}{@{}c@{}}{ $\Delta a_\mu = (29.3 \pm 9.0) \times 10^{-10}$} & \cite{NYFFELER:2014pta} & & \\
& & & & & & & & & \\
\hline
\end{tabular}
\end{center}
\caption{Current limits on the Standard Model processes with doubly charged scalar contributions (95\% CL). \label{limits_table2}}
\end{table}

\subsection{Neutrino mixing matrix and mass hierarchies within \texttt{HTM}}
\label{s:hcc_and_neutrino_oscillation}

From Eq.~(\ref{e:yukawa4}) we can see that the $H^{\pm\pm}-l-l'$ couplings depend on the neutrino oscillation 
parameters, neutrinos hierarchy and the lightest neutrino mass. Details of studies for the \texttt{HTM} model are thus very sensitive to the neutrino oscillation data, as discussed already in~\cite{Chun:2003ej} and \cite{Das:2016,Cai:2017mow,Fuks:2019clu}.   
In our analysis the following, standard parametrization of the $V_{\texttt{PMNS}}$
matrix is used:
\\
\begin{equation}
\label{e:vupms}
V_{\texttt{PMNS}} = \left[
\begin{array}{ccc}
c_{12} c_{13} e^{i\alpha_1} & s_{12} c_{13} e^{i\alpha_2} & s_{13} e^{-i\delta_{CP}} \\
(-s_{12} c_{23} - c_{12} s_{23} s_{13} e^{i\delta_{CP}})e^{i\alpha_1} & 
(c_{12} c_{23} - s_{12} s_{23} s_{13} e^{i\delta_{CP}})e^{i\alpha_2} & s_{23} c_{13} \\
(s_{12} s_{23} - c_{12} c_{23} s_{13} e^{i\delta_{CP}})e^{i\alpha_1}  &
(-c_{12} s_{23} - s_{12} c_{23} s_{13} e^{i\delta_{CP}})e^{i\alpha_2} & c_{23} c_{13} \\
\end{array}
\right],
\end{equation}
\\
where $s_{ij}$ and $c_{ij}$ denotes $\sin({\theta_{ij}})$ and $\cos({\theta_{ij}})$, respectively. Tab.~\ref{neutrino_data} shows global neutrino fits at the $2\sigma$ C.L. for neutrino parameters which are used in present analysis for two mass
orderings, defined as:
\begin{equation}
\begin{array}{ccc}
\mbox{Normal mass hierarchy:} & \mbox{Inverted mass hierarchy:} \\
& \\
\begin{array}{l}
m_{\nu_1} = m_{\nu_0}, \\
m_{\nu_2} = \sqrt{m_{\nu_0}^2 + \Delta m_{21}^2}, \\
m_{\nu_3} = \sqrt{m_{\nu_0}^2 + \Delta m_{31}^2}, \\
\end{array}
&
\begin{array}{l}
m_{\nu_1} = \sqrt{m_{\nu_0}^2 - \Delta m_{21}^2 - \Delta m_{32}^2}, \\
m_{\nu_2} = \sqrt{m_{\nu_0}^2 - \Delta m_{32}^2}, \\
m_{\nu_3} = m_{\nu_0}, 
\end{array}
\end{array}
\label{e:NH_and_IH}
\end{equation}
where $\Delta m_{ij}^2 = m_i^2-m_j^2$.
\begin{table}[H]
\begin{center}
\begin{tabular}{|c|c|c|c|c|c|c|c|c|}
\hline 
 & \multicolumn{4}{|c|}{Normal hierarchy (NH)} & \multicolumn{4}{c|}{Inverted hierarchy (IH)} \\
\cline{2-9}
 & \begin{tabular}{@{}c@{}} Best \\ fit (bf): \\ \end{tabular} 
	& $\sigma$ & bf$\pm 1\sigma$ &  bf$\pm 2\sigma$ 
	& \begin{tabular}{@{}c@{}} Best \\ fit (bf): \\ \end{tabular}  & $\sigma$ & bf$\pm 1\sigma$ &  bf$\pm 2\sigma$\\
\hline
$\sin^2{\theta_{12}}$ & 0.310 & \begin{tabular}{@{}c@{}} $+0.013$ \\ $-0.012$ \\ \end{tabular} 
	& \begin{tabular}{@{}c@{}} 0.298 \\ $\div$ \\ 0.323 \\ \end{tabular} 
	& \begin{tabular}{@{}c@{}} 0.286 \\ $\div$ \\ 0.336 \\ \end{tabular}
	& 0.310 & \begin{tabular}{@{}c@{}} $+0.013$ \\ $-0.012$ \\ \end{tabular} 
	& \begin{tabular}{@{}c@{}} 0.298 \\ $\div$ \\ 0.323 \\ \end{tabular} 
	& \begin{tabular}{@{}c@{}} 0.286 \\ $\div$ \\ 0.336 \\ \end{tabular} \\
\hline
$\sin^2{\theta_{23}}$ & 0.558 & \begin{tabular}{@{}c@{}} $+0.020$ \\ $-0.033$ \\ \end{tabular} 
	& \begin{tabular}{@{}c@{}} 0.525 \\ $\div$ \\ 0.578 \\ \end{tabular} 
	& \begin{tabular}{@{}c@{}} 0.492 \\ $\div$ \\ 0.598 \\ \end{tabular}
	& 0.563 & \begin{tabular}{@{}c@{}} $+0.019$ \\ $-0.026$ \\ \end{tabular} 
	& \begin{tabular}{@{}c@{}} 0.537 \\ $\div$ \\ 0.582 \\ \end{tabular}
	& \begin{tabular}{@{}c@{}} 0.511 \\ $\div$ \\ 0.601 \\ \end{tabular} \\
\hline
$\sin^2{\theta_{13}}$ & 0.02241 & \begin{tabular}{@{}c@{}} $+0.00066$ \\ $-0.00065$ \\ \end{tabular} 
	& \begin{tabular}{@{}c@{}} 0.02176 \\ $\div$ \\ 0.02307 \\ \end{tabular} 
	& \begin{tabular}{@{}c@{}} 0.02111 \\ $\div$ \\ 0.02373 \\ \end{tabular} &
	0.02261 & \begin{tabular}{@{}c@{}} $+0.00067$ \\ $-0.00064$ \\ \end{tabular} 
	& \begin{tabular}{@{}c@{}} 0.02197 \\ $\div$ \\ 0.02328 \\ \end{tabular} 
	& \begin{tabular}{@{}c@{}} 0.02133 \\ $\div$ \\ 0.02395 \\ \end{tabular} \\
\hline
$\delta_{CP}[^o]$ & 222 & \begin{tabular}{@{}c@{}} $+38$ \\ $-28$ \\ \end{tabular} 
	& \begin{tabular}{@{}c@{}} 194 \\ $\div$ \\ 260 \\ \end{tabular} 
	& \begin{tabular}{@{}c@{}} 166 \\ $\div$ \\ 298 \\ \end{tabular} &
	285 & \begin{tabular}{@{}c@{}} $+24$ \\ $-26$ \\ \end{tabular} 
	& \begin{tabular}{@{}c@{}} 259 \\ $\div$ \\ 309 \\ \end{tabular} 
	& \begin{tabular}{@{}c@{}} 233 \\ $\div$ \\ 333 \\ \end{tabular} \\
\hline
\hline
$\frac{\Delta m^2_{21}}{10^{-5}~\mbox{eV}^2}$ & 7.39 & \begin{tabular}{@{}c@{}} $+0.21$ \\ $-0.20$ \\ \end{tabular} 
	& \begin{tabular}{@{}c@{}} 7.19 \\ $\div$ \\ 7.60 \\ \end{tabular} 
	& \begin{tabular}{@{}c@{}} 6.99 \\ $\div$ \\ 7.81 \\ \end{tabular}
	& 7.39 & \begin{tabular}{@{}c@{}} $+0.21$ \\ $-0.20$ \\ \end{tabular} 
	& \begin{tabular}{@{}c@{}} 7.19 \\ $\div$ \\ 7.60 \\ \end{tabular} 
	& \begin{tabular}{@{}c@{}} 6.99 \\ $\div$ \\ 7.81 \\ \end{tabular} \\
\hline
$\frac{\Delta m^2_{3l}}{10^{-3}~\mbox{eV}^2}$ & +2.523 & \begin{tabular}{@{}c@{}} $+0.032$ \\ $-0.030$ \\ \end{tabular} 
	& \begin{tabular}{@{}c@{}} 2.463 \\ $\div$ \\ 2.527 \\ \end{tabular} 
	& \begin{tabular}{@{}c@{}} 2.463 \\ $\div$ \\ 2.587 \\ \end{tabular}
	& -2.509 & \begin{tabular}{@{}c@{}} $+0.032$ \\ $-0.030$ \\ \end{tabular} 
	& \begin{tabular}{@{}c@{}} -2.539 \\ $\div$ \\ -2.477 \\ \end{tabular}
	& \begin{tabular}{@{}c@{}} -2.569 \\ $\div$ \\ -2.445 \\ \end{tabular} \\
\hline
\end{tabular}
\end{center}
\caption{Neutrino oscillation data, notations as in \label{neutrino_data} \cite{NuFIT}. 
$\Delta m_{ij}^2 = m_i^2 - m_j^2$. Depending on the hierarchy, for atmospheric nutrino oscillations either 
$\Delta m_{3l}^2 = \Delta m_{31}^2 > 0$ (NH) or \mbox{$\Delta m_{3l}^2 = \Delta m_{32}^2 < 0$~(IH)}. }
\end{table}

Concerning the Dirac CP-phase $\delta_{CP}$, the global fits indicate preference for its non-zero values. Recent T2K results confirm this tendency and considered by us $2\sigma$ range of the Dirac phase covers well the best fit values given in \cite{Abe:2019vii}. In analysis we are chosing $\delta_{CP}$ data as given in Tab.\ref{neutrino_data}.
There is no direct limit on the Majorana phases $\alpha_1$, $\alpha_2$. However, in some studies there are predictions using the neutrinoless double beta decay, e.g., see \cite{Girardi:2016zwz}. 
There is no bound on the individual masses of neutrinos from the oscillation data. Therefore, the lightest neutrino mass $m_{\nu_0}$ is a free parameter and other two masses are determined through (\ref{e:NH_and_IH}). Also, there are limits on the sum of three neutrino masses from different experiments: from the tritium decay \cite{Olive:2016xmw} or neutrinoless double beta decay~\cite{Rodejohann:2011mu}, the sharpest limit comes from astrophysics and cosmology \cite{Ade:2013zuv}
\begin{equation}
\Sigma \equiv \sum_{i=i}^3 m_{\nu_i} \leq 0.23 \; \mbox{eV}.
\label{astrolimit}
\end{equation}
These limits set the upper bound on the lightest neutrino mass \cite{Barger:1999na,Czakon:2001uh}, present experimental data gives

\begin{eqnarray}
m_{\nu_0} =\left\{ \begin{array}{cc}
0.071\;\rm{eV},     &  \rm{NH}, \\
0.066\;\rm{eV},     & \rm{IH}. 
\end{array} \right.
\label{nu0limit}
\end{eqnarray}

\subsection{The triplet VEV ${v_\Delta}$ and the $\mathbf{\rho}$-parameter
}
\label{s:constraints_low_energy}

\label{rho_parameter_subsection}

As we mentioned in the introduction, the additional scalar triplet contributes to the $\rho$ parameter. It can be defined  either through a relation among massive SM gauge bosons $Z$ and $W$ and Weinberg mixing angle or relations among gauge couplings \cite{Czakon:1999ha}. In \texttt{HTM}, at the tree level, $\rho$ can be written as ~\cite{PhysRevD.21.1404}:
\begin{eqnarray}
\label{rho_parameter}
\rho = \frac{1+2\frac{v_\Delta^2}{v_\Phi^2}}{1+4\frac{v_\Delta^2}{v_\Phi^2}} \,.
\end{eqnarray}
The experimental limit on the $\rho$ parameter  \cite{PhysRevD.77.095009}:
\begin{eqnarray}
\label{rho_experiment}
\rho^{exp} = 1.00037 \pm 0.00023,
\end{eqnarray}
put the upper bound on the triplet VEV $v_\Delta$. 

Taking $\sqrt{v_\Phi^2 +2 v_\Delta^2}=v=(\sqrt{2}G_F)^{-\frac{1}{2}}$~\cite{Barger:2003rs,Kanemura:2004mg} where $G_F$ is Fermi coupling constant\\ $1.1663787(6)$~$\times$~$10^{-5}$~$\mbox{GeV}^{-2}$~\cite{Tanabashi:2018oca},~we~get 
\begin{equation}
v_\Delta \leq 1.7\;{\mbox{GeV}}, \label{gm2limit}
\end{equation}
for $\rho^{exp}$, within $2\sigma$ deviations. Let us note that the limit on $v_\Delta$ can not be obtained for $\rho^{exp}$ within $2\sigma$ deviation. It is connected with the fact that relation (\ref{rho_experiment}) has sense only for $\rho^{exp}\leq 1$, otherwise $v_\Delta$ comes out to be a complex number.
We will see in the following section that other low experimental data are more important, lowering down the scale of $v_\Delta $ in an unambiguous way to the (sub)electronvolt level.

\subsection{Relation between ${v_\Delta}$ and doubly charged scalar particles parameters in the light of low and high energy experimental limits \label{htmlow}}

\begin{figure}[h]
\begin{center}
\begin{tabular}{cc}
\includegraphics[width=0.5\textwidth,angle=0]{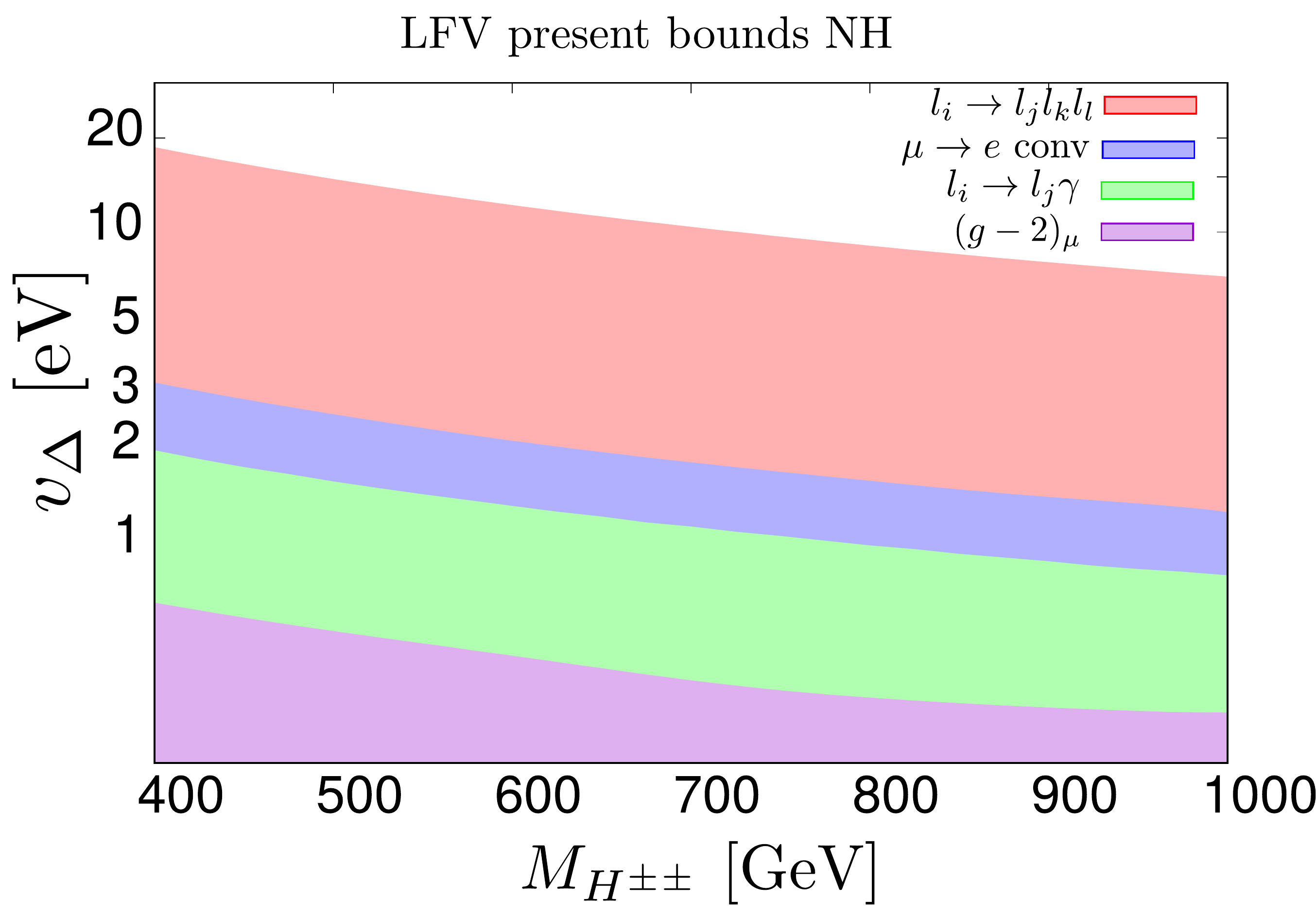} &
\includegraphics[width=0.5\textwidth,angle=0]{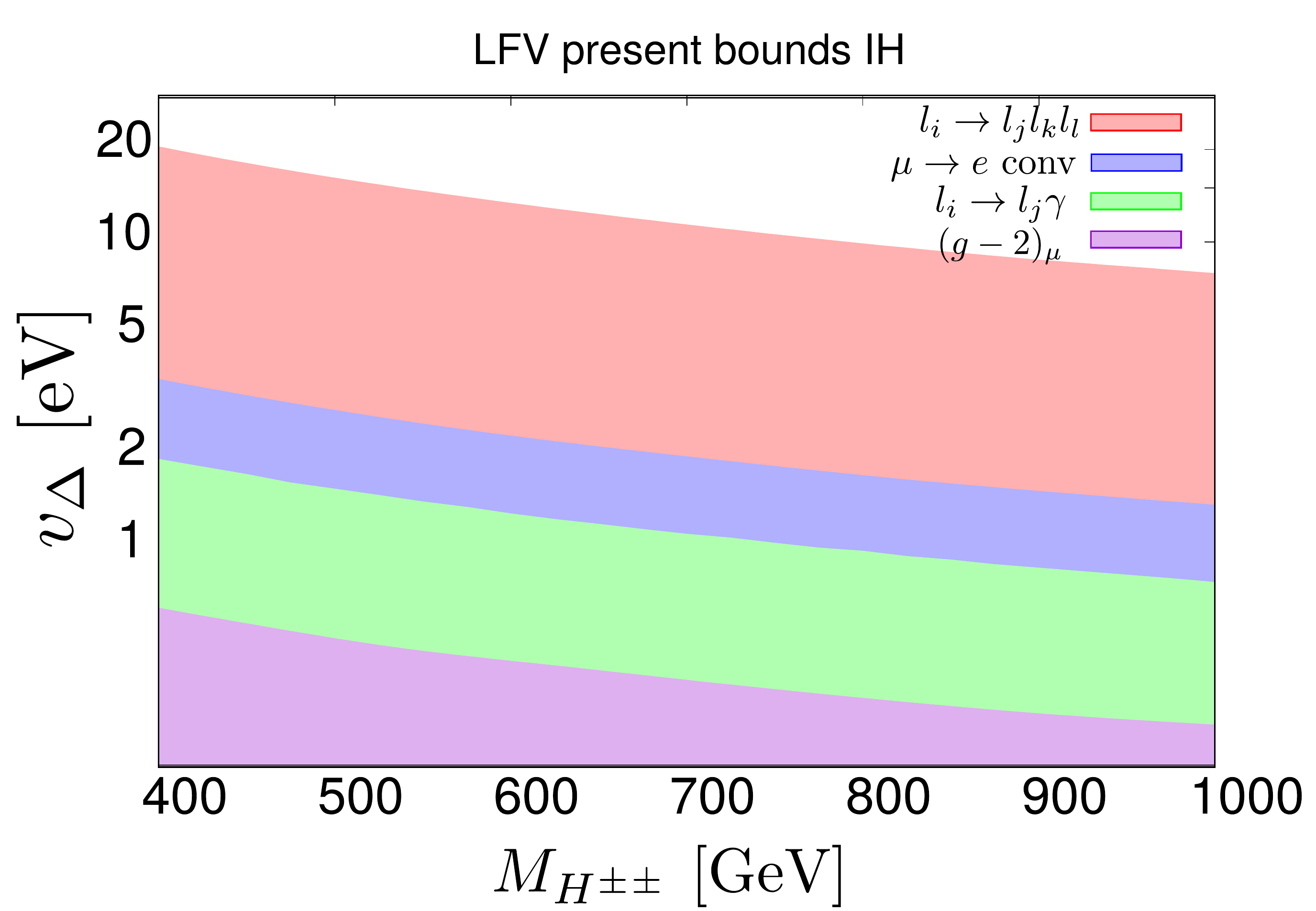}\\
 (a) & (b) \\
\end{tabular}
\caption{\label{current_future_limits} Plots for $v_\Delta$ vs $m_{H^{\pm\pm}}$ using normal and inverted hierarchy data. Shaded regions correspond to the exclusion limits coming from LFV bounds for current data and future sensitivity expectations. The neutrino oscillation data are taken in the $2\sigma$ range. In general, precision of future experiments (see Tab.~\ref{limits_table1}) will allow to get one order of magnitude better limits on $v_\Delta$.}
 \end{center}
\end{figure}

In this section, we analyze bounds on the triplet VEV $v_\Delta$ from low and high energy experiments discussed earlier (see Tables~\ref{limits_table1}~and~\ref{limits_table2}). Fig.~\ref{current_future_limits} shows excluded regions in the plane of $v_\Delta$ and $M_{H^{\pm\pm}}$ parameters' space based on current  limits  on branching ratios (for both NH and IH scenarios) for various LFV processes and $(g-2)_\mu$. The analytic formulas for the relevant quantities are collected in the Appendix. In analysis we consider  $2\sigma$ range of neutrino oscillation parameters, Tab.~\ref{neutrino_data}, Majorana phases $\alpha_1$ and $\alpha_2$ are varied in the full range (0,$2\pi$). We vary lightest neutrino mass $m_{\nu_0}$ keeping the $\Sigma$ (sum of neutrino masses) limit (\ref{astrolimit}) 
for both inverted and normal hierarchies. We assume degenerate mass for charged scalar bosons, $M_{H^{\pm \pm}}=M_{H^\pm}$, and vary them from $\sim 500$ GeV to 1000 GeV ($M_{H^{\pm\pm}}\lesssim 470$ is already excluded by the LHC, see section \ref{collidersignals} and a discussion around  Tab.~\ref{tab_BR_ll_ee_em_mm_htm}).
The shaded regions in Fig.~\ref{current_future_limits} are excluded from LFV and muon $(g-2)\mu$ limits. We use different colors to show the exclusion from individual LFV processes: radiative decay of leptons (green), three body decay of leptons(red), $\mu$-to-$e$ conversion (blue), and $(g-2)_\mu$ (violet).  
We can see that the most stringent limit is due to three body decays $l_i \rightarrow l_j l_k l_k$ specifically the $\mu \to eee$ process. We do not find any significant difference between two neutrino mass hierarchy scenarios, but for low neutrino masses the radiative decay $\mu \to e \gamma$ starts to play an important role in normal hierarchy case (see Tab.~\ref{benchmarks}).
Bounds coming from scattering processes or muonium to antimuonium conversion are at least one order of magnitude smaller than those obtain through $(g-2)_\mu$ calculation and are not included in the above plots.

In Table~\ref{benchmarks}, we collect the lower limits of $v_\Delta$ in eV for different values of Majorana phases and lightest neutrino mass assuming $m_{H^{\pm\pm}}=700$ GeV. The process which put the strongest limit is written below the numerical values. 
 For our further analysis and the \texttt{HTM} benchmark point we take $v_\Delta=15$ eV.
\begin{table}[h!]
\centering
\begin{tabular}{c|c||c|c|c||c|c|c}
\multicolumn{2}{c||}{} & \multicolumn{3}{c||}{NH} & \multicolumn{3}{c}{IH} \\
\hline
$\alpha_1$ & $\alpha_2$ & $m_{\nu_0}=0$ & $m_{\nu_0}=0.01$ & $m_{\nu_0}=0.071$ & $m_{\nu_0}=0$ & $m_{\nu_0}=0.01$ & $m_{\nu_0}=0.066$ \\
\hline
\hline
0 & 0 & 
	\begin{tabular}{c} 1.04 \\ {\small $\mu \to e \gamma$} \\ \end{tabular} &
	\begin{tabular}{c} 1.60 \\ {\small $\mu \to e e e$} \\ \end{tabular} &
	\begin{tabular}{c} 6.45 \\ {\small $\mu \to e e e$} \\ \end{tabular} &
	\begin{tabular}{c} 3.36 \\ {\small $\mu \to e e e$} \\ \end{tabular} &
	\begin{tabular}{c} 3.74 \\ {\small $\mu \to e e e$} \\ \end{tabular} &
	\begin{tabular}{c} 7.47 \\ {\small $\mu \to e e e$} \\ \end{tabular} \\ \hline
0 & $\frac{\pi}{2}$ & 
	\begin{tabular}{c} 1.04 \\ {\small $\mu \to e \gamma$} \\ \end{tabular} &
	\begin{tabular}{c} 1.15 \\ {\small $\mu \to e e e$} \\ \end{tabular} &
	\begin{tabular}{c} 7.48 \\ {\small $\mu \to e e e$} \\ \end{tabular} &
	\begin{tabular}{c} 4.92 \\ {\small $\mu \to e e e$} \\ \end{tabular} &
	\begin{tabular}{c} 4.99 \\ {\small $\mu \to e e e$} \\ \end{tabular} &
	\begin{tabular}{c} 8.09 \\ {\small $\mu \to e e e$} \\ \end{tabular} \\ \hline
$\frac{\pi}{2}$ & 0 &
	\begin{tabular}{c} 1.04 \\ {\small $\mu \to e \gamma$} \\ \end{tabular} &
	\begin{tabular}{c} 1.04 \\ {\small $\mu \to e \gamma$} \\ \end{tabular} &
	\begin{tabular}{c} 6.68 \\ {\small $\mu \to e e e$} \\ \end{tabular} &
	\begin{tabular}{c} 4.92 \\ {\small $\mu \to e e e$} \\ \end{tabular} &
	\begin{tabular}{c} 5.06 \\ {\small $\mu \to e e e$} \\ \end{tabular} &
	\begin{tabular}{c} 8.56 \\ {\small $\mu \to e e e$} \\ \end{tabular} \\ \hline
$\frac{\pi}{2}$ & $\frac{\pi}{2}$ &
	\begin{tabular}{c} 1.04 \\ {\small $\mu \to e \gamma$} \\ \end{tabular} &
	\begin{tabular}{c} 1.71 \\ {\small $\mu \to e e e$} \\ \end{tabular} &
	\begin{tabular}{c} {5.61} \\ {\small $\mu \to e e e$} \\ \end{tabular} &
	\begin{tabular}{c} {3.36} \\ {\small $\mu \to e e e$} \\ \end{tabular} &
	\begin{tabular}{c} {3.09} \\ {\small $\mu \to e e e$} \\ \end{tabular} &
	\begin{tabular}{c} {3.15} \\ {\small $\mu \to e e e$} \\ \end{tabular} \\ 
\hline	
\hline
\multicolumn{2}{c||}{
\begin{tabular}{c}
{\footnotesize Oscillations} \\
{\footnotesize \bf $\pm$ $2 \sigma$}
\end{tabular}}
& 
\multicolumn{3}{c||}{0.93 $\div $10.31} & \multicolumn{3}{c}{1.07 $\div$ 11.38} \\
\hline
\end{tabular}
\caption{\label{benchmarks}Lower bounds on the triplet vacuum expectation value $v_\Delta$ (in eV) for different values of Majorana phases and doubly charged scalar mass $M_{H^{\pm\pm}} = 700$ GeV. The most strict limit is coming from the LFV processes named under the numerical value.  
As we can see, the triplet VEV $v_\Delta$ is mainly bounded by experimental limits on $\mu \to eee$ and $\mu \to e\gamma$ dacays. First four rows present results for best fit of neutrino oscillation data. The last row shows range of the lowest possible $v_\Delta$ for oscillation parameters within $\pm 2\sigma$ range and Majorana phases within the entire $2\pi$ angle. All values in the table are in eV.
} 
\end{table}

\section{The \texttt{MLRSM} model and relevant experimental constraints on its parameters}
\label{MLRSMconstr}

We consider a left-right symmetric model based on the $SU(2)_L \otimes SU(2)_R \otimes U(1)_{B-L}$ gauge
group ~\cite{Mohapatra:1974gc,Senjanovic:1975rk,PhysRevLett.44.912,PhysRevD.44.837,Barenboim:2001vu} in its most restricted form, so-called Minimal Left-Right Symmetric
Model (\texttt{MLRSM}) which contains a bidoublet $\Phi$ and two (left and right) triplets $\Delta_{L,R}$ \cite{PhysRevD.40.1546,PhysRevD.44.837,Duka:1999uc,Bambhaniya:2013wza} see the Appendix for details.  

 \subsection{Constraints on \texttt{MLRSM} model parameters and the triplet VEV $v_R$ \label{lrvrmn}}

The heavy sector of the model is triggered by VEV $v_R$ connected with the Higgs triplet $\Delta_{R}$. All new gauge and scalar bosons are proportional to $v_R$, and $v_R \gg \kappa$, where $\kappa$ is a VEV related to the scale of the SM spontaneous symmetry breaking and to the SM gauge bosons $W_1, Z_1$,  $\kappa~\simeq~246$~GeV, see Eq.~(\ref{vev}). 

\begin{figure}[h!]
\begin{center}
\begin{tabular}{@{}c@{}c@{}c@{}}
\includegraphics[width=0.35\textwidth]{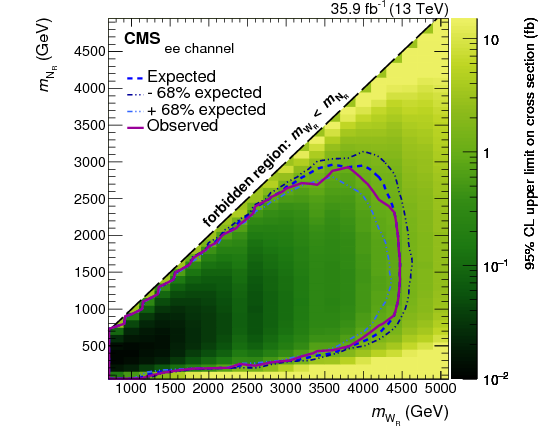}
&
\includegraphics[width=0.35\textwidth]{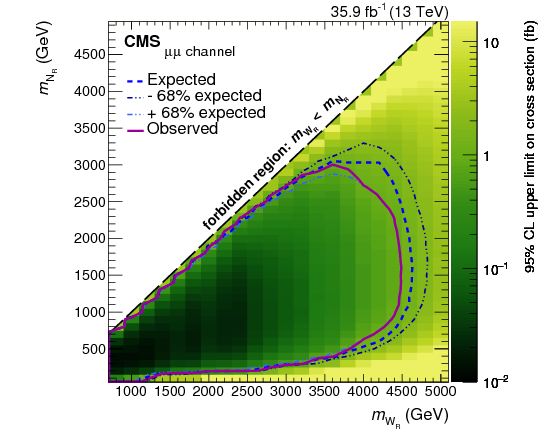}
&
\includegraphics[width=0.3\textwidth]{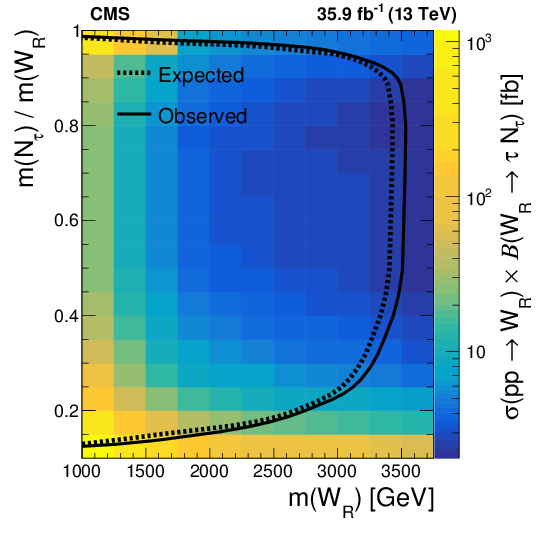}\\
a) & b) & c) \\
\end{tabular}
\end{center}
\caption{\label{CMS_data}Upper limit on the $pp \to lljj$ cross section for different $M_{W_R} \equiv M_{W_2}$ and $M_{N_R} \equiv M_N$ mass hypotheses, for the electron (a), muon (b) and taon (c) channels. The thin-dotted (blue) curves in the Fig. a) and b) indicate the region in ($M_{W_2}$, $M_{N_i}$) parameter space that is expected to be excluded at 68\% \texttt{CL} \cite{Sirunyan:2018pom,Sirunyan:2018vhk}.}
\end{figure}

Using the relation between the heavy charged gauge boson mass and the $SU(2)_R$ triplet \texttt{VEV} $v_R$

\begin{equation}
M_{W_2}^2 \simeq \frac{g^2 \: v_R^2}{2} \qquad \Rightarrow \qquad M_{W_2} \simeq 0.47 \; v_R,
\end{equation}
we can find  the parameter space for $v_R$ and heavy neutrino masses. 
In the last few years the LHC has constrained the possible $v_R$ scale very much by exploiting different channels where $W_2$ plays a crucial role, e.g., $W_2$ decays to two jets \cite{Aaboud:2017yvp}, two jets and two leptons \cite{Sirunyan:2018pom} and top-bottom quarks \cite{Sirunyan:2017ukk}. Altogether, the following bounds on $M_{W_2}$ have been obtained: (i) ATLAS - $3.6$~TeV (2017)~\cite{Aaboud:2017yvp}; 4.8 TeV ($e$-channel), 5~TeV ($\mu$-channel) for $M_{N} \in [0.4,0.5]$ TeV (2019) \cite{Aaboud:2019wfg}; (ii) CMS - $4.4$~TeV (2018)~\cite{Sirunyan:2018pom}, assuming that $SU(2)_{\rm R}$
gauge coupling $g_L$ equals the $SU(2)_{L}$ coupling  $g_R$.
These bounds can be relaxed without such an assumption
\cite{Gluza:2015goa,Dev:2015pga,Das:2017hmg,Frank:2018ifw}.
The CMS experimental data based on the $pp \to lljj$ process are presented as the $M_{W_2}-M_{N}$ exclusion plots, see Fig.~6 in \cite{Sirunyan:2018pom} and Fig.~7 in \cite{Sirunyan:2018vhk}. For convenience we repeat them here in Fig.~\ref{CMS_data}. We use these data, and analogous data from the ATLAS  collaboration  \cite{Aaboud:2018spl}, leading to restrictions on the t-channel in  Fig.~\ref{epemLR} and final signals presented in Section \ref{comparison}.  

As assumed in Fig.~\ref{CMS_data} we take $M_{W_2} \geq M_N$ and a correlation between the masses which are proportional to $v_R$  \cite{Gluza:1993gf,Chakrabortty:2016wkl}.
By passing, let us note that most of the experimental LHC analyses are based on simplified scenarios where heavy neutrinos are mass degenerate with diagonal mixings and where CP-violating effects are not taken into account. However, CP-parities of neutrinos can be different leading to destructive interference effects, relaxing limits on the $v_R$ scale, see \cite{Czakon:1999ga,Gluza:2016qqv,Dev:2019rxh}.

A simultaneous fit to the SM low energy charged and neutral currents set a rather weak bound on $M_{W_2}$, so $v_R$, namely $M_{W_2}>715$ GeV \cite{Czakon:1999ga, PhysRevD.98.030001}. However, there are additional restrictions for model's parameters coming from radiative corrections.
As far as one loop corrections are concerned and additional precision constraints on \texttt{MLRSM} parameters, there are very few studies based on LR models, i.e.,~\cite{Duka:1999uc,Czakon:1999ha,Czakon:2002wm,Czakon:1999ga}  (\texttt{MLRSM} model), the other papers are: \cite{Beall:1981ze,GagyiPalffy:1997hh} (limits on $W_2$ mass coming from the $K_L-K_S$ process (finite box diagrams, renormalization not required)), 
\cite{Pilaftsis:1995tf} (LEP physics), \cite{Ball:1999mb,Pospelov:1996fq,Kiers:2002cz,Rizzo:1994aj,Cho:1993zb} (process $b\to s \gamma$). Some interesting results are discussed also in papers \cite{Senjanovic:1978av,Senjanovic:1978ee} where the problem of decoupling of heavy scalar particles in low energy processes has been discussed. 
In \cite{Chakrabortty:2012pp} it has been shown that low-energy radiative corrections shrink non-standard parameters to very small regions, due to correlations among gauge bosons, scalars and heavy neutrino massses, though still there is a freedom connected among others with unknown scale $v_R$. 
We assume $v_R$ and windows of possible masses of heavy \texttt{MLRSM} particles allowed by low energy analysis  \cite{Chakrabortty:2012pp,Deppisch:2014zta,Borah:2017ldt,FileviezPerez:2017zwm,Borah:2016iqd}. 

Apart from experimental limits, 
due to tree-unitarity and flavor changing neutral currents (FCNC)  constraints, the scalar potential parameter $\alpha_3$ in Eq.~(\ref{VLRSM}) is restricted and masses of neutral Higgs bosons $H_0^0, A^0$ should be greater than 10 TeV. The lowest limit on $v_R$ scale is $1.3 \div 6.5$ TeV \cite{Chakrabortty:2016wkl}, depending on the mass scale of \texttt{FCNC} Higgs bosons \cite{Guadagnoli:2010sd}.  Such a relatively low (TeV) scale of the heavy sector is theoretically possible, even if gauge unification (\texttt{GUT}) is demanded, for a discussion, see~\cite{Shaban:1992he} and~\cite{Lindner:1996tf}.

\section{Collider signals, the results \label{collidersignals}}

\subsection{The $H^{\pm\pm}$ pair production at $e^+e^-$ and $pp$ colliders \label{subseepp}}

\begin{figure}[h!]
\begin{center}
\includegraphics[angle=0,width=0.8\textwidth]{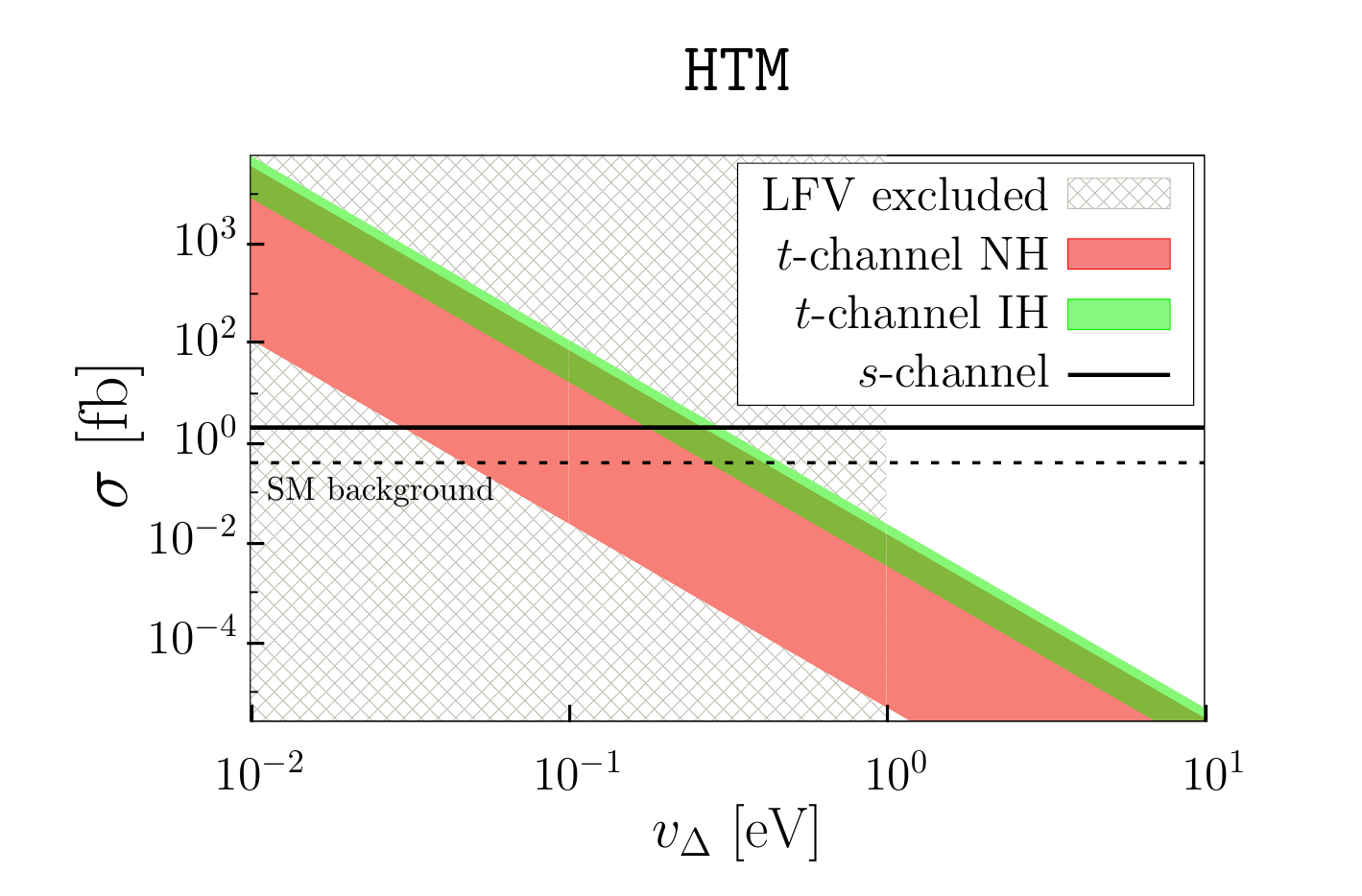} 
\caption{\label{epemhtm}Doubly charged Higgs boson pair production $e^+e^- \to H^{++} H^{--}$ for  $M_{H^{\pm\pm}} =700$ GeV and CM energy 1.5 TeV in the \texttt{HTM} model. The crossed area is excluded by the low energy data (Tab.~\ref{benchmarks}). We took the neutrino oscillation parameters within the $\pm 2 \sigma$ range (Tab.~\ref{neutrino_data}), that is why the $t$-channel is smeared.  With a dashed line, we have marked the SM background for four leptons production (electrons and muons) which is  $\sigma=0.415$ fb, see section 
\ref{secbackgr}.
}
\end{center}
\end{figure}

\begin{figure}[h!]
\begin{center}
\includegraphics[angle=0,width=0.8\textwidth]{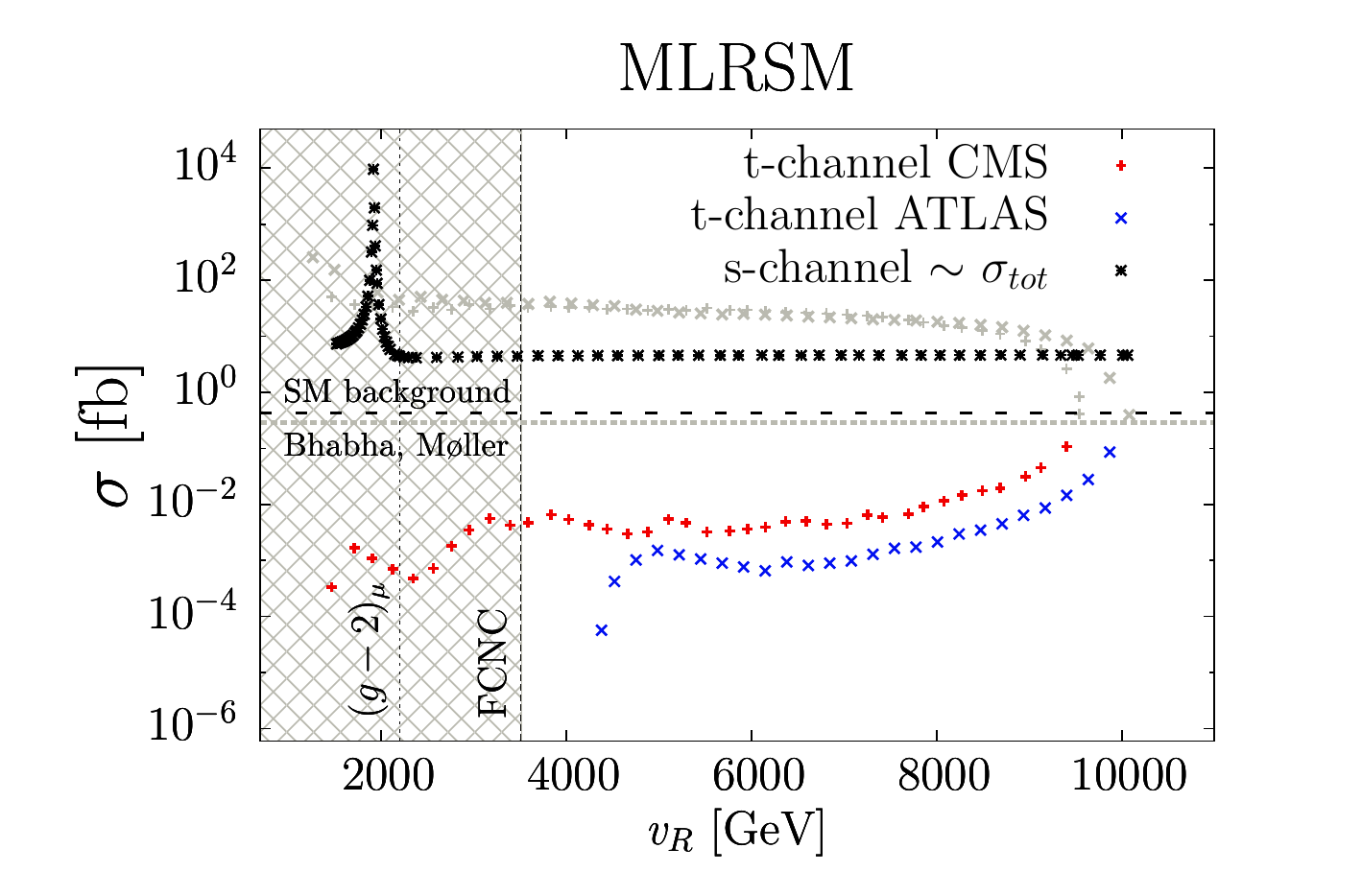}
\caption{\label{epemLR}
Doubly charged Higgs boson pair production $e^+e^- \to H_1^{++} H_1^{--}  + H_2^{++} H_2^{--}$ for  $M_{H^{\pm\pm}_{(1,2)}} =700$~GeV and CM energy 1.5 TeV in \texttt{MLRSM}. 
For the t-channel the choice $M_{W_2}-M_N$ space is restricted by LHC results best fit expected values, see Fig.~\ref{CMS_data}. 
 The crossed area on the left is excluded by $(g-2)_\mu$ and FCNC. The maximum for $v_R = 1900$ GeV comes from the $Z_2$ resonance, $M_{Z_2}=1.9$ TeV.  
 The horizontal gray dashed line "Bhabha, M\o{}ller" separates the t-channel contribution to the cross section which is still allowed by the CMS and ATLAS exclusion analysis from constraints by the Bhabha and  M\o{}ller processes (Tab.~\ref{limits_table1} and Tab.~\ref{limits_table2}). The t-channel contribution above this line is forbidden.  
The SM background (black dashed horizontal line) after applying kinematic cuts is $\sigma =0.415$ fb, see section \ref{secbackgr}.}
\end{center}
\end{figure}

As discussed in previous sections, we assume $M_{H^{\pm\pm}_{(1,2)}} =700$ GeV. This value will be further justified when the $H^{\pm \pm}$ decay branching ratios are discussed in next sections.  
Therefore, for substantial $H^{\pm\pm}$ pair production in $e^+e^-$ collisions, we need the centre mass energy $\sqrt{s}$ above 1 TeV. As discussed in Introduction such energies for $e^+e^-$ colliders are planned presently only at CLIC. 
Numerical results for $\sqrt{s}=1.5$ TeV are gathered in Fig.~\ref{epemhtm} and Fig.~\ref{epemLR} for \texttt{HTM} and \texttt{MLRSM}, respectively. 

A contribution from scalar particles in Fig.~\ref{ee_hcchcc_production} and Fig.~\ref{ee_hcchcc_production_diag} (middle diagrams) are negligible in comparison to the diagrams with the intermediate photon and $Z$ bosons, see Tab.~\ref{tabeesep}.
Within \texttt{HTM} the contribution from the heavy neutral scalar $H^0$ in the s-channel is negligible as both $l-l-H^0$ and 
$H^{++}-H^{--}-H^0$ vertices are proportional to  $\sin{\alpha}$, see Eq.~(\ref{e:alpha}), which is very small \cite{Das:2016bir}. Also, the contribution from the Standard Model Higgs boson in the s-channel is small, a few orders of magnitude lower than the contribution from the gauge bosons, because of small Yukawa  $e^+-e^--h$ coupling and heavy boson mass in the propagator.  
We have a similar situation in \texttt{MLRSM}. Even though there are some additional possible intermediating particles in the $s$ channel (scalars and the $Z_2$ gauge boson, see Fig.~ \ref{ee_hcchcc_production_diag}), they are heavy, and the couplings are small. Large Higgs boson masses in the propagators are proportional to $v_R$ (see the Appendix).  We assume that masses of both $H_1^{\pm\pm}$ and $H_2^{\pm\pm}$ are equal. $H_0^3$ does not contribute to the process, because  the $H_0^3-H_{1,2}^{\pm\pm}-H_{1,2}^{\pm\pm}$ vertex is proportional to the left-handed triplet VEV $v_L$ which is set to zero to preserve the $\rho$-parameter \cite{Bambhaniya:2014cia}.

As discussed in Section \ref{tripltesneutr}, the t-channel in \texttt{HTM}  contains the $e-l'-H^{\pm\pm}$ vertex  inversely proportional to $v_\Delta$ in Eq.~(\ref{e:yukawa4}), this diagram becomes dominant for small $v_\Delta$. However, it appears that the region where the $t$-channel can dominate is ruled out by the low energy  data  and Tab.~\ref{benchmarks}. 
The allowed  $t$-channel cross section for $e^+ e^- \to H^{++} H^{--}$ is a few orders of magnitude lower than the $s$-channel, which is equal to 2.4 fb, see a solid horizontal line in Fig.~\ref{epemhtm}.  
As it is shown, regardless of the choice of the neutrino parameters, the whole region where the $t$-channel is not negligible is excluded.
 
The $e^+e^- \to H_1^{++} H_1^{--}  + H_2^{++} H_2^{--}$ cross section in \texttt{MLRSM}, see Fig.~\ref{ee_hcchcc_production_diag} 
depends on the right-handed triplet VEV $v_R$ and heavy neutrino masses. 
The allowed space for $M_{W_2}-M_N$ parameters has been considered in section \ref{lrvrmn} and is based on limits on the heavy neutrino masses taken from the LHC CMS and ATLAS data for the $pp \to lljj$ process \cite{Sirunyan:2018vhk,Sirunyan:2018pom,Aaboud:2018spl}. This process is a collider analogue of the neutrinoless double beta decay mediated by a heavy charged boson, heavy Majorana neutrinos, and cross-sections depend strongly on masses and CP-parities of heavy neutrinos \cite{Gluza:2016qqv}. As we have in disposal CMS and ATLAS results, in calculations we assume  $M_{W_2} > M_N$ with the same CP-parities of heavy neutrinos. In Fig.~\ref{epemLR} 
we vary the $M_{W_2}$ mass from 600 GeV to 5.5 TeV and the heavy neutrino mass up to 4.8 TeV and take the best fit expected values for the LHC exclusion data. 

\begin{figure}[h!]
\begin{center}
\includegraphics[ width=0.23\textwidth]{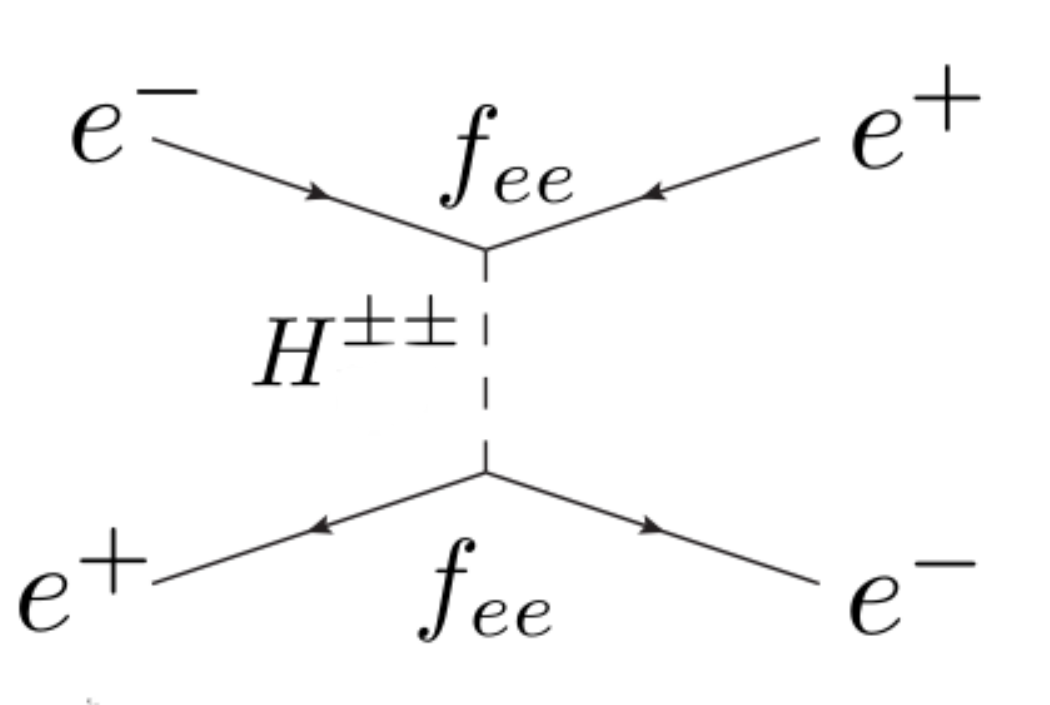}
\includegraphics[width=0.27\textwidth]{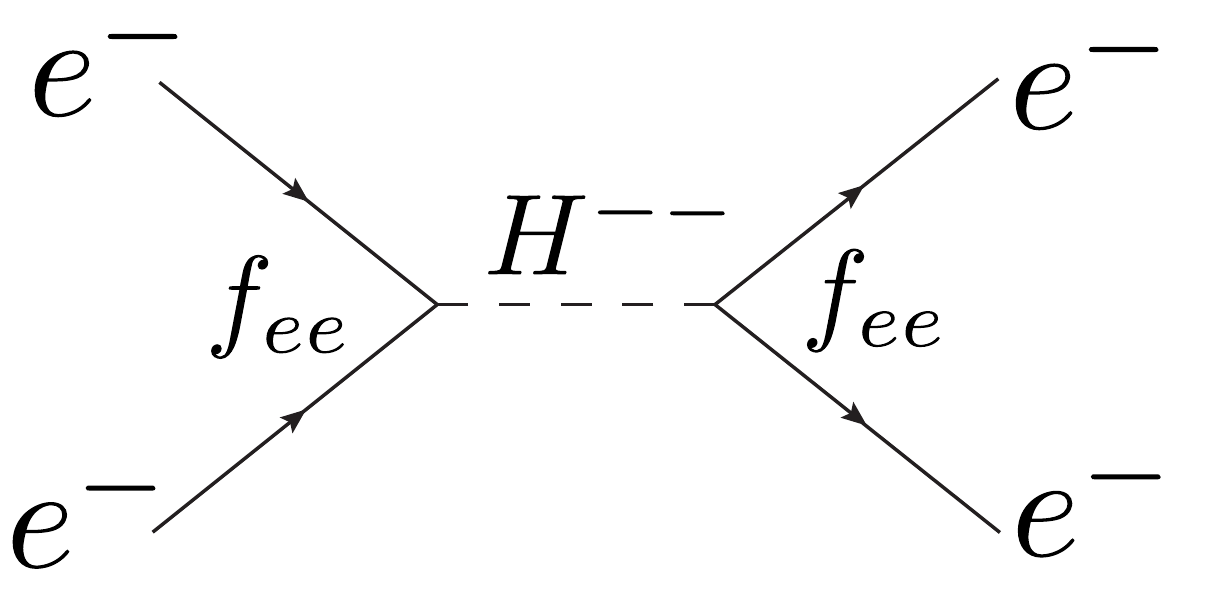}
\\
(a)\hspace{120pt} (b)
\caption{\label{LEP_T_U}The $e^+ e^- \to e^+ e^-$ (Bhabha) and $e^- e^- \to e^- e^-$ (M\o{}ller) processes at the lowest order with doubly charged Higgs bosons.}
\label{ee_4l_diagrams}
\end{center}
\end{figure}

The production through the $t$-channel is constrained by the Yukawa coupling, Eq.~ $Y_{ee}$ (\ref{hmvertex}). We assume perturbativity of the coupling $Y_{ee}\sim \mathcal{O}(1)$. From $M_N=\sqrt{2} h_M v_R$,  Eq.~(\ref{e:MRfvR}), with $h_M \lesssim 1$ we get the relation between $v_R$ and heavy neutrino masses. Since the LHC exclusion plots assume $M_N < M_{W_2}$, this condition is fulfilled automatically for the considered parameter space. 
The most strict limits comes from the Bhabha and M\o{}ller processes, see Fig.~\ref{LEP_T_U}, the doubly charged scalar particles can contribute there. In Tab.~\ref{LRSM_max_MN} we gathered region of physical masses for heavy neutrinos which arise from the discussed low energy LFV constraints.

\begin{center}
\begin{table}[h!]
\centering
\begin{tabular}{|c|c|c|}
\hline
$M_{H_{1,2}^{\pm\pm}}$ [GeV] & $v_R = 6$ TeV& $v_R=15$ TeV \\
\hline
 700  &  $M_{N_1} < 803$ GeV  &   $M_{N_1} < 2007$ GeV  \\
 \hline
1000 &  $M_{N_1} < 1147$ GeV  &  $M_{N_1} < 2867$ GeV  \\
 \hline
\end{tabular}
\caption{\label{LRSM_max_MN} Upper limits on the heavy neutrino masses for different sets of doubly charged Higgs boson and the triplet VEV $v_R$, taking into account low energy LFV constraints in Tab.~\ref{limits_table1} and  SM processes in Tab.~\ref{limits_table2}. }
\end{table}
\end{center}

As we can see in Fig.~\ref{epemLR}, the $t$-channel gray parts of the plotted lines above the long-dashed "Bhabha, M\o{}ller" line assigned with cross {\color{gray}{$\times$}}  and plus {\color{gray}{$+$}} symbols
might dominate within the whole region of the $v_R$ parameter tested by LHC.  
 However, adding the discussed Yukawa constraints on $H^{\pm \pm}$ couplings gathered in Tab.~\ref{limits_table1} and Tab.~\ref{limits_table2}, this region is eliminated  (corresponding allowed $t$-channel contributions with red and blue parts of the plotted CMS and ATLAS lines are thickened in  Fig.~\ref{epemLR}). As the Bhabha and M\o{}ller processes constraint the t-channel contribution to be below 0.3 fb, altogether with the LHC constraints, it results in a much smaller contribution than the s-channel contribution and the interference effect is small: The total cross section $\sigma_{tot}$ practically corresponds with the s-channel. Even though the mass $M_{Z_2}$ is a function of $v_R$ ($M_{Z_2} \simeq 0.78 \; v_R$), the higher resonances are suppressed since the small center mass energy is too small to observe them. For larger $v_R$ values, we are outside the s-resonance for $\sqrt{s}=1.5$ TeV and s-channel contributions are flat and small. For instance, for $v_R=6(15)$ TeV which will be used as reference values in next sections for four lepton final state analysis, and which correspond to  $M_{Z_2}=4.7 (11.7)$ TeV,  $\sigma_s \simeq 4.6$ fb.
The limits from the muon $(g-2)$ and  the $\mu^+e^-\to\mu^-e^+$ process are also taken into account, since the corresponding diagrams contain the $f_{ee}$ and $f_{\mu\mu}$ couplings, but they play no significant role.
The $(g-2)_\mu$ process restricts the $f_{\mu\mu}$ coupling, see the Appendix. It affects heavy neutrino mass bounds and for further calculations we assume that maximum $M_{N_2} = 5$ TeV, what is safe for considered values of $v_R$ (6 and 15 TeV).
Unlike in the \texttt{HTM} case, the LFV processes do not restrict further the results because we assume the LFV vertices to be negligible with no light-heavy neutrino mixings (see Section \ref{vertex_LRSM}). 
Taking into account the above constraints, the maximal cross section at the $t$-channel is $\sigma_t \sim 0.3$ fb. 

All non-standard heavy particle masses are related to the vacuum expectation value of the right-handed triplet, see Appendix \ref{s:model} and Eqs.~(\ref{h10_mass_LRSM})-(\ref{h2cc_mass_LRSM}). As discussed in \cite{Chakrabortty:2016wkl}, the combined effects of relevant Higgs potential parameters  and Higgs bosons responsible for FCNC limits regulate the lower limits of heavy gauge boson masses. In Fig.~\ref{epemLR} we put only low-energy limits on $v_R$ coming from $(g-2)_\mu$ and FCNC. We indicate $v_R \sim 3.5$ TeV, which by considering the Higgs boson mass spectrum Eqs.~(\ref{h10_mass_LRSM})-(\ref{h2c_mass_LRSM}) is the minimal $v_R$ for FCNC Higgs masses of $A_1^0,H_1^0$ scalars at the level of $\cal{O}$(10) TeV, and the minimal allowed $M_{H_{3}^0}$ for $\alpha_3$ scalar parameter to be less than 16. 
The mass limit for $A_1^0,H_1^0$ at the level of 10 TeV  is the lowest limit on FCNC Higgs boson masses
 \cite{Guadagnoli:2010sd}, one of the strongest limits has been obtained in \cite{Pospelov:1996fq} ($M_{\rm A_1^0,H_1^0} \geq 50 \,\mathrm{TeV}$).
 We can see that there are various estimates of the $v_R$ scale, see also Fig.~\ref{CMS_data}.  Apart from the dijet LHC strong limits, there are searches in the one jet and one lepton signal category \cite{Mitra:2016kov, Aaboud:2019wfg} as well as off-shell $W_2$ and $Z_2$ channels \cite{Ruiz:2017nip, Nemevsek:2018bbt}. All these studies confirm that it is not natural to expect $v_R$ to scale below 3.5 TeV. For these reasons, as $pp$ studies at HL-LHC or future FCC-hh or CEPC colliders offer investigation of heavy BSM states at higher scales, in next sections for $pp$ phenomenological studies we assume $v_R$ scale and \texttt{MLRSM} mass benchmarks corresponding to higher $v_R$ values at the level of 6 and 15 TeV.  

In Tab.~\ref{tabeesep} we show fractions  of dominating s-channel individual contributions to the doubly charged pair production cross section in $e^+e^-$ collisions.
Individual doubly charged production cross sections are:
$\sigma(e^+e^- \rightarrow H_{1}^{++} H_{1}^{--} ) = 2.46$\; fb, 
$\sigma(e^+e^-\rightarrow H_{2}^{++} H_{2}^{--} ) = 2.15$\; fb, which should be compared to 
$\sigma(e^+e^-\rightarrow H_{2}^{++} H_{2}^{--} )_{HTM} =$\; 2.4 fb in \texttt{HTM}, see the solid horizontal line in Fig.~\ref{epemhtm}.

\begin{table}[h!]
\begin{center}
\begin{tabular}{|c|c||c|c|c|c||}
\cline{3-6}
\multicolumn{2}{c||}{} & \multicolumn{4}{c||}{$\sqrt{s}=1.5$ TeV} \\
\hline
\hline
\multirow{2}{*}{Model} & Process & \multirow{2}{*}{$\gamma$} & \multirow{2}{*}{$Z_1$} & \multirow{2}{*}{$Z_2$} & \multirow{2}{*}{{\small scalars}} \\
 & $e^+e^-\to$ & & & & \\
\hline
\hline
\multirow{2}{*}{\texttt{MLRSM}} & $H_1^{++}H_1^{--}$ & 87\% & 13\% & $\ll 1\% $ & $\ll 1\% $ \\
\cline{2-6}
    & $H_2^{++}H_2^{--}$ & 90\% & 10\% & $\ll 1\%$ & $\ll 1\%$ \\
\hline
\texttt{HTM} & $H^{++}H^{--}$  &88\% & 12\% & --- &  $\ll 1\%$\\ 
\hline
\end{tabular}
\caption{\label{tabeesep} Individual s-channel contributions to the doubly charged pair production in electron-positron collision for $\sqrt{s}=1.5$ TeV c.m. energy (CLIC) in the \texttt{HTM} and \texttt{MLRSM} models.}
\end{center}
\end{table}

Let us proceed to the hadron colliders and pair production of $H^{\pm\pm}$ Higgs bosons. Basic tree-level diagrams for considered models are given in Fig.~\ref{fig_pp_hcchcc}.

\begin{figure}[h]
\begin{center}
\includegraphics[width=0.8\textwidth]{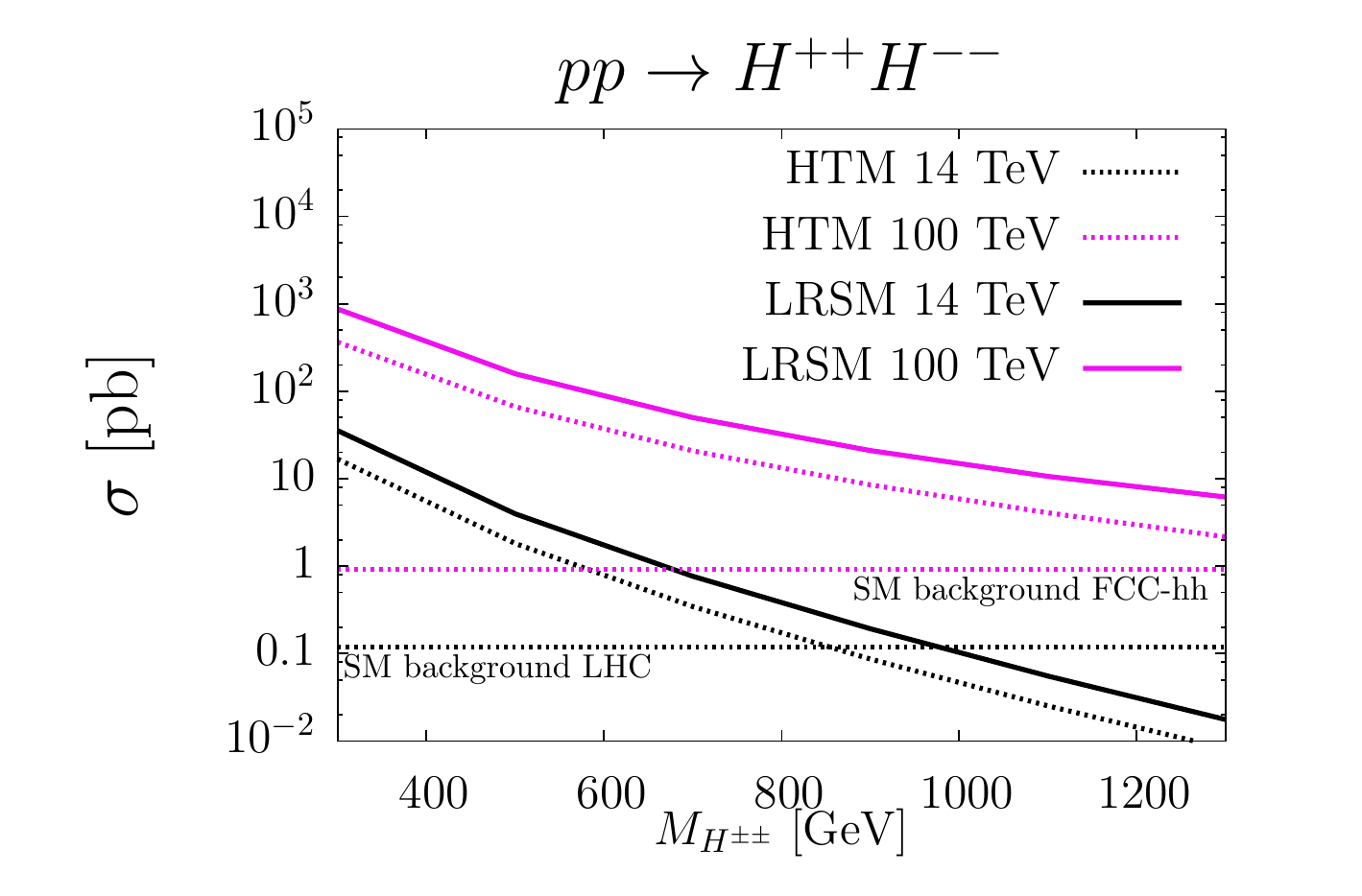}
\caption{The $H^{\pm\pm}$ pair production $pp \to H^{\pm \pm} H^{\pm \pm}$ within the \texttt{HTM} and \texttt{MLRSM} models for LHC and FCC-hh center-of-mass energies. Horizontal dashed lines give the SM background for the process $pp \to 4l$, Tab.~\ref{sm__pp_4l_background_tab}, with kinematic cuts defined in section \ref{secbackgr}. The QCD NLO $H^{\pm\pm}$ pair production  k-factors are taken into account, see the main text.}
\label{sec_pp_hcchcc}
\end{center}
\end{figure}
Fig.~\ref{sec_pp_hcchcc} shows the plot for the $p p \rightarrow H^{++} H^{--}$ cross sections both in the \texttt{HTM} and the \texttt{MLRSM} models. The cross sections are comparable in both models with slightly larger values for \texttt{MLRSM}. 
Typically, for $M_{H_1^{++}} = M_{H_2^{++}} = 1000$ GeV:
\begin{equation}
\sigma(pp\rightarrow (H_{1}^{++} H_{1}^{--}+H_{2}^{++} H_{2}^{--})\rightarrow {\ell_i}^+{\ell_i}^+{\ell_j}^-{\ell_j}^-) = 0.063\; (13.02)~fb,
\end{equation}
for $\sqrt{s}=14(100)\;{\rm TeV}$.
The individual $H_1^{\pm \pm}$ and 
$H_2^{\pm \pm}$ contributions to the cross section  for $\sqrt{s}=14(100)\;{\rm TeV}$ are:
\begin{eqnarray}
\sigma(pp\rightarrow H_{1}^{++} H_{1}^{--} ) &=& 0.046\; (7.64)~fb, 
\label{eqhpp814_600}\\
\sigma(pp\rightarrow H_{2}^{++} H_{2}^{--} ) &=& 0.017\; (5.38)~fb,\\
\sigma(pp\rightarrow H_{}^{++} H_{}^{--} )_{HTM} &=& 0.044\; (5.13)~fb.
\label{eqhpp814_600_a}
\end{eqnarray}

The \texttt{HTM} production process is about 70\% (40\%) of that in \texttt{MLRSM} for $\sqrt{s}=14(100)$ TeV, respectively. We can see that $\sigma(pp\rightarrow H_{}^{++} H_{}^{--})_{HTM} \simeq \sigma(pp\rightarrow H_{1}^{++} H_{1}^{--})$, especially for the HL-LHC case. 
In Tab.~\ref{tabppsep} we sum up fractions of particle contributions to the process coming from individual channels. As we can see from Eqs.~(\ref{eqhpp814_600})-(\ref{eqhpp814_600_a}) and Tab.~\ref{tabppsep}: (i) production of the $H_2^{\pm \pm}$ is smaller than the $H_1^{\pm \pm}$ one, its contribution increases with c.m. energy; (ii) the $\gamma$ channel dominates for both $H_1^{\pm \pm}$ and $H_2^{\pm \pm}$ pair production at HL-LHC c.m.~energies while for FCC-hh/CEPC option, $Z_2$-channel starts to be important. Due to the shown differences between \texttt{MLRSM} and \texttt{HTM} models we can expect a higher number of events for 4-lepton final states in \texttt{MLRSM} when masses of doubly charged Higgs bosons are the same. However, it does not have to be the case as the final results depend strongly on branching ratios which we will consider in the next section.

\begin{table}
\begin{center}
\begin{tabular}{|c|c||c|c|c|c||c|c|c|c||}
\cline{3-10}
\multicolumn{2}{c||}{} & \multicolumn{4}{c||}{14 TeV} & \multicolumn{4}{c||}{100 TeV} \\
\hline
\hline
\multirow{2}{*}{Model} & Process & \multirow{2}{*}{$\gamma$} & \multirow{2}{*}{$Z_1$} & \multirow{2}{*}{$Z_2$} & \multirow{2}{*}{{\small scalars}} & \multirow{2}{*}{$\gamma$} & \multirow{2}{*}{$Z_1$} & \multirow{2}{*}{$Z_2$} & \multirow{2}{*}{{\small scalars}} \\
 & $pp\to$ & & & & & & & & \\
\hline
\hline
\multirow{2}{*}{\texttt{MLRSM}} & $H_1^{++}H_1^{--}$ & 63\% & 36\% & $< 1\%$ & $\ll 1\% $ & 43\% & 27\%\% & 30\% & $\ll 1\%$ \\
\cline{2-10}
    & $H_2^{++}H_2^{--}$ & 74\% & 25\% & $\sim 1\%$ & $\ll 1\%$ & 68\% & 9\% & 23\% & $\ll 1\%$ \\
\hline
\texttt{HTM} & $H^{++}H^{--}$  &65\% & 35\% & --- &  $\ll 1\%$ & 62\% & 38\% & --- & $\ll 1\%$ \\ 
\hline
\end{tabular}

\caption{\label{tabppsep} Individual channel contributions to the doubly charged pair production $\sqrt{s}=14$ TeV c.m. energy  (HL-LHC) and $\sqrt{s}=100$ TeV c.m. energy (FCC-hh/CEPC) hadron colliders in \texttt{HTM} and the \texttt{MLRSM} models.}
\end{center}
\end{table}

The QCD contributions to the doubly charged Higgs boson pair production increase the cross section at the NLO level.  
The role of the QCD effects  in the hadronic processes of $H^{\pm \pm}$ pair   production has been considered  in \cite{Fuks:2019clu}. A similar situation with positive contribution of QCD at the NLO and higher levels has been observed also for other processes in models which include triplet Higgs bosons and heavy neutral leptons \cite{Fuks:2019clu,Das:2017pvt,Ruiz:2015zca,Padhan:2019jlc,Gallinaro:2020cte}. The corresponding k-factors (which measure ratios of higher order QCD effects to the tree level cross section) do not change considerably with the $H^{\pm \pm}$ mass and centre of mass energies,  k-factor $ \in (1.15 \div 1.20$). Due to different ratios of $H_{1}^{++}$ and $H_{2}^{++}$ pair production processes (see Eqs.(\ref{eqhpp814_600})-(\ref{eqhpp814_600_a}) and Tab.~\ref{tabppsep}), for $m_{H^{\pm \pm}} = 1$ TeV, the k-factor in \texttt{HTM} is 1.15 and is smaller than k-factors in \texttt{MLRSM}, which are $ \simeq 1.6\; (1.85)$ for HL-LHC (FCC-hh/CEPC) centre of mass energies, respectively.
There are various QCD contributions at the NLO level to the considered process, in which the s-channels $\gamma/Z_1/Z_2$ dominate over the gluon and photon fusion mechanisms, both for HL-LHC and FCC-hh/CEPC. Concerning potential contributions beyond NLO, the $N^3LL$ terms are found to be about three times larger  than NLO terms. However, this is connected mainly with gluon fusion which is subdominant for the considered $H_{1}^{++}$ masses in the s-channel \cite{Fuks:2019clu}. {\color{black}{As the doubly charged pair production signals are dominated by the exchange of the SM particles in $e^+e^-$ collisions (see Tab.~\ref{tabeesep}),  differences between doubly charged pair production signals in both models are small. A better estimation of QCD corrections, evaluating the NNLO terms, would resolve expected signals better. In the $pp$ collision case, the production difference between the models for the considered benchmark points is much larger. NLO QCD corrections seem to be enough to discriminate the models, though we should note that the production difference between both models will decrease above $v_R ={15}$ TeV, which is the upper limit for the $v_R$ value considered in the present work.
In scenarios with $v_R > {15}$ TeV a knowledge of NNLO QCD corrections will be also useful in the $pp$ collisions.
Anticipating final four-lepton results, the above conclusions do not change for the considered benchmark points and kinematic cuts. Namely, ratios of \texttt{MLRSM} ($v_R= 15$ TeV) to \texttt{HTM} four-lepton signals can be as large as 1.7 (34 and 43)  at $e^+e^-$ and $pp$ (HL-LHC and CEPC/FCC-hh) colliders, respectively (see $4\mu$ signals, Tab.~\ref{tab_fin_ee} and Tab.~\ref{tab_fin_pp}). Then NLO QCD k-factors should be enough to distinguish the \texttt{HTM} and \texttt{MLRSM} signals in $pp$ collisions, unless the $v_R$ scale is  too large and the ${Z_2}$ gauge boson contribution becomes at the NNLO QCD level.}}
 
To summarize, the QCD contributions to the considered production processes at the NLO level are substantial in both models and must be taken into account in the analysis. To discriminate both models, evalauation of higher order QCD terms may be needed for higher $v_R$ scales. 

\subsection{\texttt{HTM}, a choice of benchmark parameters and ${H^{\pm\pm}}$ decay scenarios}
\label{results_decay_HTM}

In \texttt{HTM} the doubly charged scalar has nine possible decay channels, depending on the scalar boson mass
\begin{itemize}
\item[(i)]	$H^{\pm\pm} \to l_i \: l_j,$ \;\;\;$i,j={e,\mu,\tau}$, 
\item[(ii)]	$H^{\pm\pm} \to W^\pm \: W^\pm$,
\item[(iii)]	$H^{\pm\pm} \to H^\pm \: W^\pm$,
\item[(iv)]	$H^{\pm\pm} \to H^\pm \: H^\pm$.
\end{itemize}
In this paper we focus on the first channel (i) and present a case study for  pair production of a doubly charged scalar boson and its subsequent leptonic decays, considered also in \cite{Agrawal:2018}. It is a very clean channel which provides  a unique signature for colliders signal with a pair of the same sign leptons \cite{Bambhaniya:2013wza}.
Scenarios (iii) and (iv) require non-degenerate masses for charged scalar particles: $M_{H^{\pm\pm}} > M_{H^\pm} + M_W$ and  
$M_{H^{\pm\pm}} > 2 M_{H^\pm}$, respectively. 
\begin{figure}[h!]
\begin{center}
\begin{tabular}{cc}
\includegraphics[width=0.47\textwidth,angle=0]{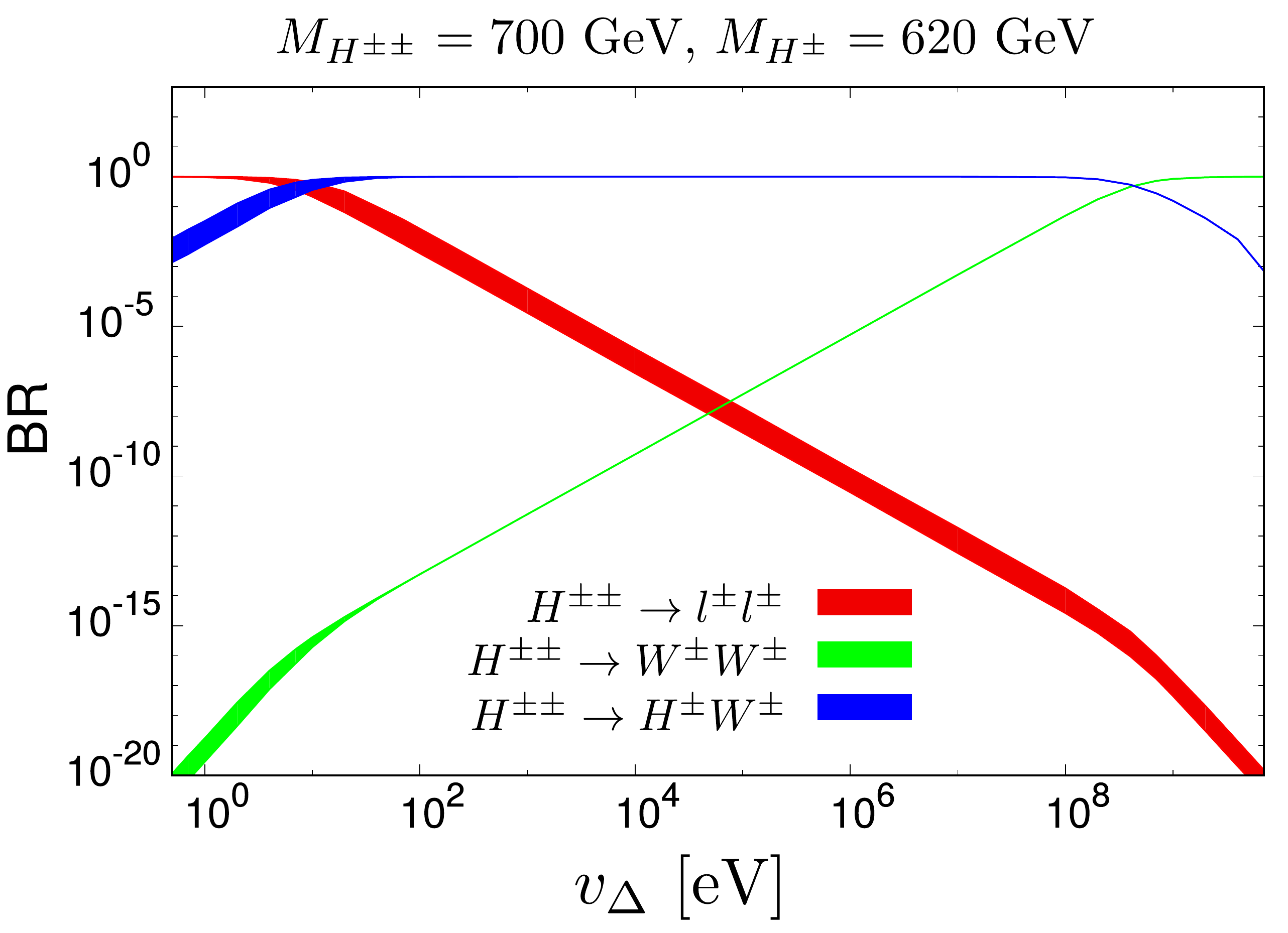} &
\includegraphics[width=0.49\textwidth,angle=0]{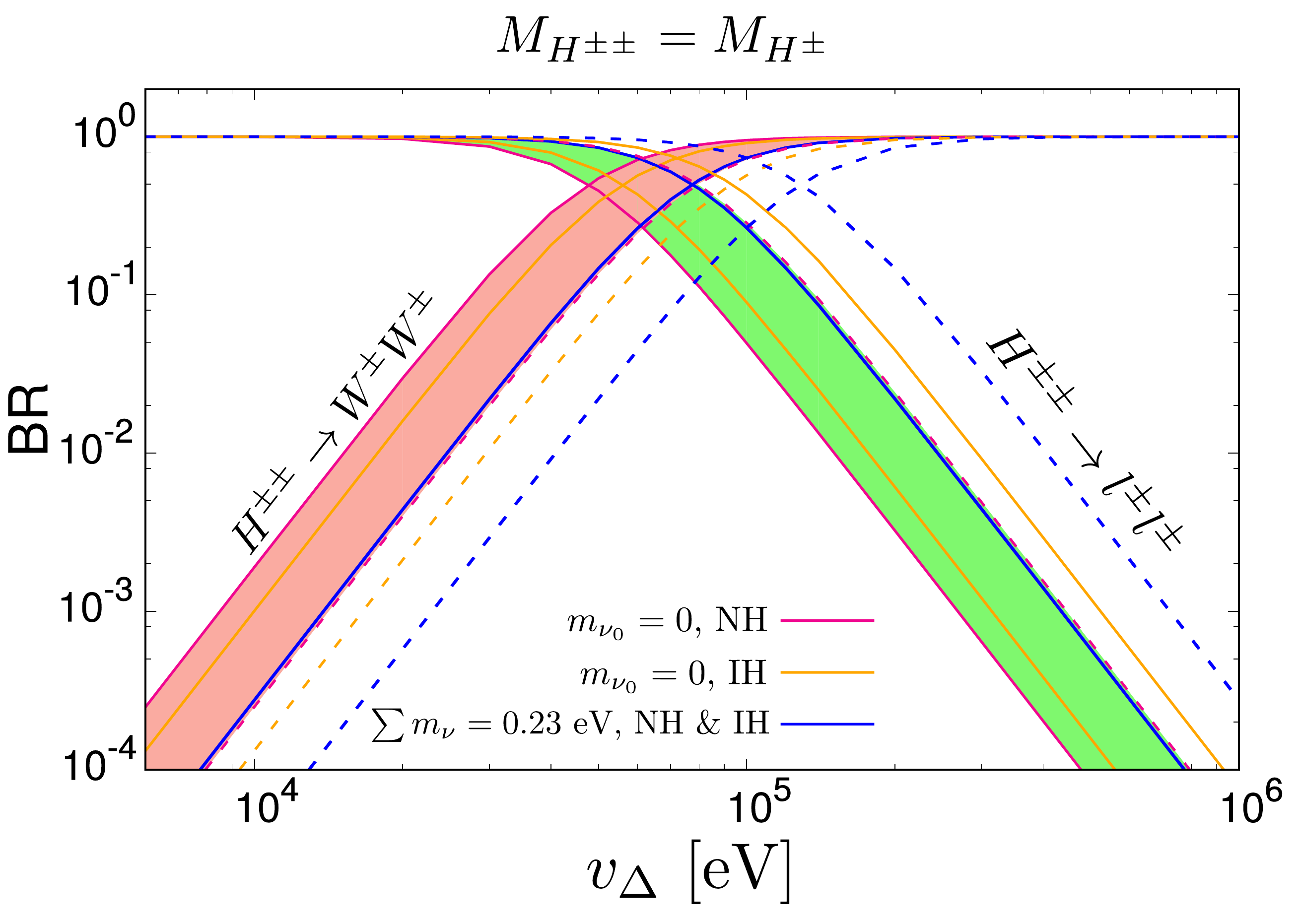}\\
 (a) & (b) \\
\end{tabular}
\caption{\label{BR_all}Branching ratios for $H^{\pm\pm}$ in \texttt{HTM} for a non-degenerated case (a) with $m_{H^{\pm\pm}}=700$ GeV and $m_{H^\pm}=620$ GeV 
and for a degenerate case (b) when   $m_{H^{\pm\pm}}=m_{H^{\pm}}=700$ and $300$ GeV are assumed. The shaded regions correspond to IH and NH neutrino mass hierarchies
with $m_{\nu_0}$ limited by $\Sigma$ in (\ref{astrolimit}) and $M_{H^{\pm\pm}}=700$ GeV. Dashed lines in case (b) describes the
branching ratios for $M_{H^{\pm\pm}} = M_{H^\pm} = 300$ GeV. The oscillation data are taken in the $2\sigma$ range.}
\end{center}
\end{figure}
\\
In  Fig.~\ref{BR_all} we show a variety of branching ratios as a function of  $v_\Delta$ for various $H^{\pm \pm}$ decay channels. On the left plot we show the following decay modes:  leptonic (red), $W^\pm$ gauge bosons (green) and $H^{\pm} W^\pm$ (blue). On right we give a variation of leptonic and pair of gauge boson decay branching ratios  for a degenerate mass of $H^{\pm\pm}$. There are  two cases there: the solid line is for $M_{H^{\pm\pm}} =M_{H^{\pm}}= 700$ GeV  and the dashed line is for a charged scalar boson mass of 300 GeV (this mass is already excluded by LHC, we left it for comparison with previous work, see Fig. 4 in \cite{Perez:2008ha}).
The shaded region is connected with the lightest neutrino mass and mass hierarchy, within  2$\sigma$ oscillation parameter range. This region does not change the result substantially. We can see that
the cross-cut point is shifting with charged scalar boson mass, but in the interesting mass region, the lepton channel dominates till $v_\Delta$ reaches values in range of $10^4 \div 10^5$ eV.  In Fig.~\ref{BR_all} (a) we take a mass gap $M_{H^{\pm\pm}} - M_{H^\pm} = 80$ GeV, in Fig.~\ref{BR_all} (b)  there is no mass gap and both $H^{\pm\pm} \to H^\pm W^\pm$ and $H^{\pm\pm} \to H^\pm H^\pm$
channels are suppressed. It has been shown in~\cite{Das:2016} and \cite{Gluza:2020icp}
that there are limits on the mass gap $|M_{H^{\pm\pm}} - M_{H^\pm}|$ in order to preserve the oblique T-parameter, unitarity and potential stability condition. 
For recent work on vacuum stability conditions of Higgs potentials in various variants of \texttt{HTM} models, see \cite{Moultaka:2020dmb}.
From electroweak precision data and limits from the $h \to \gamma \gamma$ process \cite{Chun:2012jw,Shen_2015_EPL_2} the dominant contributions are in the degenerate mass case. Therefore only leptonic and $W$ gauge boson decay channels are possible. 
However, the $H^{\pm\pm}-W^\mp-W^\mp$ vertex is proportional to the triplet VEV $v_\Delta$ while the Yukawa coupling in the $H^{\pm\pm}-l^\mp-l^\mp$ vertex
is proportional to $\frac{1}{v_\Delta}$, so the lepton channels dominate strongly over the scenario (ii) for the triplet VEV $v_\Delta < 10^5$ eV. 

For VEV $v_\Delta$ in a range of eV, the cumulative leptonic channel dominates in that region regardless of the neutrino masses
and oscillation parameters as well as doubly charged scalar boson masses. So, our final conclusion is that when $H^\pm W^\pm$  and $H^\pm H^\pm$ channels are suppressed, the leptonic decays dominate for low~$v_\Delta$. 
 \begin{center}
\begin{table}[h!]
\centering
\begin{tabular}{|c||c|c|c|c|}
\hline
 BR & $ll$ & $ee$& $e\mu$ & $\mu\mu$ \\
\hline
\hline
0.01	&	-	&	249.2	&	216.3	&	309.7	\\
0.02	&	-	&	310.9	&	300.0	&	335.7	\\
0.03	&	-	&	323.7	&	316.6	&	367.5	\\
0.04	&	-	&	333.9	&	329.5	&	418.2	\\
0.05	&	-	&	342.5	&	339.5	&	434.1	\\
0.1	&	473.7	&	478.5	&	473.7	&	480.7	\\
0.2	&	493.5	&	613.7	&	573.1	&	557.9	\\
0.3	&	518.1	&	638.9	&	648.0	&	683.4	\\
0.4	&	645.4	&	658.4	&	671.7	&	714.6	\\
0.5	&	662.7	&	691.5	&	690.0	&	734.0	\\
0.6	&	679.6	&	-	&	-	&	-	\\
0.7	&	695.6	&	-	&	-	&	-	\\
\hline
\end{tabular}
\caption{ \label{tab_BR_ll_ee_em_mm_htm}Lowest limits on a mass of the doubly charged scalar boson $M_{H^{\pm\pm}}$ for different branching ratios \cite{Aaboud:2017qph}. We removed data which corresponds to the branching ratio region beyond what has been obtained in Fig. \ref{fig_BR_ll_ee_em_mm} 
within 2$\sigma$ range of the neutrino oscillation parameters.  }
\end{table}
\end{center}
The sharpest limit from ATLAS on  $M_{H^{\pm\pm}}$ is that the ${H^{\pm\pm}}$ mass should be larger than 870 GeV for the left-handed triplet doubly charged scalar boson field, assuming the 100\% branching ratio for the $H^{\pm\pm} \to l^\pm l^\pm$ decay ($l^\pm = e^\pm, \mu^\pm$). However, it is possible to lower down the limit to 450 GeV for a 10\% leptonic decay branching ratio (see Fig.~13 d in \cite{Aaboud:2017qph}). On the other hand, the decays into a $\tau$ lepton are not considered in the above analysis. In Tab.~\ref{tab_BR_ll_ee_em_mm_htm} we present branching ratios for those channels and the result
for the $ee$, $e\mu$ and $\mu\mu$ decays, within the $\pm 2 \sigma$ range of the oscillation parameter space. For other channels including the $\tau$ we refer to  \cite{Garayoa:2007fw}. The strength of lepton decay channels depends strongly on the neutrino masses, their hierarchies and oscillation parameters. It is possible to find the parameter space where the branching ratio for the particular lepton channel is small regardless  $v_\Delta$ even if the cumulative lepton channel dominates over the~$W$ boson channel (the relative lepton decay contributions $\Gamma(H^{\pm\pm}~\to~l~l')/\sum~\Gamma(H^{\pm\pm}~\to~l_i~l_j)$ do not depend on the triplet VEV $v_\Delta$).

\begin{center}
\begin{figure}[h!]
\begin{tabular}{cc}
\includegraphics[width=0.49\textwidth,angle=0]{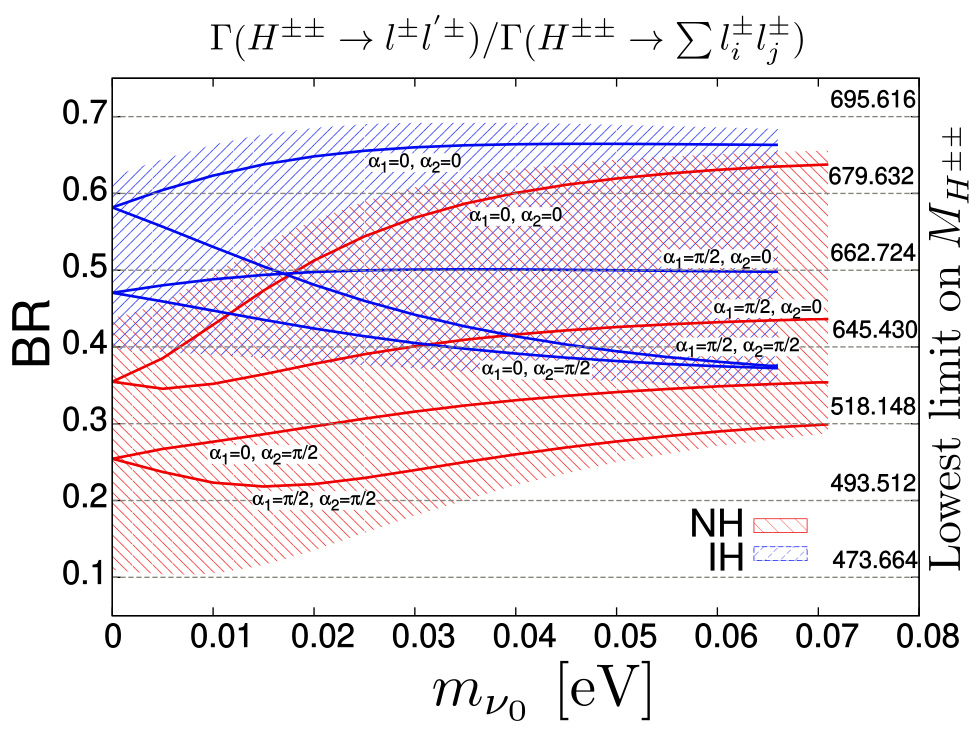} &
\includegraphics[width=0.49\textwidth,angle=0]{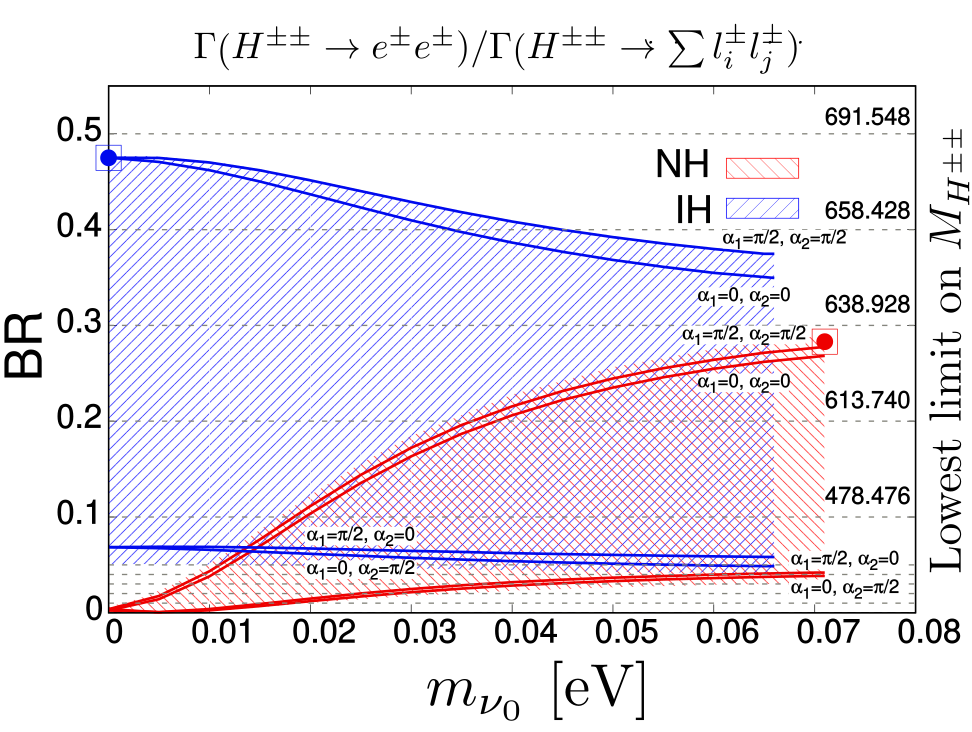}\\
 (a) & (b) \\
\includegraphics[width=0.49\textwidth,angle=0]{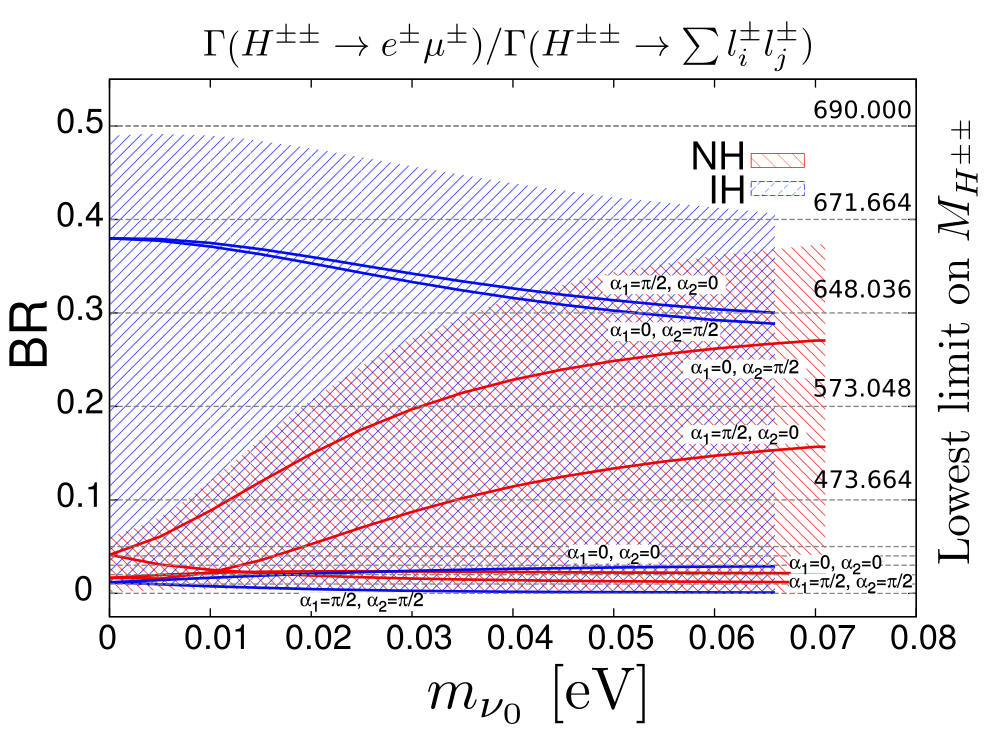} &
\includegraphics[width=0.49\textwidth,angle=0   ]{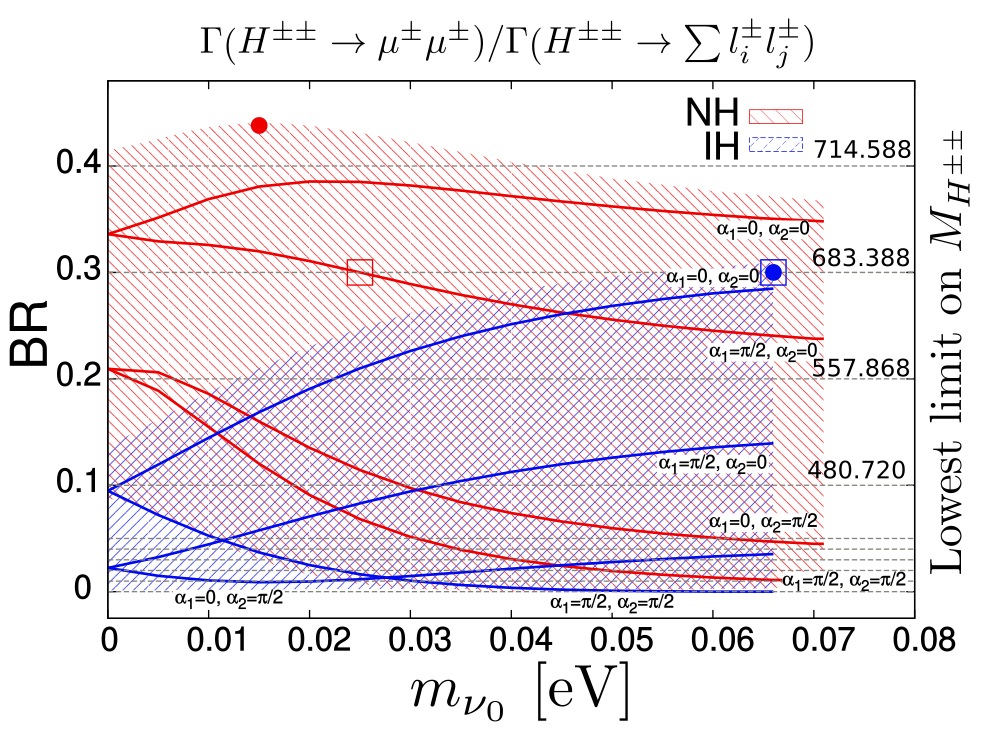}\\
 (c) & (d) \\
\end{tabular}
\caption{\label{fig_BR_ll_ee_em_mm} $H^{\pm\pm}$ decay branching ratios, $l$,$l'$ = $e$,$\mu$ within the \texttt{HTM} model, with corresponding lower limits on the doubly charged scalar particle's masses \cite{Aaboud:2017qph}. Neutrino parameters are within the $\pm 2 \sigma$ range,  Tab.~\ref{neutrino_data}. 
Solid lines present the result for best fit of neutrino parameters and particular values of Majorana phases. We have marked the points used for further calculations with $\msquare$ and $\bullet$ which satisfy the following conditions: $\bullet$ gives maximum possible BR for NH and IH cases with $M_{H^{\pm\pm}}$ = 1000 GeV; $\msquare$ gives lower BR values which allows for  
$M_{H^{\pm\pm}} = 700$~GeV.}  
\end{figure}
\end{center}

 We combine the data both from the LHC limits \cite{Aaboud:2017qph} and neutrino parameters within the~$\pm 2\sigma$ range given in Tab.~\ref{neutrino_data} and compute the lowest limit on the  doubly charged scalar boson mass\footnote{There is in principle a subtlety in the fact that the branching ratios are not directly measured. Instead, the rate for $4l$ production is measured. Here we rely on basic analysis and outcome given by the ATLAS collaboration.}.  
 In Tab.~\ref{tab_BR_ll_ee_em_mm_htm} we removed the BR values that are forbidden due to the neutrino oscillation parameters. Another interesting conclusion from this table is that within the \texttt{HTM} the doubly charged scalar boson cannot be lighter than 473~GeV for the normal neutrino mass scenario (and 518 GeV for the inverted mass hierarchy), see~Fig.~\ref{fig_BR_ll_ee_em_mm}~(a). Finally, the lowest mass limit on  $M_{H^{\pm\pm}}$ within  \texttt{HTM}  is 473.7 GeV for NH and 645.4 GeV for IH with
BR$(H^{\pm\pm} \to ll') = (\Gamma (H^{\pm\pm} \to e^\pm e^\pm + e^\pm \mu^\pm + ~\mu^\pm \mu^\pm))/\Gamma (H^{\pm\pm}~\to~\sum_{i,j} l_i^\pm l_j^\pm) \ge 0.1$ and $0.4$, respectively,  where  
$l_{i,j}=e, \mu, \tau$.
The most severe limit at 734 GeV comes from the same sign muon channel when BR is 50\%. 
{In conclusion, when assuming the complete scenarios with ${H^{\pm\pm}}$ decays to all the leptons, still  $M_{H^{\pm\pm}}$ can be relatively light.}
\\

Our main aim is to analyse the final four lepton ($4l$) signals which can be potentially seen at the colliders. The dominant signatures are $e^+e^+e^-e^-$ and $\mu^+\mu^+\mu^-\mu^-$ final states within both \texttt{HTM} and \texttt{MLRSM} models. In \texttt{MLRSM} they are not bounded by the neutrino oscillation parameters since the $H_{1,2}^{\pm\pm}-l-l$ vertex is related to the heavy right-handed neutrino masses and parameters, as discussed in section \ref{vertex_LRSM}. Within the \texttt{HTM} model these $4l$ contributions are restricted by the light neutrino oscillation data. Using branching ratios shown in Fig. \ref{fig_BR_ll_ee_em_mm} we compute two parameter sets (for normal and inverted hierarchy) for which the branching ratio for $e^\pm e^\pm$ and $\mu^\pm \mu^\pm$ are the highest. We collect the chosen parameters in Tab.~\ref{mass_limit_phases_htm}. We choose two benchmark masses for the collider analyses: $M_{H^{\pm\pm}}$ = 700 GeV (which can be probed at very high energies in $e^+ e^-$ collision, when available, see section \ref{subseepp}) and $M_{H^{\pm\pm}}$~=~1000~GeV (this higher mass range can be probed without problems at the HL-LHC and FCC-hh, see Fig.~\ref{sec_pp_hcchcc}). For the $e^\pm e^\pm$ decay channel we chose the same neutrino parameters because within the whole neutrino parameter space $M_{H^{\pm\pm}} = 700$ GeV and $M_{H^{\pm\pm}} = 1000$ GeV are not excluded. For the $\mu^\pm \mu^\pm$ channel we chose the maximum possible BR for $M_{H^{\pm\pm}} = 1000$ GeV and BR=0.3 for $M_{H^{\pm\pm}} = 700$ GeV to keep the bound on the doubly charged scalar particle's mass lower than 700 GeV.
\begin{center}
\begin{table}[h!]
\centering
\begin{tabular}{|c|c|c|c|c|c|}
\hline
\multirow{2}{*}{$M_{H^{\pm\pm}}$} &
\multirow{2}{*}{$H^{\pm\pm}\to XX$} & \multicolumn{4}{c|}{\texttt{HTM}}  \\
\cline{3-6}
& & \multicolumn{2}{c|}{NH} & \multicolumn{2}{c|}{IH} \\
 \hline
\multirow{6}{*}{700 GeV ($\msquare$)} &  &
 \multirow{3}{*}{BR=0.283} & $\alpha_1 = \frac{\pi}{2}$ & \multirow{3}{*}{BR=0.475}  & $\alpha_1 = \frac{\pi}{2}$ \\
 & $ee_{\mbox{max}}$ & & $\alpha_2 = \frac{\pi}{2}$ & & $\alpha_2 = \frac{\pi}{2}$ \\
 & \scriptsize{$BR < 0.5$} & & $m_{\nu_0}= 0.071$ eV & & $m_{\nu_0}=0$ \\
 \cline{2-6}
 &  &
 \multirow{3}{*}{BR=0.3} & $\alpha_1 = \frac{\pi}{2}$ & \multirow{3}{*}{BR=0.3} & $\alpha_1 = 0$ \\
 & $\mu\mu_{\mbox{max}}$ & & $\alpha_2 = 0$ & & $\alpha_2 = 0$ \\
 & \scriptsize{$BR < 0.3$} & & $m_{\nu_0}= 0.025$ eV & & $m_{\nu_0}=0.066$ eV \\
  \hline
\multirow{6}{*}{1000 GeV ($\bullet$)} & \multirow{3}{*}{$ee_{\mbox{max}}$} &
 \multirow{3}{*}{BR=0.283} & $\alpha_1 = \frac{\pi}{2}$ & \multirow{3}{*}{BR=0.475}  & $\alpha_1 = \frac{\pi}{2}$ \\
 & & & $\alpha_2 = \frac{\pi}{2}$ & & $\alpha_2 = \frac{\pi}{2}$ \\
 & & & $m_{\nu_0}= 0.071$ eV& & $m_{\nu_0}=0$ \\
 \cline{2-6}
 & \multirow{3}{*}{$\mu\mu_{\mbox{max}}$} &
 \multirow{3}{*}{BR=0.438} & $\alpha_1 = 0$ & \multirow{3}{*}{BR=0.3} & $\alpha_1 = 0$ \\
 & & & $\alpha_2 = 0$ & & $\alpha_2 = 0$ \\
 & & & $m_{\nu_0}= 0.015$ eV& & $m_{\nu_0}=0.066$ eV\\
  \hline
\end{tabular}
\caption{\label{mass_limit_phases_htm} Chosen parameter set for maximum branching ratios BR$(H^{\pm\pm} \to ee)$ and BR$(H^{\pm\pm} \to \mu\mu)$ and  for the best fit neutrino parameters in Tab.~\ref{neutrino_data}. Corresponding benchmark points are marked in Fig. \ref{fig_BR_ll_ee_em_mm} (b) and (d) with $\msquare$ ($M_{H^{\pm\pm}}$ =~700~GeV) and $\bullet$ ($M_{H^{\pm\pm}}$ =~1000~GeV).}
\end{table}
\end{center}

\subsection{\texttt{MLRSM}, a choice of benchmark parameters and ${H^{\pm\pm}_{1,2}}$ decay scenarios} 
\label{results_decay_LRSM}

Contributing vertices to the non-leptonic decay channels stem from the kinetic term and scalar potential (see Eqs. 19 and 25 in \cite{Duka:1999uc}). Relevant decay modes of doubly charged scalar bosons and respective strength of couplings are gathered in Tab.~\ref{H12cc_decays}. The emboldened processes in the table dominate for $v_L=\rho_4=0$ and $\xi \to 0$ \cite{Bambhaniya:2014cia,Chakrabortty:2016wkl,Dekens:2014ina}, see the Appendix for details. Apart from the values of vertices, we need to take into account the mass spectrum. 
To suppress the FCNC processes some of neutral scalar particles have to be heavier than 10 TeV. As a consequence, the mass of $H_2^\pm$ should above 10 TeV, (\ref{h10_mass_LRSM}) and (\ref{h2c_mass_LRSM}). Therefore we can neglect the $H_2^{\pm \pm}$ decay to the $H_2^\pm$ scalar boson for CLIC and LHC energies.
From (\ref{h10_mass_LRSM}) it is easy to find that the triplet VEV should fulfil an inequality: $v_R > \sqrt{2} \: 10^3 / \/\sqrt{\alpha_3} $ [GeV]. Because $\alpha_3$ is a quartic coupling (four-scalar interaction) it contributes to the  $2\to 2$ scattering and the unitarity condition requires  $\alpha_3  <8 \pi$ \cite{Chakrabortty:2016wkl}. The triplet VEV $v_R$ has to be higher than $\sim 2800$ GeV that translate to $M_{W_2} > 1325$ GeV. So we can neglect the doubly charged scalar bosons pair production with the subsequent decay to the heavy gauge boson pair $H_2^{\pm\pm} \to W_2^\pm + W_2^\pm$ for energies 
 lower than $2 M_{W_2}$. 
\begin{table}[h!]
\centering
\begin{tabular}{@{}|cccl@{\hspace{1cm}}cccl|}
\hline
$H_1^{\pm\pm}$ & $\to$ &  $W_1 + W_1$ & $\sim \cos^2{\xi}\;v_L$ & $H_2^{\pm\pm}$ & $\to$ & $W_1 + W_1$ & $\sim \sin^2{\xi}\;v_R$ \\
$H_1^{\pm\pm}$ & $\to$ & $W_1 + W_2$ & $\sim \cos{\xi}\sin{\xi}\;v_L$ & $H_2^{\pm\pm}$ & $\to$ & $W_1 + W_2$ & $\sim \cos{\xi}\sin{\xi}\;v_R$ \\
$H_1^{\pm\pm}$ & $\to$ & $W_2 + W_2$ & $\sim \sin^2{\xi}\;v_L$ & \boldmath{$H_2^{\pm\pm}$} & \boldmath{$\to$} & \boldmath{$W_2 + W_2$} & \boldmath{$\sim \cos^2{\xi}\;v_R$} \\
\boldmath{$H_1^{\pm\pm}$} & \boldmath{$\to$} & \boldmath{$H_1^\pm + W_1$} & \boldmath{$\sim \cos{\xi}\;g_L$} & $H_2^{\pm\pm}$ & $\to$ & $H_2^\pm + W_1$ & $\sim \sin{\xi}\;g_R$ \\
$H_1^{\pm\pm}$ & $\to$ & $H_1^\pm + W_2$ & $\sim \sin{\xi}\;g_L$ & \boldmath{$H_2^{\pm\pm}$} & \boldmath{$\to$} & \boldmath{$H_2^\pm + W_2$} & \boldmath{$\sim \cos{\xi}\;g_R$} \\
$H_1^{\pm\pm}$ & $\to$ & $H_1^\pm + H_1^\pm$ & $\sim \rho_2\;v_L$ & $H_2^{\pm\pm}$ & $\to$ & $H_1^\pm + H_1^\pm$ & $\sim \rho_4\;v_R$ \\
$H_1^{\pm\pm}$ & $\to$ & $H_2^\pm + H_2^\pm$ & $\sim \rho_4\;v_L$ & \boldmath{$H_2^{\pm\pm}$} & \boldmath{$\to$} & \boldmath{$H_2^\pm + H_2^\pm$} & \boldmath{$\sim \rho_2\;v_R$} \\
$H_1^{\pm\pm}$ & $\to$ & $H_1^\pm + H_2^\pm$ & $\sim \kappa_2$ & 
$H_{1,2}^{\pm\pm}$ & $\to$ & $H_{2,1}^{\pm\pm} + H_0^0 $  & $\sim \rho_4\;v_L$ \\
$H_{1,2}^{\pm\pm}$ & $\to$ & $H_{2,1}^{\pm\pm} + H_1^0 $  & $\sim \rho_4\;v_L$ & 
$H_{1,2}^{\pm\pm}$ & $\to$ & $H_{2,1}^{\pm\pm} + H_2^0 $  & $\sim \rho_4\;v_L$ \\
$H_{1,2}^{\pm\pm}$ & $\to$ & $H_{2,1}^{\pm\pm} + H_3^0 $  & $\sim \rho_4 v_R$ & 
$H_{1,2}^{\pm\pm}$ & $\to$ & $H_{2,1}^{\pm\pm} + A_2^0 $  & $\sim \rho_4 v_R$ \\
\hline
\end{tabular}
\caption{\label{H12cc_decays}Doubly charged scalar boson decay channels  to scalar and gauge bosons in  \texttt{MLRSM}. We have listed all possible vertices, thickening the dominating processes assuming that the left triplet VEV $v_L$ is equal to zero and keeping in mind experimental limits on the $W_1-W_2$ mixing angle $\xi < 10^{-2}$ \cite{Czakon:1999ga,Tanabashi:2018oca} and setting the $\rho_4$ parameter to zero \cite{Bambhaniya:2014cia,Chakrabortty:2016wkl}. The leptonic decays are analysed separately.}
\end{table}
%

In Tab.~\ref{H12cc_decays} we present the other possible decay channels of $H_{1,2}^{\pm\pm}$ and corresponding vertices. Most of them are negligible due to model's consistency \cite{Bambhaniya:2014cia,Chakrabortty:2016wkl}, only the bold decay channels can be sunstantial. The $H_{1,2}^{\pm\pm}$ decay to $H^\pm_2$ is not possible for CLIC and LHC energies because of the FCNC limits (\ref{h2c_mass_LRSM}). Vertices contributing to the $H_{1,2}^{\pm\pm}$ decays to $W_1$, $W_2$ can be large and are included in analysis leading to final four lepton signals.
  
Regarding $H_1^{\pm\pm}$, its decay to $H_1^\pm + W_1^\pm$ is limited by Higgs potential parameters and, as proved  analytically in \cite{Bambhaniya:2015wna}, the allowed split $\Delta M_H = M_{H_1^{\pm\pm}}-M_{H_1^\pm}$
can not exceed value 65.3 GeV. 
 
We choose the benchmark points for $v_R=6$ TeV and $15$ TeV. The first value falls in energy range of LHC with $pp \to W_2 \to l N_l \to ll W_2^* \to ll q\bar{q}'$, assuming that $M_{{N}_i} < M_{W_2}$. Corresponding experimental results can be found in \cite{Sirunyan:2018pom,Sirunyan:2018vhk}.  
 We assume that the doubly charged scalar masses are degenerate and choose two benchmark points: $700$ GeV and $1000$ GeV. For the  $M_{H_{1,2}^{\pm\pm}}$~=~700~GeV case we keep the leptonic branching ratio limits as given in Tab.~\ref{tab_BR_ll_ee_em_mm_htm}, that means BR$(H^{\pm\pm}_{1,2} \to ee)~<~0.5$ and BR$(H^{\pm\pm}_{1,2} \to \mu\mu)~<~0.3$. {\color{black}{Tab. \ref{BR_vs_mass_HL_HR} presents the maximum possible branching ratios for $M_{H^{\pm\pm}_{1,2}}=700$ GeV. 
 }} 
 For doubly charged Higgs boson mass of 1000 GeV there is no relevant experimental limits and the maximum branching ratios for $ee$ or $\mu\mu$ decays can reach 100\% also in a case of ${H_{1}^{\pm\pm}}$. 
 Here the situation is different than in $\texttt{HTM}$ where upper bounds for ${H^{\pm \pm}}$ branching ratios are given. As discussed in Sections \ref{tripltesneutr} and \ref{HTMconstr},  neutrino Yukawa couplings can be rewritten in terms of oscillation parameters and $v_\Delta$ and experimental data restricts possible branching ratios in a substantial way.
  In addition, depending on the branching ratios, the lowest limit on the mass of a doubly charged scalar can be obtained.  However, in the context of \texttt{MLRSM}, 
leptonic branching ratios for ${H^{\pm \pm}}$ depend in addition on $v_R$ scale and heavy neutrino masses and couplings. This freedom makes it possible to reach full leptonic decays for $M_{H^{\pm \pm}}$, as given in Tab. \ref{BR_vs_mass_HL_HR} and Tab. \ref{mass_limit_phases_lr}. 
\begin{center}
\begin{table}[h!]
\centering
\begin{tabular}{|c|c|c|c|}
\cline{2-4}
\multicolumn{1}{c|}{} & $ee$ & $\mu\mu$ & $ee+\mu\mu$ \\
\hline
$\mbox{BR}_{H_1^{\pm\pm}}$ & 0.5 & 0.3 & 0.7\\
\hline
$\mbox{BR}_{H_2^{\pm\pm}}$ & 1.0 & 0.8 & 1.0\\
\hline
\end{tabular}
\caption{\label{BR_vs_mass_HL_HR} Maximum branching ratios for $H_{1,2}^{\pm\pm} \to XX$ and $M_{H_{1,2}^{\pm\pm}}$= 700 GeV. Results for $H_1^{\pm\pm}$   
coincides with the \texttt{HTM} case in Tab.~\ref{tab_BR_ll_ee_em_mm_htm}. Branching ratios for $H_2^{\pm\pm}$  are due to right-handed leptonic couplings as analysed in \cite{Aaboud:2017qph}.
}
\end{table}
\end{center}

Tab.~\ref{mass_limit_phases_lr} shows chosen, allowed values of heavy neutrino masses for given maximal branching ratios. They are consistent with assumption $M_{W_2} \geq M_N$ discussed in section \ref{MLRSMconstr}, and a correlation between the masses which are proportional to $v_R$  \cite{Gluza:1993gf,Chakrabortty:2016wkl}. In section \ref{results_decay_HTM} we have obtained the lowest limits for $M_{H^{\pm \pm}}$  as a function of the lightest neutrino mass for a given  $H^{\pm \pm}$ branching ratio, arguing that $M_{H^{\pm \pm}}$ at the level of 700 GeV is still possible within \texttt{HTM}. The lowest limits on masses of ${H^{\pm \pm}_{1,2}}$ Higgs bosons have been obtained in \cite{Bambhaniya:2014cia} by analyzing restrictions on the scalar potential.  
\begin{table}[h!]
\centering
\begin{tabular}{|c|c|c|c|c|c|}
\hline
\multirow{2}{*}{$M_{H_{1,2}^{\pm\pm}}$} &
\multicolumn{4}{c|}{\texttt{MLRSM}} &  \multirow{2}{*}{\small{$H_{1,2}^{\pm\pm} \to$}} \\
\cline{2-5}
&  \multicolumn{2}{c|}{$v_R = 6$ TeV} & \multicolumn{2}{c|}{$v_R = 15$ TeV } &\\
 \hline
\multirow{6}{*}{700 GeV} &  
 \multirow{6}{*}{ $\mbox{BR}^{ee,\mu \mu}_{H_{1,2}^{\pm \pm}}=0.123$} &\multirow{6}{*}{\begin{tabular}{@{}c@{}}
$ M_{N_1} = 250$ \\ $M_{N_2} = 250$  \\ $M_{N_3} = 620$ \\  \end{tabular} } & \multirow{3}{*}{ \begin{tabular}{@{}c@{}}
 $\mbox{BR}^{ee}_{H_{1,2}^{\pm \pm}}=0.5$ \\ $\mbox{BR}^{\mu \mu}_{H_{1,2}^{\pm \pm}}=0.25$ \end{tabular}}  & \multirow{3}{*}{\begin{tabular}{@{}c@{}} $M_{N_1} = 1300$ \\$M_{N_{2,3}}=918$ \end{tabular}} &\multirow{3}{*}{$4e$} \\
  & & & & & \\
  & & & & & \\
 \cline{4-6}
 & & & \multirow{3}{*}{\begin{tabular}{@{}c@{}}
 $\mbox{BR}^{\mu\mu}_{H_{1,2}^{\pm\pm}}=0.3$ \\ 
 $\mbox{BR}^{ee}_{H_{1,2}^{\pm\pm}}=0.4$ \end{tabular}} & \multirow{3}{*}{\begin{tabular}{@{}c@{}}
 $M_{N_1} = 1300$ \\ $M_{N_{2,3}} = 1130$  \end{tabular}}  & \multirow{3}{*}{$4\mu$}   \\
 & & & & & \\
 & & & & & \\ 
  \hline 
  \multirow{6}{*}{1000 GeV} &  
 \multirow{6}{*}{
 $\mbox{BR}^{ee,\mu\mu}_{H_{1,2}^{\pm\pm}}=0.123$}  &\multirow{6}{*}{\begin{tabular}{@{}c@{}} $M_{N_1} = 250$  \\ $M_{N_2} = 250$   \\ $M_{N_3} = 620$ \end{tabular} } & \multirow{3}{*}{\begin{tabular}{@{}c@{}}
  $\mbox{BR}^{ee}_{H_{1,2}^{\pm\pm}} \sim 1$  
 \end{tabular}}  & \multirow{3}{*}{\begin{tabular}{@{}c@{}}   $M_{N_1} = 2867 $  \\ $M_{N_{2,3}}=300$ \end{tabular}}  & \multirow{3}{*}{$4e$} \\
  & & & & & \\
  & & & & & \\
 \cline{4-6}
 & & & \multirow{3}{*}{\begin{tabular}{@{}c@{}}
 $\mbox{BR}^{\mu \mu}_{H_{1,2}^{\pm\pm}}\sim 1$  \end{tabular}} & \multirow{3}{*}{\begin{tabular}{@{}c@{}}
 $M_{N_2} = 5000$ \\ $M_{N_{1,3}} = 300$
 \end{tabular}} & \multirow{3}{*}{$4\mu$}  \\
 & & & & & \\
 & & & & & \\
  \hline
\end{tabular}
\caption{\label{mass_limit_phases_lr} \texttt{MLRSM} parameters which maximize separately the  branching ratios BR$(H^{\pm\pm}\to ee)$ and BR$(H^{\pm\pm}\to\mu\mu)$ for $v_R=6$ TeV and $v_R=15$ TeV. A scenario with $v_R=6$ TeV has been covered already by the LHC analysis, and branching ratios are due to Tab.~\ref{BR_vs_mass_HL_HR}, based on    \cite{Sirunyan:2018pom,Sirunyan:2018vhk}. The heavy neutrino masses for $v_R=6$ TeV    fulfill the low energy constraints given in Tab.~\ref{LRSM_max_MN}.  $M_{N_1}$ is mostly restricted by the M\o{}ller scattering, while $M_{N_2}$ is bounded by $(g-2)_\mu$.}
\end{table} 
Before we present final results we will discuss the SM background for the considered leptonic final states.
\subsection{The four leptons background in $pp$ and $e^+e^-$ collisions \label{secbackgr}}
\label{s:ee_processes}

We are interested at estimation of  the Standard Model background for $pp \to l^+ l^- l^+ l^-$ and $e^+ e^- \to l^+ l^- l^+ l^-$ processes, where $l^\pm = e^\pm, \mu^\pm$. The four leptons production at LHC is discussed in~\cite{Bambhaniya:2013yca,Bambhaniya:2013wza}. The most relevant processes which contribute to the background are $t\bar{t}(Z/\gamma^*)$ and $(Z/\gamma^*)(Z/\gamma^*)$ production. 
To optimize the collider non-standard effects (decreasing SM tri- and four- lepton SM background and reducing the efficiency of misidentification of b-jets as leptons),  we use the following criteria and selection cuts 
\begin{enumerate}
\item[C1.] Lepton identification criteria: transverse momentum $p_T \ge 10$ GeV, pseudorapidity $|\eta|< 2.5$.
\item[C2.] Detector efficiency for electron (muon): 70\% (90\%). \label{eeEffic}
\item[C3.] Lepton-lepton separation: $\Delta R_{ll} \ge 0.2$.
\item[C4.] Lepton-photon separation $\Delta R_{l\gamma} \ge 0.2$ with $p_{T_\gamma}>10$ GeV.
\item[C5.] Lepton-jet separation $\Delta R_{lj} \ge 0.4$.
\item[C6.] Hadronic activity cut - within the cone of radius 0.2 around the lepton the hadronic activity should fulfill the inequality: $\sum p_{T_{hadron}} \ge 0.2 \times p_{T_l}$.
\item[C7.] Z-veto - the invariant mass of any same flavour and opposite charge lepton should satisfy the condition: $|m_{l_1 l_2} - M_{Z_1}| \ge 6 \: \Gamma_{Z_1}$.
\item[C8.] Hard $p_T$ cuts: $p_{T} (l_1) > 30$ GeV, $p_T (l_2) > 30$ GeV, $p_T (l_3) > 20$ GeV, $p_T (l_4) > 20$ GeV.
\item[C9.] Parton Distribution Functions (PDFs):  CTEQ6L1 \cite{Pumplin:2002vw,Hou:2019efy}.
\end{enumerate}

\begin{figure}[h]
\begin{center}
\includegraphics[ width=0.9\textwidth]{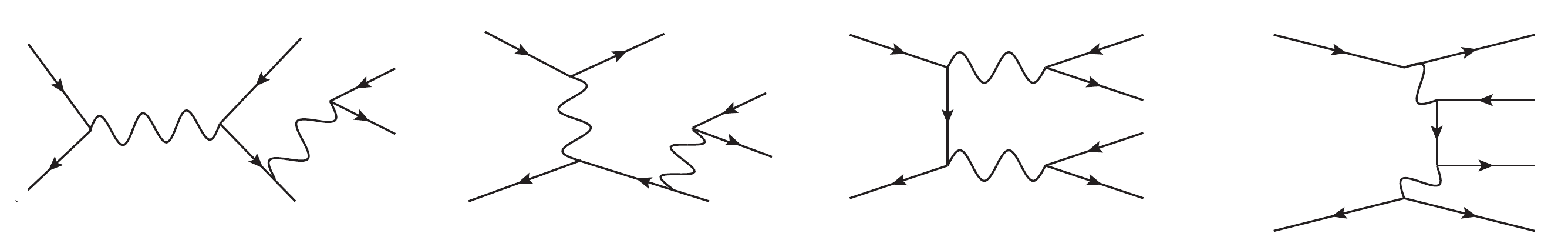}
\\
\hspace{10pt}(a)\hspace{90pt} (b)\hspace{90pt} (c)\hspace{90pt} (d)
\caption{Four lepton background diagrams in electron-positron colliders: $e^+ e^- \to e^+ e^-$ with FSR $e^+e^-$ pair emission  (a) and (b); with $Z/\gamma^*$ production (c) and with  multiperipheral processes (d).}
\label{ee_4l_diagrams_per}
\end{center}
\end{figure}

The results are gathered in Tab.~\ref{sm__pp_4l_background_tab}.
\begin{table}[h!]
\begin{center}
\begin{tabular}{|c|c||c|c|c|}
\hline
Process & Energy & $t\bar{t}(Z/\gamma^*)$ & 
 $(Z/\gamma^*)(Z/\gamma^*)$ & TOTAL \\
 \hline
\multirow{2}{*}{$\sigma (pp \to 4l)$ [fb]} & 14 TeV &  0.060  &  0.054  &  0.114  \\
\cline{2-5}
& 100 TeV &   0.58  &  0.20  &  0.78  \\
\hline 
\end{tabular}
\caption{ \label{sm__pp_4l_background_tab}Dominant Standard Model background contributions to four-lepton signals at the LHC with $\sqrt{s}= 14$ TeV and FCC-hh with  $\sqrt{s}= 100$ TeV after applying cuts given in the text. For the inclusive $t\bar{t}$ process the QCD NLO k-factor is 2.2 \cite{Cacciari:2008zb}, accordingly, for $t\bar{t}(Z/\gamma^*)$ it is $k=1.6$ \cite{Alwall:2014hca}, for $(Z/\gamma^*)(Z/\gamma^*)$ it is $k=1.5$ \cite{Chiesa:2020ttl}. Cross section values are given in fb.}
\end{center}
\end{table}
For the $e^+ e^-$ collision we consider scattering and annihilation channels with photon radiation, $(Z/\gamma^*)( Z/\gamma^*)$ production and multiperipheral processes in Fig.~\ref{ee_4l_diagrams_per}. The most relevant are diagrams b) and d). For $\sqrt{s} = 1500$ GeV we get $\sigma = 4.465$ 
fb before and $\sigma = 0.415$ fb after applying the cuts defined above. 

\subsection{Final four lepton signals within the \texttt{HTM} and \texttt{MLRSM} models, a comparison}
\label{comparison}
\label{seccomp}

The final signals depends on subsequent $H^{\pm \pm}$ decays ($\to 4e, 4 \mu, 2e 2\mu$) and suitable kinematic cuts.  In the \texttt{HTM} model we take benchmark points for the model connected with maximal $4e$ and $4\mu $ signals as given in Fig.~\ref{fig_BR_ll_ee_em_mm} (plots on right). Analogous parameters for \texttt{MLRSM} are given in Tab.~\ref{mass_limit_phases_lr}.

\begin{table}[h!]
\begin{center}
\begin{tabular}{|c|c||c|c||c|c|}
\hline
\multicolumn{6}{|l|}{SM background: $e^+e^- \to 4l$} \\
\hline
\multirow{2}{*}{$4e$} &\multicolumn{5}{|l|}{No cuts: $\sigma = 2.1$ fb } \\
 & \multicolumn{5}{|l|}{After cuts: $\;\;\sigma = 0.13$ fb, \;\; N = {\bf 200}}\\
\hline
\multirow{2}{*}{$4\mu$} &\multicolumn{5}{|l|}{No cuts: $\sigma = 0.07$ fb } \\
 & \multicolumn{5}{|l|}{After cuts: $\;\;\sigma = 0.005$ fb, \;\; N = {\bf 8}}\\
\hline
\hline
\multicolumn{2}{|c||}{\multirow{2}{*}{BSM signal: $e^+e^-\to H^{++} H^{--} \to 4l$}} & \multicolumn{2}{c||}{\texttt{HTM}} & \multicolumn{2}{c|}{\texttt{MLRSM}} \\
\cline{3-6}
 \multicolumn{2}{|c||}{} & NH & IH & $v_R$ = 6 TeV & $v_R$ = 15 TeV \\
\hline
\hline
\multirow{3}{*}{4$e$} & No cuts: & 0.19 fb & 0.53 fb & 
0.06 fb & 0.924 fb \\
\cline{2-6}
 & \multirow{2}{*}{After cuts:} & 
 0.02 fb & 
 0.06 fb & 
 0.007 fb  & 0.113 fb \\
 &  & N={\bf 30} & N={\bf 90} & N={\bf 10} & N={\bf 169} \\
 \hline
\multirow{3}{*}{4$\mu$} & No cuts: & 0.22 fb & 0.19 fb &  
0.06 fb  & 0.33 fb \\
\cline{2-6}
 & \multirow{2}{*}{After cuts:} & 
 0.08 fb & 
 0.08 fb & 
 0.03 fb  & 0.137 fb \\
 & & N={\bf 120} & N={\bf 120} & N={\bf 38} & N={\bf 205} \\
\hline
\end{tabular}
\caption{Four lepton signals at lepton colliders for the doubly charged scalar pair production with subsequent decays , $e^+e^-\to H^{++} H^{--} \to 4l$ for  $M_{H^{\pm\pm}}$ = 700 GeV and $\sqrt{s}=1.5$ TeV. In order to maximize signals in electron and muon channels we have applied different parameter sets from Tab.~\ref{mass_limit_phases_htm} (for \texttt{HTM}) and Tab.~\ref{mass_limit_phases_lr} (for \texttt{MLRSM}), see the main text for details. "N" estimates a number of final events with the assumed luminosity  $L=1500$ $fb^{-1}$.}
\label{tab_fin_ee}
\end{center}
\end{table}

The 4-lepton signals obtained for the $e^+e^-$ case are gathered
in Tab.~\ref{tab_fin_ee}. In section \ref{secbackgr} we defined the kinematic cuts which maximise the 4-lepton signals.
With assumed total luminosity, we can see that the SM background is comfortable small for muons and  the maximal $4\mu$ signal's prediction in \texttt{HTM} can be significant, which is not true in the case of electrons. The difference is enhanced by assumed detector efficiency for electrons (muons), see the cut C2 in section \ref{secbackgr}. For \texttt{MLRSM} chosen parameters in Tab.~\ref{mass_limit_phases_lr} and $v_R=6$ TeV the signals are small when compared to the SM background, especially for electrons. For muons a signal is $\sim 3$ times smaller than in \texttt{HTM}.   However, for $v_R=15$ TeV the signals for muons detection can be larger in \texttt{MLRSM},
since for that value of $v_R$, $M_N$ and $M_{W_2}$ values of parameters lie outside the region examined by the LHC (Fig \ref{CMS_data}).
In this case independent, maximal  branching ratios for ${H_{1,2}^{\pm\pm}} \to e^\pm e^\pm$ and ${H_{1,2}^{\pm\pm}} \to \mu^\pm \mu^\pm$ can reach 100\%, (Tab.~\ref{mass_limit_phases_lr}), which is not possible for \texttt{HTM} (Tab.\ref{mass_limit_phases_htm}). 

It should be noted that the $e^+e^- \to 4 l$ results in \texttt{MLRSM} depends strongly on interference effects and the chosen heavy neutrino parameters as the LHC exclusion data affects directly the t-channel contributions. In fact, comparing the \texttt{HTM} results with the \texttt{MLRSM} results for $v_R=6$ TeV, we  can see that the $4 l$  signals can be larger in \texttt{HTM} where the t-channel is negligible for all allowed parameters space (Fig.~\ref{epemhtm}), while in the \texttt{MLRSM} model the t-channel effects can still be large and comparable to the s-channel contributions (Fig.~\ref{epemLR}). However, in both models the signals are much below the SM background level. 

The maximal significance value $S\equiv S'/\sqrt{S'+B}$ where  $S'$ and $B$ are the total number of signal and background events is $S=14$ for $4\mu$ signals in \texttt{MLRSM} with $v_R$=15 TeV. For \texttt{HTM}, $S=11$ for both NH and IH neutrino mass scenarios and the $4 \mu$ signal. 
  The goal for HL-LHC is to deliver
about $L=0.25 \;ab^{-1} = 250$ $fb^{-1}$  per year with the aim of integrating a total luminosity in the range of 3 to
4.5 $ab^{-1}$ by the late 2030s
\cite{Strategy:2019vxc}. For the FCC-hh, defined by the target of 100 TeV proton-proton collisions,
a total integrated luminosity of 20-30~$ab^{-1}$ is considered \cite{Benedikt:2018csr}.

In Tab.~\ref{tab_fin_pp} the results are given for the final $4l$ signals. This time we consider higher ${H^{\pm\pm}}$ mass of 1 TeV. The kinematic cuts are defined in section \ref{secbackgr}.

\begin{table}[h!]
\begin{center}
\begin{tabular}{|c|c||c|c||c|c|}
\hline
\multicolumn{6}{|l|}{SM background: $pp \to 4l$} \\
\hline
\multirow{2}{*}{$4e$} &\multicolumn{5}{|l|}{No cuts:   $\sigma = 9.1$ [102.6] fb  } \\
 & \multicolumn{5}{|l|}{After cuts:   $\;\;\sigma = 0.0071$ [0.153] fb,  \;\;  N = {\bf   28 [3825]  }}\\
\hline
\multirow{2}{*}{$4\mu$} &\multicolumn{5}{|l|}{No cuts:   $\sigma = 9.1$   [100.6] fb } \\
 & \multicolumn{5}{|l|}{After cuts:   $\;\;\sigma = 0.022$    [0.62] fb, \;\; N = {\bf   88    [15\;167]}}\\
\hline
\hline
\multicolumn{2}{|c||}{\multirow{2}{*}{BSM signal: $pp \to H^{++} H^{--} \to 4l$}} & \multicolumn{2}{c||}{\texttt{HTM}} & \multicolumn{2}{c|}{\texttt{LRSM}} \\
\cline{3-6}
 \multicolumn{2}{|c||}{} & NH & IH & $v_R$ = 6 TeV & $v_R$ = 15 TeV \\
\hline
\hline
\multirow{6}{*}{4$e$} & \multirow{2}{*}{No cuts:} &   0.0038 fb  &  0.0109 fb  &    0.0029    fb &   0.136 fb    \\
& &   [0.39 fb]    &   [1.11 fb]    &    [0.87 fb]     &   [19.6 fb]    \\
\cline{2-6}
 & \multirow{4}{*}{After cuts:} 
 &   0.00032 fb    &   0.00092 fb    &    0.00026 fb     &    0.0116 fb    \\
 &  &   N={\bf 1.3}    &   N={\bf 3.7}    &   N={\bf 1.1}    &   N={\bf 45}    \\
 &&   [0.020 fb]    &   [0.059 fb]    &   [0.0407 fb]     &    [0.98 fb]    \\
 &  &   [N={\bf 484 }]    &   [N={\bf 1459}]    &   [N={\bf 1032}]    &   N={\bf [24\:492]}    \\
 \hline
\multirow{6}{*}{4$\mu$} & \multirow{2}{*}{No cuts:} &   0.0092     &   0.0039 fb    &   0.0029 fb    &   0.136 fb    \\
& &   [1.086 fb]    &   [0.48 fb]    &    [0.87 fb]     &    [19.6 fb]    \\
\cline{2-6}
 & \multirow{4}{*}{After cuts:} &   0.0031    &   0.00132 fb    &   0.001 fb     &   0.048 fb    \\
 &      &    N={\bf 11.5}    &   N={\bf 5.3}    &   N={\bf 4}    &   N={\bf 180}    \\
 &&   [0.202 fb]    &   [0.090 fb]    &   [0.181 fb]     &   [3.9 fb]    \\
 &  &   [N={\bf 5057}]    &    [N={\bf 2262}]    &   [N={\bf 4509}]    &   N=[{\bf 97\:199]}    \\
\hline
\end{tabular}
\caption{Four lepton signals for doubly charged scalar pair production with subsequent decays $pp \to H^{++} H^{--} \to 4l$ for  $M_{H^{\pm\pm}}$ = 1000 GeV and $\sqrt{s}=14\;  [100]$ TeV. In order to maximize signals in electron and muon channels we have applied different parameter sets from Tab.~\ref{mass_limit_phases_htm} (for \texttt{HTM}) and Tab.~\ref{mass_limit_phases_lr} (\texttt{MLRSM}), see the main text for details. "N" estimates the number of final events with assumed luminosity  $L=4 \;ab^{-1} = 4000$ $fb^{-1}$ for HL-LHC \cite{Strategy:2019vxc} and $L=25\; ab^{-1} = 25000$ $fb^{-1}$ for FCC-hh \cite{Benedikt:2018csr}.
}
\label{tab_fin_pp}
\end{center}
\end{table}

For $pp$ collisions the $4e$ channel gives comparable to the background signals in \texttt{MLRSM} with $v_R=15$ TeV. 
In $pp$ collision the lowest order t-channel is not present, so no destructive interference with the s-channel is possible. 
As given in Tab.~\ref{tab_fin_pp}, the maximal significance value is $S=11$ [290] for $4\mu$ in \texttt{MLRSM} with $v_R$=15 TeV for HL-LHC and FCC-hh, respectively. For \texttt{HTM} in the same $4\mu$ channel  $S<1$  both in NH and IH neutrino mass scenarios. So a detection of $4\mu$ signals above the background level at HL-LHC and FCC-hh would give a clear indication for the \texttt{MLRSM} model with high values of $v_R$.

	\begin{figure}[b!]
		    \centering
		    \includegraphics[angle=270,width=0.45\textwidth]{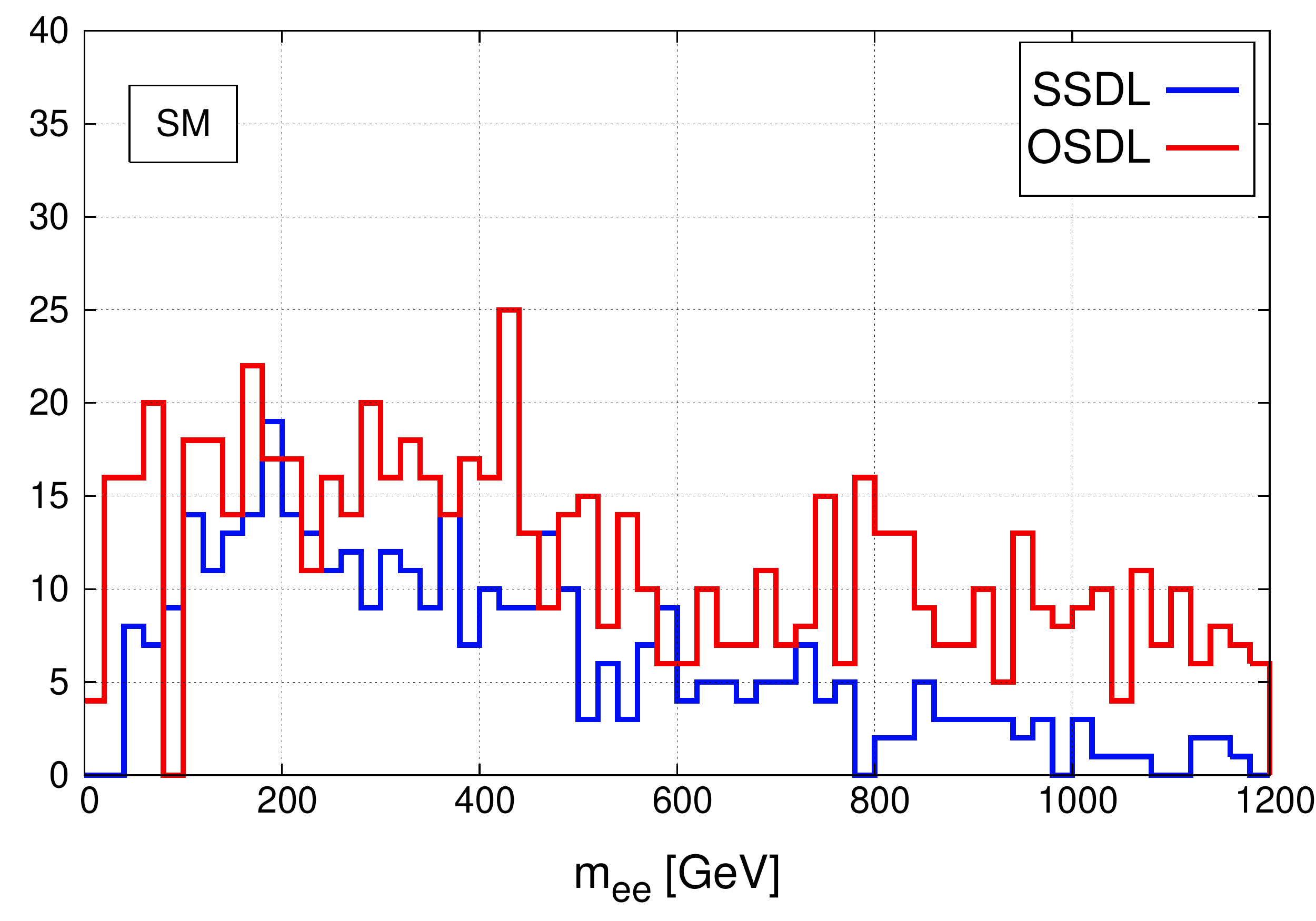}  \includegraphics[angle=270,width=0.45\textwidth]{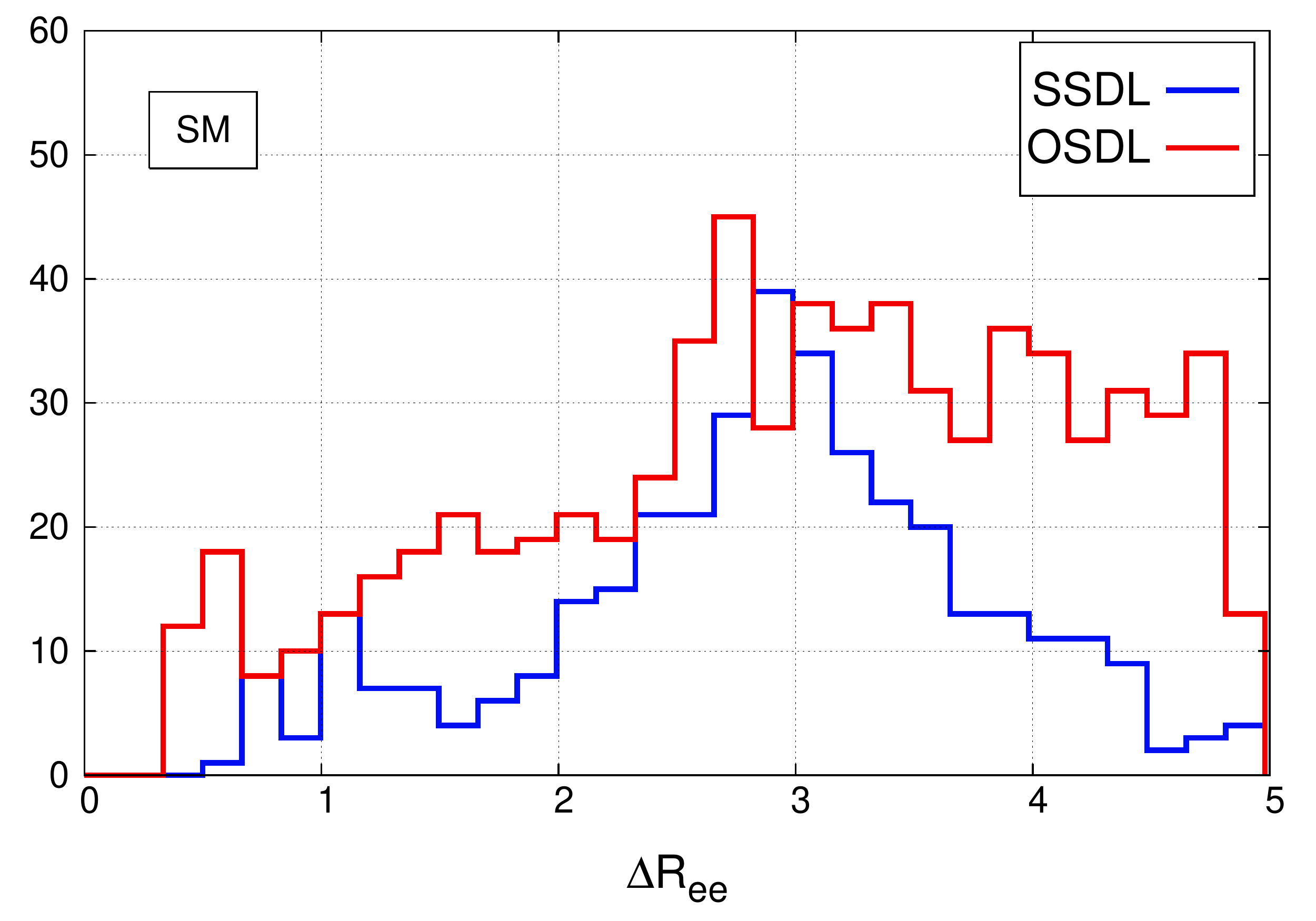} \\
		    \includegraphics[angle=270,width=0.45\textwidth]{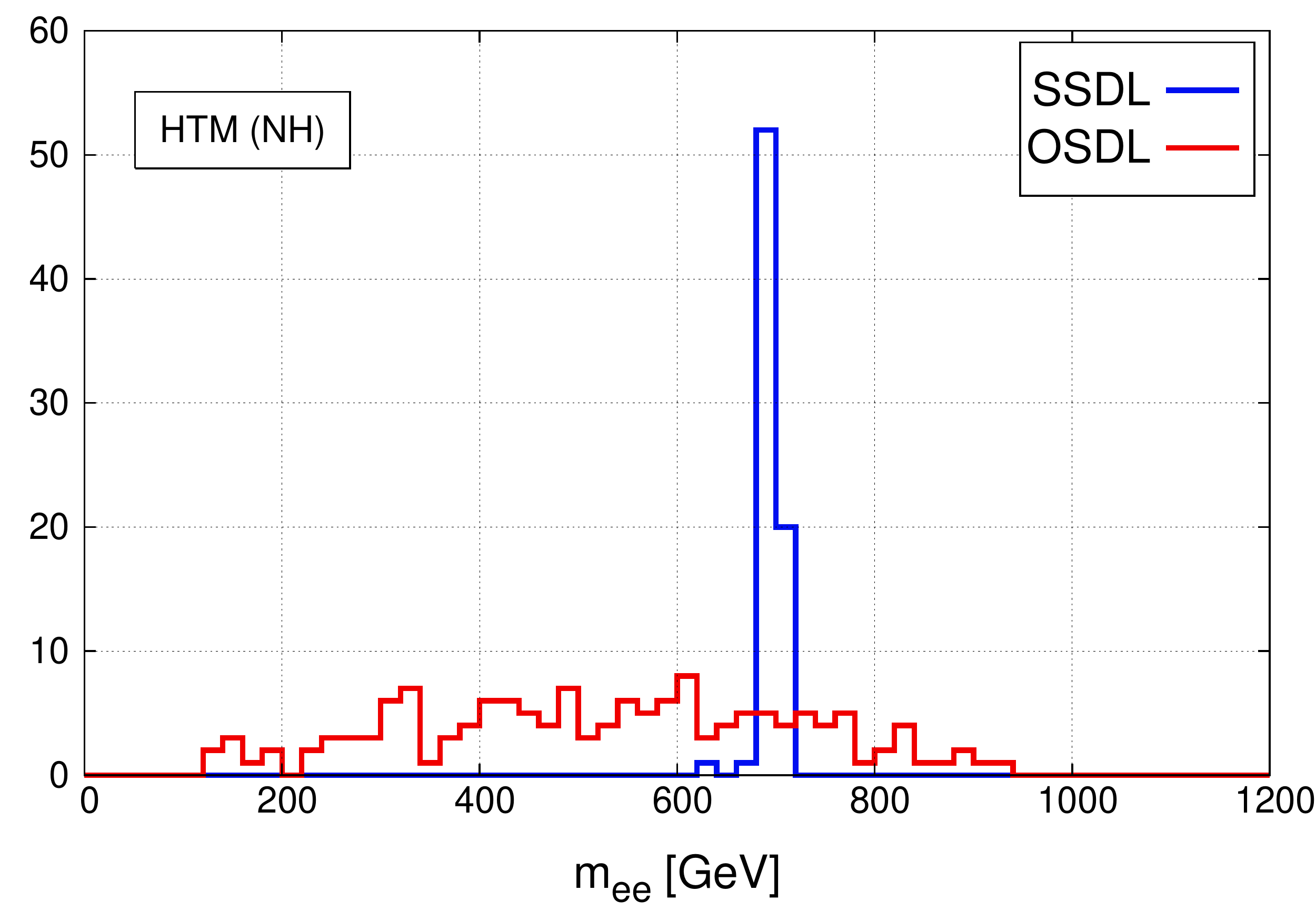}  \includegraphics[angle=270,width=0.45\textwidth]{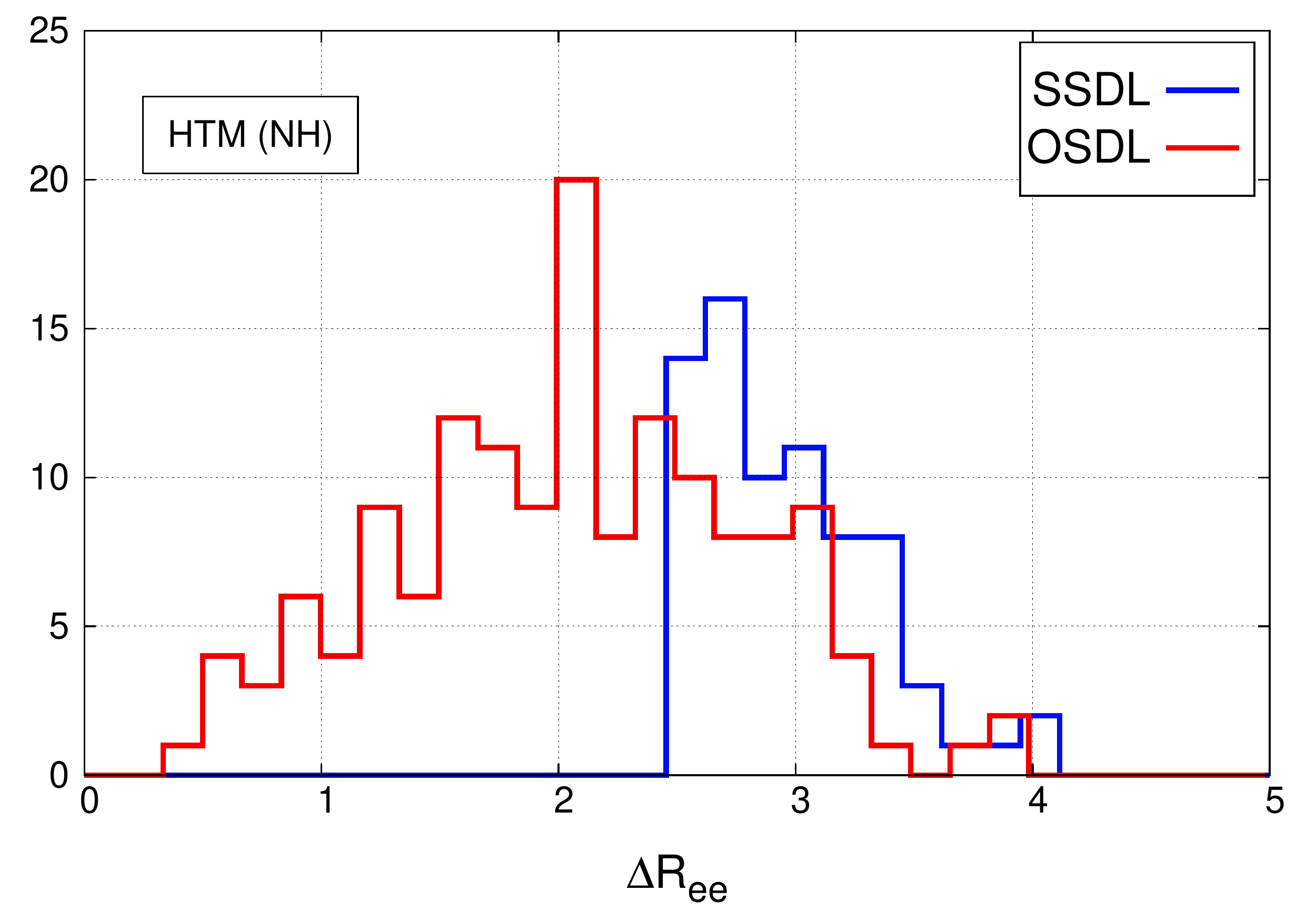} \\

		    \includegraphics[angle=270,width=0.45\textwidth]{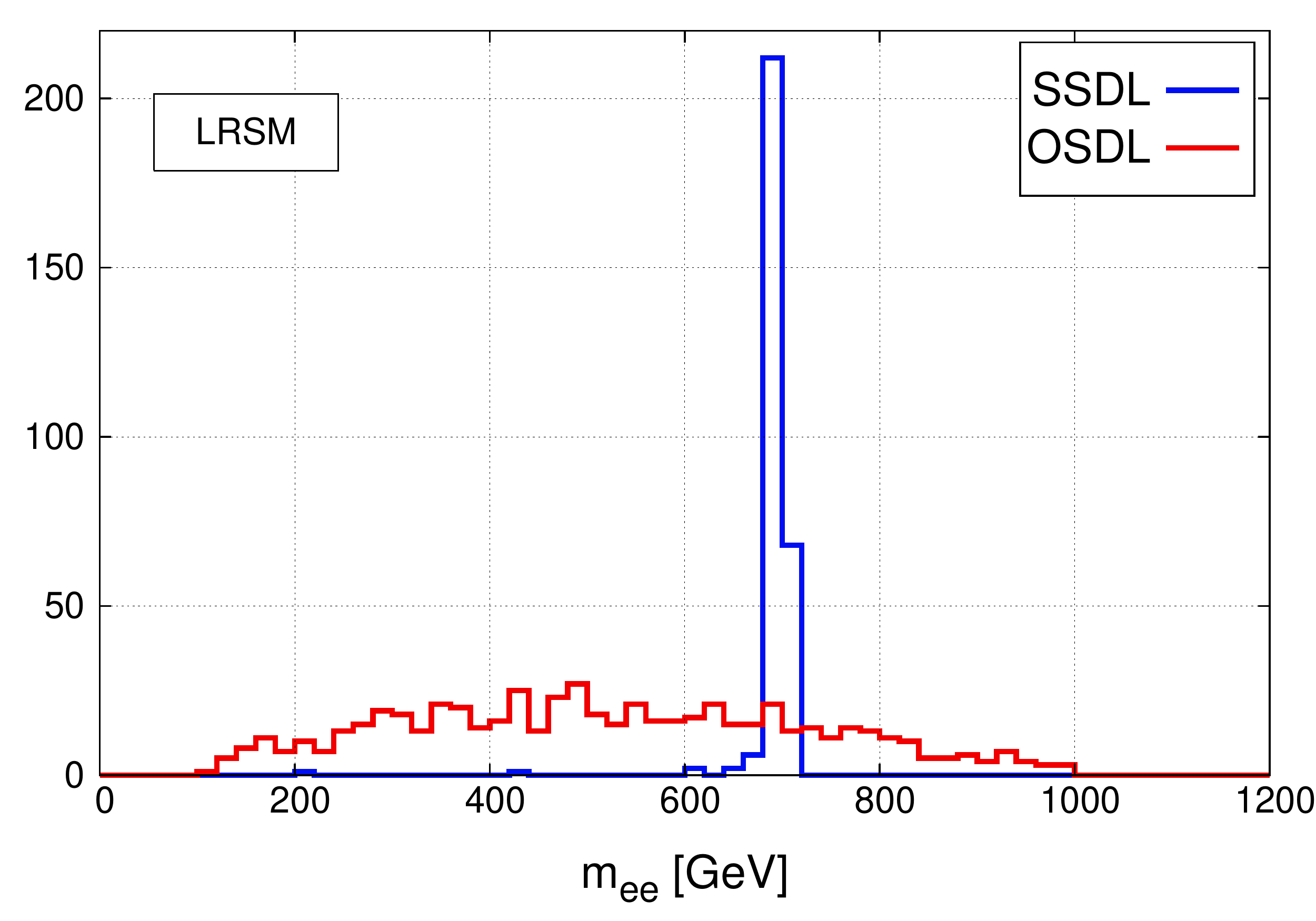}  \includegraphics[angle=270,width=0.45\textwidth]{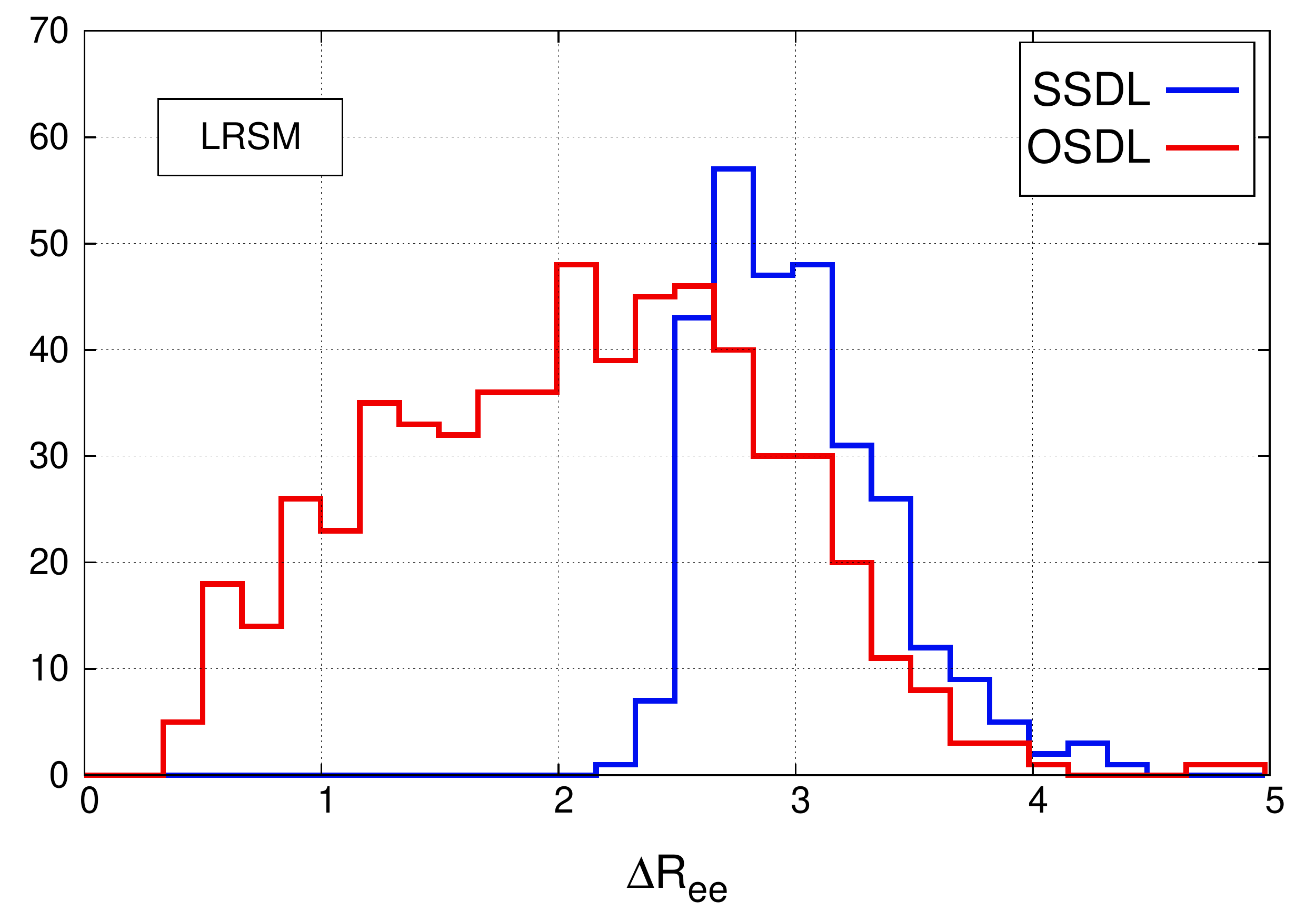}
		    
		    \includegraphics[angle=270,width=0.45\textwidth]{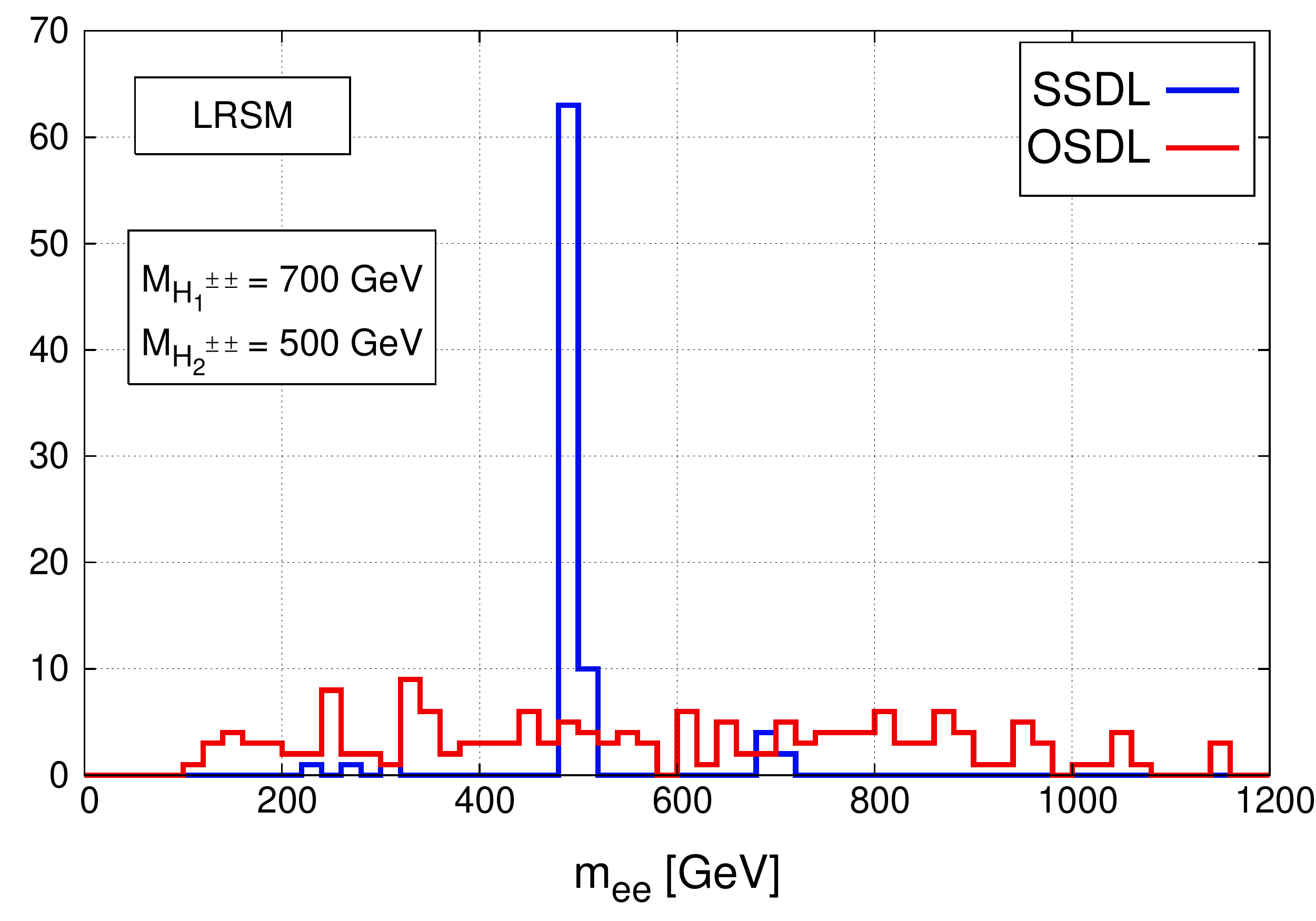}
		    \includegraphics[angle=270,width=0.45\textwidth]{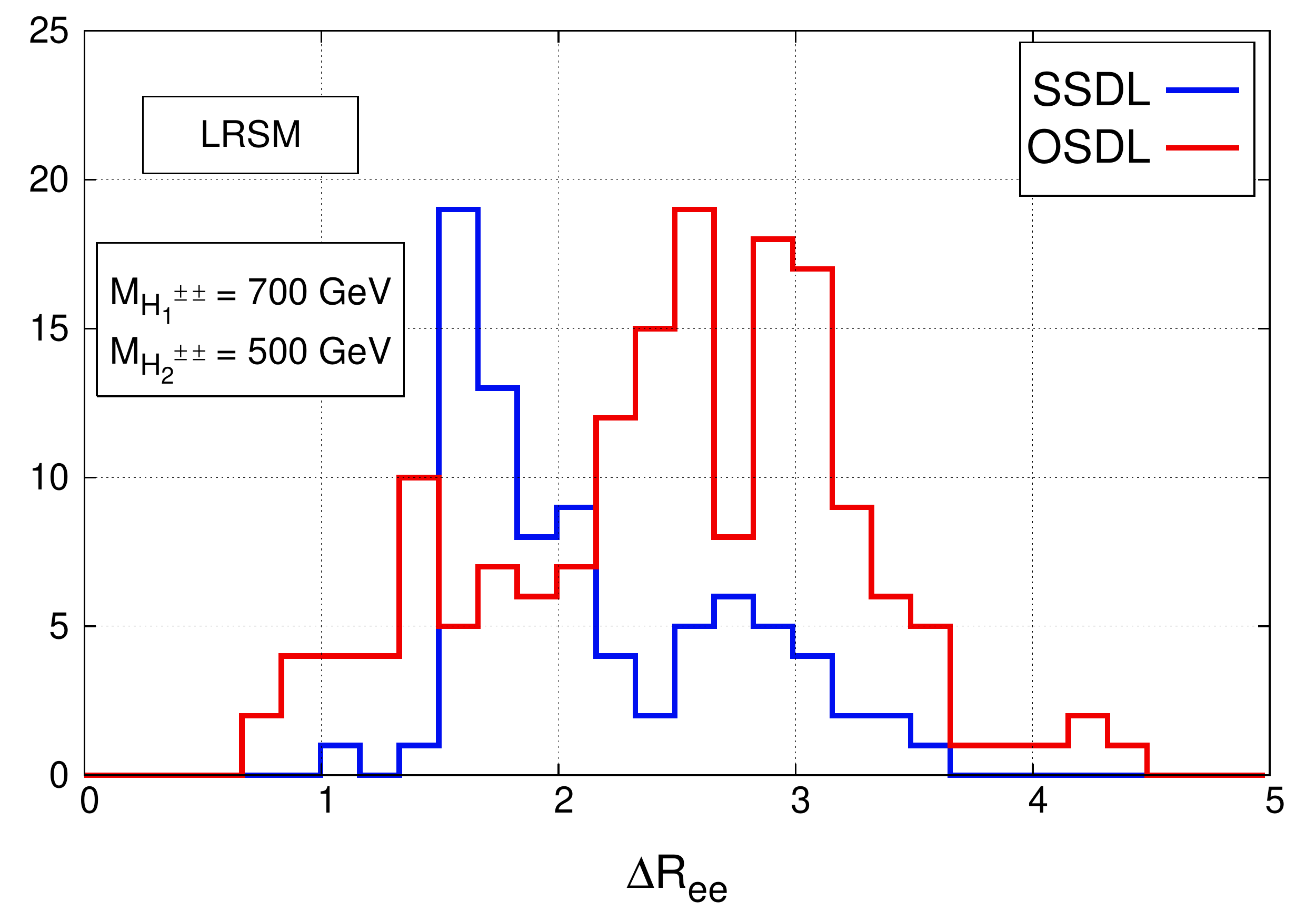}

		    \caption{Dilepton distributions for $e^+e^- \to 4e$. In the left column $m_{ee}$ distributions are shown for the SM background (top figure); \texttt{HTM} model with {NH} scenario and benchmark parameters as given in Tab.\ref{tab_fin_ee}  
		    (second figure); \texttt{MLRSM} with benchmark parameters for $v_R=6$ TeV as given in Tab.\ref{tab_fin_ee} (third figure). The bottom, last row plots are for \texttt{MLRSM} with $m_{{H_1^{\pm \pm}}}=700$ GeV and $m_{{H_2^{\pm \pm}}}=500$ GeV. On right, analogous figures for the $e-e$ separation observable $\Delta R_{ee}$ are given. }
		    \label{fig:distrib}
		\end{figure}

So far, we have focused on comparisons of the two models looking for specific signals for the total four charged lepton production rates and compared it to the background processes.
In this way, we can present clear differences in a prediction for the leptonic signals in both models. In particular, as shown in Tab.~\ref{tab_fin_ee} and Tab.~\ref{tab_fin_pp}, there are cases where the SM background is comparable or exceeds the BSM signal for electrons and positrons in the finals state. The question is if in such cases, dilepton distributions can help to identify small BSM signals and discriminate further BSM models.  In Fig.~\ref{fig:distrib} the SM background and BSM dilepton distributions for $e^+e^- \to 4e$  are given. We consider distributions of pairs of electrons/positrons (the same charge leptons) and electron-positron pairs, assigned as SSDL and OSDL, respectively. As we can see, the SM invariant mass SSDL and OSDL distributions are quite uniform, in opposite to the BSM signals where clear peaks are present for SSDL signals. The reason is obviously that the same sign dileptons originate from the same doubly charged particle. As we can see, in this case, though  $4e$  signals in Tab.~\ref{tab_fin_ee} are below the SM background, both \texttt{HTM} and \texttt{MLRSM} dileptons can be identified. It is less visible for the lepton-lepton separation $\Delta R_{ee}$ though the SSDL (OSDL) signals are enhanced for higher (lower) values of $\Delta R_{ee}$, respectively, in both considered models. Let us note that $m_{ee}$ and $\Delta R_{ee}$ distributions are very similar. This conclusion does not change for dimuon   distributions or hadron colliders. For non-degenerate doubly charged masses in \texttt{MLRSM} in Fig.~\ref{fig:distrib}, maximum number of same di-lepton events are with an invariant mass peak around $m_{{H_2^{\pm \pm}}}=500$ GeV  and that around $m_{{H_1^{\pm \pm}}}=700$ GeV is much smaller, as expected. Comparing \texttt{HTM} plots (second row) with non-degenerate \texttt{MLRSM} plots (bottom row), we can see that SSDL signals are shifted between both cases for both $m_{ee}$ and $\Delta R_{ee}$ distributions.
  
\section{Conclusions and outlook}
The doubly charged Higgs bosons $H^{\pm\pm}$ pair production at $e^+e^-$ and $pp$ colliders, with their subsequent decays to four charged leptons can give a very clear signal when searching for non-standard scalar particles effects without missing energy. We discuss a relation between vacuum expectation value of the triplet $v_\Delta$ and $H^{\pm\pm}$ couplings with leptons, taking into account constraints on $v_\Delta$
coming from low energy studies connected with the $\rho$-parameter, muon $(g-2)_\mu$, lepton flavor violation,  $e^+e^-$, LHC  processes,  and neutrino oscillations (normal and inverse mass scenarios).  
The low energy experiments rule out $v_\Delta$ below 10 eV (for $M_{H^{\pm\pm}} \sim 700$ GeV) both for  normal and inverted hierarchy, the strongest limit for non-zero mass of the lightest neutrino comes from LFV $\mu \to 3e$, see Fig.~\ref{current_future_limits} and Tab.~\ref{benchmarks}. As the Yukawa $H^{\pm \pm}-l-l'$ couplings are inversely proportional to  $v_\Delta$, the t-channel $e^+ e^- \to H^{++} H^{--}$ process could be enhanced, however, neutrino oscillation data makes it very small, and the s-channel dominates over allowed  $v_\Delta$, see Fig.~\ref{epemhtm}. Similarly, Yukawa $H^{\pm \pm}_{1,2}-l-l'$ couplings in \texttt{MLRSM} could dominate the cross section for $e^+ e^- \to H^{++}_{1,2} H^{--}_{1,2}$, however, $e^+e^-$ Bhabha and M\o{}ller processes makes it below the s-channel contribution, see Fig.~\ref{epemLR}. 
These two cases show nicely how important are present SM and LFV experimental data, allowing to  predict properly BSM signals in colliders studies. 
Altogether, $H^{\pm\pm}$ pair production processes in \texttt{HTM} and \texttt{MLRSM} are comparable, larger in \texttt{MLRSM}. The contributions of individual $H_{1}^{\pm \pm}$ and $H_{2}^{\pm \pm}$ pair production channels in $\texttt{MLRSM}$ and $H^{\pm \pm}$ in \texttt{HTM} are discussed. The contributions change with HL-LHC and FCC-ee/CEPC center of mass energies. QCD NLO k-factors are discussed and taken into account in the $H^{\pm \pm}$ pair production and four lepton processes.  
Taking into account present bounds on \texttt{MLRSM} parameters, additional contributions from both the right-handed current and extra scalar particles within \texttt{MLRSM} do not make much difference.  

Still, assuming non-universality of leptonic decays, and due to fields richness of \texttt{MLRSM}, branching ratios for the $H^{\pm \pm}$ decays can be very different in both models, leading to different final signals. 

We discuss the same $H^{\pm\pm}$ masses in both models. Taking into account all leptonic decays, we show that LHC experimental data still allow for  $H^{\pm\pm}$  mass as small as 700 GeV. We take it as the first scenario, the second is for $H^{\pm\pm}$ mass equal to 1 TeV.

We discuss carefully possible decay channels and finally, we make predictions for the complete process $pp \to H^{++} H^{--} \to 4l$.
In both models, we optimised parameters to maximise separately $e^+e^-(pp) \to 4e$ and $e^+e^- (pp) \to 4 \mu$ signals, at the same time being in agreement with all experimental constraints coming from other considered processes. 

The results are gathered in Tab.~\ref{tab_fin_ee} and Tab.~\ref{tab_fin_pp}. 
There are many interesting conclusions that we can draw from them, as discussed in section \ref{seccomp}. In general, due to kinematic cuts and chosen parameters, $4\mu$ signals dominate over $4e$. The latter signals are in most cases at best at the level of the SM background, both for lepton and hadron colliders. This situation gives a way to discriminate the two models. In fact, the most interesting situation in which $v_R$ in \texttt{MLRSM} is relatively large, above sensitivity of LHC (we took $v_R=15$ TeV) does not give too strict constraints on the model parameters, and the discovery signals can be large for $e^+e^- (pp) \to 4 \mu$. In particular, for the \mbox{HL-LHC} and FCC-hh cases, detectable signals which would exceed the SM background are possible only for \texttt{MLRSM}. This conclusion is rather stable over changes of model parameters, for considered kinematic cuts. Though analysis of dilepton distributions can help further in detection of small BSM signals which are comparable or below the SM background, they are similar in patterns for both models and does not help in discrimination between \texttt{HTM} and \texttt{MLRSM}. 

With the most straightforward setups, relying only on the production and decay total counting of events, we can discriminate models, and show in which channels we should look for that. We think that our work is an exemplary case study and from the minimal considerations, more sophisticated approaches can follow.    
As an outlook for further studies,   a discussion of $e^\pm e^\pm \mu^\mp \mu^\mp$ and $e^\pm e^\mp \mu^\pm \mu^\mp$ channels might also be enjoyable, as well as four-lepton signal analysis with final state polarisation.  It will be also interesting to investigate for chosen benchmark points processes with single produced $H^{\pm \pm}$ or single charged Higgs scalars, and associated gauge bosons.
For such cases the SM background will be much larger but it does not exclude positive BSM signals. 
 
\section*{Acknowledgements} We thank Joydeep Chakrabortty for useful discussions and Dipankar Das for his help with the FeynRules model. The research has been supported by the Polish National Science Center (NCN) under grant 2015/17/N/ST2/04067, COST (European Cooperation in Science and Technology) Action CA16201 PARTICLEFACE  and the research activities co-financed by the funds granted under the Research Excellence Initiative of the University of Silesia in Katowice.

\section{Appendix. }
\label{s:appendixB}

\noindent

\subsection{The \texttt{HTM} scalar potential and fields}
\label{s:model}
The Higgs Triplet Model extends the Higgs sector of the SM by adding one scalar $SU(2)_L$ triplet~($\Delta$) with hypercharge $Y=2$ to the Standard Model doublet $\Phi$ (following the convention $Q=\frac{1}{2}Y+T_3$). 

The most general scalar potential is given by~\cite{Arhrib:2011uy}
\begin{eqnarray}
\label{e:potential}
V &=& -m_{\Phi}^2\left(\Phi^\dagger\Phi\right) + \frac{\lambda}{4}\left(\Phi^\dagger\Phi\right)^2 + M_\Delta^2\Tr\left(\Delta^\dagger\Delta\right) +\left[\mu \left(\Phi^Ti\sigma_2\Delta^\dagger\Phi\right) + {\rm h.c.} \right] 
 \nonumber \\
&& +\lambda_1\left(\Phi^\dagger\Phi\right)\Tr\left(\Delta^\dagger\Delta\right) +\lambda_2\left[\Tr\left(\Delta^\dagger\Delta\right)\right]^2 +\lambda_3\Tr\left[\left(\Delta^\dagger\Delta\right)^2\right] +\lambda_4 \Phi^\dagger\Delta \Delta^\dagger\Phi  \,.
\end{eqnarray}
Without loss of generality we can take all the parameters to be real \cite{Gunion:1989ci,Dey:2008jm}. Denoting by 
$v_{\Delta}$ and $v_\Phi$ the vacuum expectation values (VEV's) of the doublet and triplet
\begin{subequations}
\label{e:min}
\begin{eqnarray}
m_{\Phi}^2&=& \frac{\lambda}{4} v_\Phi^2 +\frac{(\lambda_1+\lambda_4)}{2} v_{\Delta}^2 -\sqrt{2}\mu \: v_{\Delta} \,,  \\
M_\Delta^2 &=& -(\lambda_2+\lambda_3) \: v_{\Delta}^2 - \frac{(\lambda_1+\lambda_4)}{2} v_\Phi^2 
			+\frac{\mu}{\sqrt{2}} \frac{v_\Phi^2}{v_{\Delta}} \,.
\label{M2}
\end{eqnarray}
\end{subequations}
We represent the scalar multiplets in the following way
\begin{eqnarray}
\Phi = \frac{1}{\sqrt{2}} \begin{pmatrix} \sqrt{2} w_\Phi^+ \\ v_\Phi+h_\Phi+iz_\Phi \end{pmatrix} \,, &&
\Delta = \frac{1}{\sqrt{2}} \begin{pmatrix}  w_\Delta^+ & \sqrt{2}\delta^{++} \\ v_{\Delta}+h_\Delta+iz_\Delta & -w_\Delta^+ \end{pmatrix} \,.
\end{eqnarray}
The triplet VEV $v_{\Delta}$ is expected to be at most at the order $\order(1)~{\rm GeV}$ to keep the electroweak
$\rho$-parameter $\sim$~1 \cite{Georgi:1981pg,Perez:2008ha,Melfo:2011nx,Arhrib:2011uy,Kanemura:2012rs} (see section \ref{rho_parameter_subsection}
for more details). The electroweak VEV is then given by
\begin{eqnarray}
\label{vev_dependence}
v = \sqrt{v_\Phi^2+2v_{\Delta}^2} \simeq 246~{\rm GeV} \, .
\end{eqnarray}
 \\
The Yukawa sector contains the complete SM Yukawa Lagrangian along with an extra part for the triplet
\begin{eqnarray}
\label{e:tripY}
{\mathscr L}_Y^{\rm \Delta} = \frac{1}{2}\mathcal{Y}_{\ell\ell'} L_\ell^T C^{-1} i\sigma_2 \Delta L_{\ell'} + {\rm h.c.} \,,
\end{eqnarray}
where, $C$ is the charged conjugation operator, $\mathcal{Y}_{\ell\ell'}$ is the symmetric Yukawa matrix and
\begin{eqnarray}
L_\ell = \begin{pmatrix} \nu_\ell \\ \ell \end{pmatrix}_L \,, ~~~ \left[\ell= e,\mu,\tau \right] \,,
\end{eqnarray}
are the left handed $SU(2)$ doublets for the three lepton generations. After spontaneous symmetry breaking (SSB), the Yukawa
couplings in Eq.~(\ref{e:tripY}) will lead to the Majorana mass matrix for the left handed neutrinos. The same term in the Lagrangian is responsible for the interaction between doubly charged scalar particles and charged leptons. 
The $H^{\pm\pm}-l^{\mp}-l'^{\mp}$ vertex breaks the lepton number (see sections \ref{tripltesneutr},\ref{HTMconstr}).
\noindent The fields, $\delta^{\pm\pm} = H^{\pm\pm}$, represent the doubly charged  scalar with the mass
\begin{eqnarray}
\label{doubly_charged_scalar_mass}
M_{H^{\pm\pm}}^2 = \frac{\mu v_\Phi^2}{\sqrt{2}v_{\Delta}} -\frac{\lambda_4}{2}v_\Phi^2 -\lambda_3 v_{\Delta}^2 \,.
\end{eqnarray}
To get physical states for neutral and singly charged particles,  appropriate rotation of fields in the CP-odd and CP-even sectors must follow
\begin{eqnarray}
\begin{pmatrix} G_0 \\ A \end{pmatrix} = \begin{pmatrix} \cos\beta' & \sin\beta' \\ -\sin\beta' & \cos\beta' \end{pmatrix} \begin{pmatrix} z_\Phi \\ z_\Delta \end{pmatrix} \,, ~~~ {\rm with} ~~ \tan\beta' = \frac{2v_{\Delta}}{v_\Phi} \,,
\end{eqnarray}
\begin{eqnarray}
\label{e:alpha}
\begin{pmatrix} h \\ H \end{pmatrix} = \begin{pmatrix} \cos\alpha & \sin\alpha \\ -\sin\alpha & \cos\alpha \end{pmatrix} \begin{pmatrix} h_\Phi 
\label{e:alpha1}\\ 
h_\Delta \end{pmatrix} \,, && ~ {\rm with} ~~ \tan2\alpha =\frac{\sqrt{2}\mu v_\Phi -
(\lambda_1+\lambda_4)v_{\Delta} v_\Phi}{\frac{\mu v_\Phi^2}{2\sqrt{2}v_{\Delta}}+(\lambda_2+\lambda_3)v_{\Delta}^2 
-\frac{\lambda v_\Phi^2}{4}} \,.
\end{eqnarray}
Further, we use an approximation~$\sin{\alpha}~\sim~2\frac{v_\Delta}{v_\Phi}~\rightarrow~0$~\cite{Das:2016},  neutral scalar masses becomes 
\begin{eqnarray} 
M_A^2 & = & \frac{\mu}{\sqrt{2}v_{\Delta}}(v_\Phi^2+4v_{\Delta}^2), \\
M_h^2 & = &\lambda v_\Phi^2 \cos^2\!\alpha + \left(\frac{\mu v_\Phi^2}{\sqrt{2}v_{\Delta}} 
	+ 2 v_{\Delta}^2 (\lambda_2+\lambda_3) \right) \sin^2\!\alpha 
 + 2 \left( v_\Phi v_{\Delta} (\lambda_1+\lambda_4) - \sqrt{2}\mu v_\Phi \right) \cos{\alpha}\sin{\alpha}, \nonumber	\\
&&\\
M_{H}^2 & = &\lambda v_\Phi^2 \sin^2\!\alpha + \left(\frac{\mu v_\Phi^2}{\sqrt{2}v_{\Delta}} 
	+ 2 v_{\Delta}^2 (\lambda_2+\lambda_3) \right) \cos^2\!\alpha  - 2 \left( v_\Phi v_{\Delta} (\lambda_1+\lambda_4) - \sqrt{2}\mu v_\Phi \right) \cos{\alpha}\sin{\alpha}. \nonumber \\
	&&
\end{eqnarray}
In the singly charged sector rotation of fields and masses are the following
\begin{eqnarray}
\label{beta_definition}
\begin{pmatrix} G^\pm \\ H^\pm \end{pmatrix} = \begin{pmatrix} \cos\beta & \sin\beta \\ -\sin\beta & \cos\beta \end{pmatrix} \begin{pmatrix} w_\Phi^\pm \\ w_\Delta^\pm \end{pmatrix} \,, ~~~ {\rm with} ~~ \tan\beta = 
		\frac{\sqrt{2}v_{\Delta}}{v_\Phi} \,,
\end{eqnarray}
to obtain the charged Goldstone~($G^\pm$) along with a  singly charged scalar~($H^\pm$) with
mass
\begin{eqnarray}
\label{e:singly_charged_scalar_mass}
M_{H^{\pm}}^2= \frac{(2\sqrt{2}\mu-\lambda_4v_{\Delta})}{4v_{\Delta}}(v_\Phi^2+2v_{\Delta}^2) \,.
\end{eqnarray}
The $H^\pm$ and $H^{\pm\pm}$ scalar's squared masses (\ref{e:singly_charged_scalar_mass}) and (\ref{doubly_charged_scalar_mass}) contain terms proportional to $v_\Phi^2$ and are inversely proportional to the triplet VEV $v_{\Delta}$, which should be less than $\order(1~{\rm GeV})$ (see section~\ref{rho_parameter_subsection}). That means that $M_{H^\pm}$, $M_{H^{\pm\pm}}$ can be at the level of a few hundred GeV or higher. Latest LHC bounds on the doubly charged scalar masses  vary from 450 to 870 GeV, depending on the decay modes, assuming that 
$\mbox{BR}\left(H^{\pm\pm} \to l^\pm l^\pm \right) \ge 10\%$ \cite{Aaboud:2017qph}. Photon-photon fusion studies  \cite{Babu:2016rcr} set a~bound on $M_{H^{\pm\pm}}$ at the level of 748\;GeV. Limits coming from $e^+ e^-$ colliders are significantly lower, from L3 Collaboration (LEP) it is about 100~GeV \cite{Achard:2003mv}. This bound comes with assumption that the t-channel is negligible (Fig.~\ref{ee_hcchcc_production}) as suppressed by the low $H^{\pm\pm}-l-l$ coupling.
For singly charged scalar masses the mass bound  is even  lower, $M_{H^\pm}$~=~80~GeV \cite{Olive:2016xmw}.

In this paper we assume that the neutral and charged scalars' masses are degenerated$^{ }$\footnote{Even though the mass split  $M_H \equiv M_{H^{\pm\pm}}-M_{H^\pm}$ is proportional to $v_\Phi^2$, the electroweak precision data ($h\rightarrow \gamma\gamma$) gives a limit $|M_{H^{\pm\pm}}-M_{H^\pm}| \le 40 \; \mbox{GeV}$ \cite{Chun:2012jw,Akeroyd:2012ms,Shen_2015_EPL_2}.}, that means  $M_{H^{\pm\pm}}~=~M_{H^\pm}~=~M_{H}~=~M_A$. That choice protects proper ranges of the T-parameter and  potential unitarity for $v_{\Delta} \lesssim 1$ GeV \cite{Das:2016,Gluza:2020icp}.

\subsection{The \texttt{MLRSM} scalar potential and fields}
\label{s:model}
 
The spontaneous symmetry breaking occurs in two steps: $SU(2)_R \otimes U(1)_{B-L} \to U(1)_Y$, and $SU(2)_L \otimes U(1)_Y \to U(1)_{em}$. 
To achieve this symmetry breaking we choose a traditional spectrum of Higgs sector multiplets
with a bidoublet and two triplets  \cite{Mohapatra:1980yp,Gunion:1989in,PhysRevD.40.1546,PhysRevD.44.837,Duka:1999uc,Bambhaniya:2013wza}.
 
\begin{equation}
\phi =\left( 
\begin{array}{lr}
\phi_1^0 \;&\; \phi_1^+\\
\phi_2^- & \phi_2^0
\end{array} 
\right)\equiv [2,2,0], 
\end{equation}  
\begin{equation}  
\Delta_{L(R)}=\left( 
\begin{array}{cc}
\delta_{L(R)}^+/\sqrt{2} & \delta_{L(R)}^{++}\\     
\delta_{L(R)}^0 & -\delta_{L(R)}^+/\sqrt{2}
\end{array} 
\right) \equiv [3(1),1(3),2],
\end{equation}
\\
where the quantum numbers in square brackets are given for 
$SU(2)_L$, $SU(2)_R$ and  $U(1)_{B-L}$ groups, respectively. 

The vacuum expectation values (VEVs) of the scalar fields can be recast in the following form:
\begin{equation}
\left< \phi \right>  =\left( \begin{array}{cc}
                        \kappa_1/\sqrt{2} & 0\\
                        0      \;&\; \kappa_2/\sqrt{2}
                       \end{array}\right) , \hskip 2pt                       
\left< \Delta_{L,R} \right>  =\left( \begin{array}{lr}
                        0   & 0\\
                        v_{L,R}/\sqrt{2}\; & \;0
                       \end{array}\right).                       
\label{vev}
\end{equation}

VEVs of the right-handed triplet ($\Delta_R$) and the bi-doublet ($\phi$), propel the respective symmetry breaking:  
$SU(2)_R \otimes U(1)_{B-L}\to U(1)_Y$, and $SU(2)_L \otimes U(1)_Y \to U(1)_{em}$. As $v_L\ll \kappa_{1,2}\ll v_R$, we take safely $v_L=0$.

The full scalar potential includes left and right-handed triplets \cite{Gunion:1989in,Deshpande:1990ip,Duka:1999uc}:
\begin{eqnarray}\label{VLRSM}
 & & V(\phi,\Delta_L,\Delta_R) =   \nonumber \\
    &+& \lambda_1\bigg\{\Big(\Tr\big[\phi^\dagger \phi\big]\Big)^2\bigg\} +
    \lambda_2\bigg\{ \Big(\Tr\big[\tilde{\phi}\phi^\dagger\big]\Big)^2+\Big(\Tr\big[\tilde{\phi}^\dagger \phi\big]\Big)^2 \bigg\} \nonumber\\
    &+& \lambda_3\bigg\{\Tr\big[\tilde{\phi}\phi^\dagger\big]\Tr\big[\tilde{\phi}^\dagger \phi\big] \bigg\} \nonumber  \\
    &+& \lambda_4 \bigg\{ \Tr\big[\phi^\dagger \phi\big]\Big(\Tr\big[\tilde{\phi}\phi^\dagger\big]
    +\Tr\big[\tilde{\phi}^\dagger \phi\big]\Big) \bigg\}\nonumber\\
    &+& \rho_1 \bigg\{ \Big(\Tr\big[\Delta_L \Delta_L^\dagger\big]\Big)^2+\Big(\Delta_R \Delta_R^\dagger\Big)^2 \bigg\} \nonumber \\
    &+& \rho_2 \bigg\{\Tr\big[\Delta_L \Delta_L\big]\;\Tr\big[\Delta_L^\dagger \Delta_L^\dagger\big]
    +\Tr\big[\Delta_R \Delta_R\big]\;\Tr\big[\Delta_R^\dagger \Delta_R^\dagger\big]  \bigg\}\nonumber\\
    &+& \rho_3 \bigg\{\Tr\big[\Delta_L\Delta_L^\dagger\big]\;\Tr\big[\Delta_R\Delta_R^\dagger\big]\bigg\} \nonumber \\
    &+& \rho_4  \bigg\{\Tr\big[\Delta_L\Delta_L \big]\;
   \Tr\big[\Delta_R^\dagger \Delta_R^\dagger\big] + \Tr\big[\Delta_L^\dagger \Delta_L^\dagger \big]\;
    \Tr\big[\Delta_R \Delta_R \big] \bigg\} \nonumber \\
  &  + & \alpha_1 \bigg\{\Tr\big[\phi^\dagger \phi\big]\Big(\Tr\big[\Delta_L\Delta_L^\dagger\big]
    +\Tr\big[\Delta_R\Delta_R^\dagger\big]\Big)\bigg\} \nonumber  \\
    &+& \alpha_2
    \bigg\{\Tr\big[\phi\tilde{\phi}^\dagger\big]\Tr\big[\Delta_R\Delta_R^\dagger\big] 
    + \Tr\big[\phi^\dagger\tilde{\phi}\big]\Tr\big[\Delta_L\Delta_L^\dagger\big]\bigg\} \nonumber \\
    &+& \alpha_2^*
    \bigg\{\Tr\big[\phi^\dagger\tilde{\phi}\big]\Tr\big[\Delta_R\Delta_R^\dagger\big] 
    + \Tr\big[\tilde{\phi}^\dagger\phi\big]\Tr\big[\Delta_L\Delta_L^\dagger\big]\bigg\} \nonumber  \\
    &+& \alpha_3 \bigg\{ \Tr\big[\phi \phi^\dagger \Delta_L \Delta_L^\dagger\big]
    +\Tr\big[\phi^\dagger \phi \Delta_R \Delta_R^\dagger\big]\bigg\}\nonumber\\    
    &-&\mu_1^2\Tr[\phi^\dag\phi]-\mu_2^2(\Tr[\widetilde{\phi}\phi^\dag]+\Tr[\widetilde{\phi}^\dag\phi])\nonumber\\
    &-&\mu_3^2(\Tr[\Delta_L\Delta_L^\dag]+\Tr[\Delta_R\Delta_R^\dag]).
\end{eqnarray}
Though in \texttt{HTM} and \texttt{MLRSM} we have left-handed triplets, \texttt{HTM} is not a simple subset of \texttt{MLRSM} as the scalar potentials, SSB mechanism, VEVs and underlying physics which follows are different. 
The scalar potential (7.18) in \texttt{MLRSM} is much more complicated than its counterpart in \texttt{HTM}: 
in \texttt{MLRSM} the triplet $\Delta_L$ is intertwined with right-handed multiplet $\Delta_R$ and bidoublet $\phi$.
It makes relations among physical and unphysical Higgs boson fields rather complex in $\texttt{MLRSM}$.
Here significant are relations between the $\alpha_3$ scalar potential parameter (which includes a mixture of a bidoublet and triplet fields) and $\rho_1,\rho_3$ scalar potential parameters for doubly charged Higgs boson masses given in Eq.~(\ref{h1cc_mass_LRSM})  and Eq.~(\ref{h2cc_mass_LRSM}) below. In correlation with experimental constraints for singly charged and neutral scalar fields, these parameters give the lowest limits for doubly charged Higgs masses, as discussed in \cite{Bambhaniya:2014cia}.
Moreover, due to Yukawa couplings of left- and right-handed leptons with bidoublet in Eq.~(\ref{e:LRSM_yukawa}), the doubly charged Higgs bosons in both models couple differently to leptons. Consequently, both model neutrino mass relations are different, in \texttt{HTM} restricted directly by neutrino oscillation data, as discussed in the main text. 

After spontaneouss symmetry breaking of the potential Eq.~(\ref{VLRSM}), the mass matrix which includes $M_{H_0^0}$ can be written in the following form
(for details, see \cite{Gunion:1989in})

\begin{equation}
M=\left(
\begin{array}{ccc}
 2 \epsilon ^2 \text{$\lambda_1$} & 2 \epsilon ^2 \text{$\lambda_4$} & \text{$\alpha_1$} \epsilon  \\
 2 \epsilon ^2 \text{$\lambda_4$} & \frac{1}{2} \left[4 (2 \text{$\lambda_2$}+\text{$\lambda_3$}) \epsilon ^2+\text{$\alpha_3$}\right] &
   2 \text{$\alpha_2$} \epsilon  \\
 \text{$\alpha_1$} \epsilon  & 2 \text{$\alpha_2$} \epsilon  & 2 \text{$\rho_1$} \\
\end{array}
\right).
\end{equation}
Expanding eigenvalues of this matrix in a small $\epsilon = \sqrt{\kappa_1^2+\kappa_2^2}/v_R$ parameter we get 

\begin{equation}
    M_{H_0^0}^2  =  2 \left( \lambda_1 - \frac{\alpha_1^2}{4\rho_1} \right)(\kappa_1^2+\kappa_2^2) \qquad \qquad \qquad \;  \simeq (125 \; \mbox{GeV})^2, \label{higgs_mass_LRSM}	\\
\end{equation}
The analytic mass formulas for other  scalar bosons in \texttt{MLRSM} {\color{black}as a function of quartic couplings and $v_R$} can be written as \cite{Chakrabortty:2016wkl}
\begin{eqnarray} 
M_{H_1^0}^2 & = & \frac{1}{2} \alpha_3 v_R^2  \qquad \qquad \qquad \qquad \qquad \qquad \qquad > (10 \; \mbox{TeV})^2, \label{h10_mass_LRSM}\\
M_{H_2^0}^2 & = & 2 \rho_1 v_R^2,  \\
M_{H_3^0}^2 & = & \frac{1}{2} (\rho_3-2\rho_1) v_R^2 \qquad \qquad \qquad \qquad \qquad \; \; > (55.4 \; \mbox{GeV})^2, \label{h30_mass_LRSM}\\
M_{A_1^0}^2 & = & \frac{1}{2} \alpha_3 v_R^2 - 2(\kappa_1^2+\kappa_2^2) (2\lambda_2-\lambda_3) \quad \quad \qquad > (10 \; \mbox{TeV})^2, \\
M_{A_2^0}^2 & = & \frac{1}{2} (\rho_3 - 2 \rho_1) v_R^2 \\
M_{H_1^\pm}^2 & = &  \frac{1}{2} (\rho_3 - 2 \rho_1) v_R^2 + \frac{1}{4}\alpha_3 (\kappa_1^2+\kappa_2^2),\label{h1c_mass_LRSM}\\
M_{H_2^\pm}^2 & = & \frac{1}{2} \alpha_3 v_R^2 + \frac{1}{4} \alpha_3 (\kappa_1^2+\kappa_2^2)  \quad \qquad \qquad \qquad > (10 \; \mbox{TeV})^2, \label{h2c_mass_LRSM}\\
M_{H_1^{\pm\pm}}^2 & = & \frac{1}{2} (\rho_3-2\rho_1) v_R^2 +\frac{1}{2}\alpha_3 (\kappa_1^2+\kappa_2^2), \label{h1cc_mass_LRSM}\\
M_{H_2^{\pm\pm}}^2 & = & 2 \rho_2 v_R^2 + \frac{1}{2}\alpha_3 (\kappa_1^2+\kappa_2^2), \label{h2cc_mass_LRSM} \\
\nonumber
\end{eqnarray}
\noindent where $\kappa_1,\kappa_2$ are VEVs of the bidoublet and $\sqrt{\kappa_1^2+\kappa_2^2}$ has to be equal to the electroweak symmetry breaking scale $v$, see  (\ref{vev_dependence}). We assume that $\kappa_1 = v = 246$ GeV and $\kappa_2 \to 0$. Some explicit masses of Higgs bosons relevant for $H^{\pm \pm}$ branching ratios in section \ref{results_decay_LRSM} comes from restrictions discussed in \cite{Chakrabortty:2016wkl}. 

In \texttt{MLRSM} relations among physical and unphysical fields (``G" stands for Goldstone modes) are
\begin{eqnarray}
\phi_1^0&\simeq&\frac{1}{\sqrt{2}}\left[
H_0^0+i \tilde{ G_1^0} \right] ,
\\
\phi_2^0 &\simeq& \frac{1}{\sqrt{2}}\left[
H_1^0-i A_1^0 \right], \label{phi20}
\\
\delta_R^0 &=& \frac{1}{\sqrt{2}}\left( H_2^0+iG_2^0 \right),\;\; 
\delta_L^0 = \frac{1}{\sqrt{2}} \left( H_3^0+iA_2^0 \right),\\
\delta_L^+ &=& H_1^+,\;\;\;\delta_R^+ \simeq G_R^+ ,\\
\phi_{1}^+ &\simeq& H_2^+, \;\;\;
\phi_{2}^+ \simeq G_L^+, \\
 \delta_R^{\pm \pm} &=& H_1^{\pm \pm},\;\;\; \delta_L^{\pm \pm} = H_2^{\pm \pm}. 
\end{eqnarray}

The structure of Higgs potential in a general framework of left-right symmetric models has been discussed in details in  \cite{Gunion:1989in,Deshpande:1990ip}.
We adopted it in our studies. In particular, to retain the invariant Majorana Yukawa couplings of the leptons to the Higgs triplet,  the potential  does not include  
the terms with all multiplets (bidoublet, two triplets) present simultaneously, e.g. $Tr[\phi \Delta_R\phi^\dagger \Delta_L^\dag]\;+h.c.$  (in \cite{Gunion:1989in,Deshpande:1990ip} denoted as the $\beta_i$-type terms). In the limit of vanishing $\beta_i$ terms  the doubly charged Higgs scalar $2 \times 2 $ mass matrix is diagonal and does not include the mixed mass terms $\delta_L^{\pm \pm}\delta_R^{\mp \mp}$. These restrictions simplify a form of doubly charged mass terms, as given in the manuscript, Eqs. (7.28) and (7.29). It means that in \texttt{MLRSM} (and other extensions when gauge couplings $g_L \neq g_R$), doubly charged Higgs triplets in $\Delta_{L,R}$ are physical fields. There is no mixing angle between two doubly charged Higgs bosons in \texttt{MLRSM} and the mass matrix which appears there for unphysical fields is diagonal from the very beginning. This no-mixing feature is also true in general where the $\beta_i$-type terms are allowed, in the limit $v_R \gg  \kappa_{1,2}$.

\subsection{$H_1^{\pm\pm}$ and $H_2^{\pm\pm}$ couplings with leptons in \texttt{MLRSM}}
\label{vertex_LRSM}

In \texttt{MLRSM} due to additional heavy states the neutrino sector and Yukawa couplings are more complicated than in \texttt{HTM}.  
Here we argue that due an energy scales difference between $v_R$ and the low-energy bidoublet VEV $\kappa \equiv \sqrt{\kappa_1^2+\kappa_2^2}$, $\kappa \ll v_R$,
the see-saw mechanism is possible and low energy LFV signals are suppressed due to high $v_R$ and heavy neutrino masses. To see, this,   
 the most general doubly charged couplings to leptons, which takes into account mixing matrices, reads \cite{Duka:1999uc}
\begin{eqnarray}\label{delrll}
\delta_R^{++}\bar{l}_L^{\prime c}h_Ml^\prime_R +h.c. &=& 
 \frac{1}{\sqrt{2}v_R}\sum\limits_{l,k}
\left\{
\delta_R^{++}
\left[
l^T_{l}C
\left(
K^T_R \left( M_\nu \right)_{diag} K_R
\right)_{lk}
P_R
l_{k}
\right] \right. \nonumber \\
&+& \left. 
\delta_R^{--}
\left[
\bar{l}_{l}
\left(
K^\dagger_R \left( M_\nu \right)_{diag} K_R^\ast
\right)_{lk}
P_L
C\bar{l}^T_{k}
\right]
\right\},\label{HRllcoupl} \\
\delta_L^{++}\bar{l}_R^{\prime c}h_Ml^\prime_L +h.c. &=&
\frac{1}{\sqrt{2}v_R}\sum\limits_{l,k}
\left\{
\delta_L^{++}
\left[
l^T_{k}C
\left(
K_L^T X K_L^\ast
\right)_{kl}
P_L l_{l} \right] \right. \nonumber \\
&+& \left. 
\delta_L^{--}
\left[
\bar{l}_{k}
\left(
K_L^T X^\ast K_L^\ast
\right)_{kl}
P_R
C\bar{l}^T_{l}
\right]
\right\}
\label{HLllcoupl}
\end{eqnarray}
where  $X=(K_L^\ast K_R^T) \left( M_\nu \right)_{diag} (K_R K_L^\dagger)$.\\
These couplings originate from the Yukawa part of the Lagrangian for additional scalar triplets and a bidoublet $\phi$:
\begin{equation}
- \bar{L}_L \left[ h_l \phi +\tilde{h}_l \tilde{\phi}  \right] L_R 
- i\bar{L}_R^c \sigma_2 \Delta_L h_M L_L 
- i\bar{L}_L^c \sigma_2 \Delta_R h_M L_R + h.c.
\label{e:LRSM_yukawa}
\end{equation}
\noindent

Uniqueness of left- and right-handed couplings for positively and negatively charged doubly charged Higgs bosons to leptons in Eq.~(\ref{delrll}) and Eq.~(\ref{HLllcoupl}) is due to the Feynman rules and flow of the charged currents in vertices, as explained in length in \cite{Gluza:1991wj}.
The relations between physical and unphysical scalar, gauge and fermion fields are embedded in our FeynRules package to calculate branching ratios and cross sections. 
We also mentioned that for neutral scalars, due to the bidoublet coupling in Eq.~(7.38) to both left- and right-handed leptons, there is a mixture of scalar fields coming from left and right triplets, however, as given in Eqs.~(7.30)-(7.35), for $v_R \gg \kappa_{1,2}$, most of the mixings are negligible.

Diagonalization of the resulting neutrino mass matrix
\begin{equation}
M_\nu = \left( \begin{array}{cc} 0 & M_D\\ M_D^T & M_R \end{array} \right),
\qquad M_R = \sqrt{2} h_M v_R,
\label{e:MRfvR}
\end{equation}
goes with the help of a unitary 6$\times$6 matrix $U$
\begin{displaymath}
U^T M_\nu U = (M_\nu)_{diag}, \qquad U = \left( \begin{array}{c} 
K_L^*
\\ K_R
\\ \end{array} \right).
\end{displaymath}
This procedure leads to the introduction of the $K_L$ and $K_R$ submatrices  in Eq.~(\ref{HRllcoupl}) and Eq.~(\ref{HLllcoupl}) \cite{Gluza:1993gf,Gluza:1995ky}.\\
The charged lepton mass matrix is diagonalised by ${V^l_{L,R}}_{3\times 3}$
\begin{displaymath}
{V^l_L}^\dagger M_l V^l_R = (M_l)_{diag}.
\end{displaymath}
Apart from charged lepton and neutrino mass terms,  Lagrangian Eq.~(\ref{e:LRSM_yukawa}) contains scalar-lepton interactions too. 

$\left( M_\nu \right)_{diag}$ contains 3 light neutrinos, there contribution to the couplings Eq.~(\ref{HLllcoupl}) and Eq.~(\ref{HRllcoupl}) are negligible. To see amount of heavy neutrinos contributions to Eq.~(\ref{HLllcoupl}) and Eq.~(\ref{HRllcoupl}), we note that structure of $K_L$ and $K_R$ mixing matrices are the following \cite{Gluza:1993gf,Gluza:1995ky}
\begin{equation}
{(K_{L})}_{l_i \nu_j}=
\left( \begin{array}{c} 
e\;\; \mu \;\; \tau \;\;\;\;\;\;\;\;\;\;\;\;\;\;\;\;\;\;\;\;\;\;\;\;\;\; \\
  \left. \left( { \begin{array}{ccc}
            \cdot & \cdot & \cdot \\
            \cdot & \cdot & \cdot \\
            \cdot & \cdot & \cdot
            \end{array} } \right) 
  \right\} {\rm light\;\; neutrinos} \\
   \left. \left( { \begin{array}{ccc}
            \cdot & \cdot & \cdot \\
            \cdot & \cdot & \cdot \\
            \cdot & \cdot & \cdot
            \end{array} } \right) 
   \right\} {\rm heavy\;\; neutrinos}
\end{array} \right) ~\sim \left( \begin{array}{c}
\\
{\cal{O}}{(1)}\\
\\ \\
{\cal{O}}{\left(\frac{1}{m_N}\right) } \\ \\
\end{array} \right), \label{KLstr}
\end{equation}
\begin{equation}
{(K_{R})}_{l_i \nu_j}=
\left( \begin{array}{c} 
e\;\; \mu \;\; \tau \;\;\;\;\;\;\;\;\;\;\;\;\;\;\;\;\;\;\;\;\;\;\;\;\;\; \\
  \left. \left( { \begin{array}{ccc}
            \cdot & \cdot & \cdot \\
            \cdot & \cdot & \cdot \\
            \cdot & \cdot & \cdot
            \end{array} } \right) 
  \right\} {\rm light\;\; neutrinos} \\
   \left. \left( { \begin{array}{ccc}
            \cdot & \cdot & \cdot \\
            \cdot & \cdot & \cdot \\
            \cdot & \cdot & \cdot
            \end{array} } \right) 
   \right\} {\rm heavy\;\; neutrinos}
\end{array} \right) ~\sim \left( \begin{array}{c}
\\
{\cal{O}}{\left(\frac{1}{m_N}\right) } \\
\\ \\
{\cal{O}}{(1)} \\ \\
\end{array} \right). \label{KRstr}
\end{equation}
Off-diagonal elements for heavy neutrinos couplings in $K_R$ are typically also of the order of inverse heavy neutrino mass scale, that is why LFV couplings of leptons with doubly charged Higgs bosons are strongly suppressed, $\delta_{L,R}-l-l$ off-diagonal lepton couplings are suppressed by $1/m_N^2$ when comparing to diagonal cases. 

For reasons discussed in \cite{Czakon:2002wm} and more extensively in \cite{Gluza:2002vs}, we take seesaw  diagonal light-heavy neutrino mixings. 
It means that $W_1$ couples mainly to light
neutrinos, while $W_2$ couples to the heavy ones. 

To summarize, unlike in the HTM case,  the $H^{\pm \pm}-l^\mp-l^\mp$ vertex does not depend on the light neutrino mixing. With $v_L=0$ the \texttt{MLRSM} realizes the seesaw type-I mechanism and the light neutrino mass is due to the existence of additional heavy neutrino states and $v_R$ scale.

We should note that it is not natural and very hard to create non-decoupling mixings for non-diagonal $K_L$ and $K_R$ matrix elements, even when some symmetries are considered in type-I seesaw models \cite{Gluza:2002vs}.

\subsection{Supplemental material for phenomenological studies of $H^{\pm\pm}$ scalar particles}

\noindent Diagrams in Fig.~\ref{LFV-Mu}  present the contribution from the singly and doubly charged scalar particles to lepton flavour violating processes and to muon $(g-2)_\mu$. Those diagrams contain  vertices  $H^{\pm\pm}-l_i-l_j$ and $H^\pm-l_i-\nu_j$ which origin from the Yukawa part of the Lagrangian, combining the Standard Model Yukawa term with the triplet part Eq.~(\ref{e:tripY})

\begin{equation}
{\mathscr L}_Y = {\mathscr L}_Y^\Phi+{\mathscr L}_Y^{\rm \Delta}= -\: {y}_{ij} \: \overline{L^i_L} \: \Phi \: {l^j}_R \;
+ \mathcal{Y}_{ij} \: \overline{{L^i_L}^c} \: i\sigma_2 \: \Delta \: L^j_L \; + \; {\rm h.c.}
\label{whole_yukawa}
\end{equation}

From Eq.~(\ref{whole_yukawa}) we obtain the interaction between charged leptons and a doubly charged scalar and the interaction of a singly charged scalar with a charged lepton and neutrino. Taking into account Eq.~(\ref{beta_definition}) and Eq.~(\ref{vev_dependence}) and keeping in mind that
$y_{ij} \propto \frac{1}{v_\Phi}$ is a SM diagonal matrix  

	\begin{equation}
	\label{vertex_doubly}
	\mathcal{V}^{\pm\pm} = \left\{
	\begin{array}{lll}
	l_i^+ - l_j^+ - H^{--} &=& i \: \left( \mathcal{Y}_{ij} + \mathcal{Y}_{ji} \right) \\
	l_i^- - l_j^- - H^{++} &=& i \: \left( \mathcal{Y}_{ij}^* + \mathcal{Y}_{ji}^* \right) \\
	\end{array}
	\right.,
	\end{equation}

	\begin{equation}
	\label{LNV_interaction}
	\mathcal{V}_\Delta^\pm = \left\{		
	\begin{array}{c}
	\tilde{\nu}_i - l_j^+ - H^{-} = \frac{i}{\sqrt{2}} \: \cos{\beta} \: \left( \mathcal{Y}_{ij} + \mathcal{Y}_{ji} \right) \\
	\nu_i - l_j^- - H^{+} = \frac{i}{\sqrt{2}} \: \cos{\beta} \: \left( \mathcal{Y}_{ij}^* + \mathcal{Y}_{ji}^* \right) \\
	\end{array}
	\propto  \frac{\sqrt{v_\Phi^2-2v_\Delta^2}}{v_\Phi\cdot v_\Delta}
	\right.,
	\end{equation}
	
	\begin{equation}
	\label{non_LNV_interaction}
	\mathcal{V}_\Phi^\pm = \left\{
	\begin{array}{c}
	\tilde{\nu_i} - l_i^- - H^{+} = i \: \sin{\beta} \: y_{i} \\
	\nu_i - l_i^+ - H^{-} = i \: \sin{\beta} \: y_{i}
	\end{array}
	\propto	\frac{\sqrt{2} v_\Delta}{v_\Phi\sqrt{v_\Phi^2-2 v_\Delta^2}}
	\right. .
	\end{equation}

Vertices $\mathcal{V}_\Delta^\pm$ and $\mathcal{V}_\Delta^{\pm\pm}$ comes from the same part of the Lagrangian and they break the lepton flavour. Vertex $\mathcal{V}_\Phi^\pm$ is proportional to $v_\Delta$ while vertex $\mathcal{V}_\Delta^\pm$ is inversely proportional to the triplet VEV and dominates\ up to $v_\Delta \sim 10^6$ eV. Since we are interested in lower regions of $v_\Delta$ values, its effect is negligible. So, with a good approximation for low values of $v_\Delta$

\begin{equation}
\label{Vpm_and_Vpmpm}
	\mathcal{V}_\Delta^\pm \equiv \mathcal{V}^\pm \simeq \frac{1}{\sqrt{2}} \mathcal{V}^{\pm\pm}.
\end{equation}

As we discussed in section~\ref{HTMconstr}, the branching ratios of the radiative and  $\mu$-to-$e$ conversion depends on the one-loop form factors. From Eq.~\eqref{eq:htm_formfac}, we can read them explicitly
\begin{eqnarray}
A_L(q^2) &= &-\frac{(\mathcal{Y}^*)_{ei} (\mathcal{Y})_{\mu i} }{24\sqrt{2} G_F \pi^2 }\Big(\frac{1}{12 M_{H^{\pm}}^2 }+\frac{f(r,s_i)}{M_{H^{\pm\pm}}^2}\Big),
\nonumber \\
A_R&= &-\frac{(\mathcal{Y}^* \mathcal{Y})_{e\mu }}{192\sqrt{2} G_F\pi^2 }\Big(\frac{1}{8 M_{H^{\pm}}^2}+\frac{1}{M_{H^{\pm\pm}}^2}\Big),
\label{eq:al_ar}
\end{eqnarray}
where $f(r,s_i)$ is given by:
\begin{eqnarray}
 f(r,s_i)&=&\frac{4 s_i}{r}+\log(s_i)+\Big(1-\frac{2 s_i}{r}\Big) \sqrt{1+\frac{4s_i}{r}}\log \Big(\frac{\sqrt{r}+\sqrt{r+4s_i}}{\sqrt{r}-\sqrt{r+4s_i}}\Big),
\\
 r&=&\frac{-q^2}{m_{H^{\pm\pm}}^2}  \mbox{,} \quad s_i= \frac{m_i^2}{m_{H^{\pm\pm}}^2}.
\end{eqnarray}
We have checked with earlier literature and the analytic forms for the CLFV processes are given as~\cite{Chakrabortty:2015zpm,Chun:2003ej,Kakizaki:2003jk,Akeroyd:2009nu,Dinh:2012bp}: 

\subsubsection*{Radiative lepton decay $\boldsymbol{l_i \to l_j \gamma}$:}
The branching ratios of radiative decay processes can be given by:
\begin{equation*}
\text {BR}(l_i\rightarrow l_j \gamma)=384 \pi^2 (4\pi \alpha_{em})|A_R|^2~\text{BR}(l_i\rightarrow l_j\nu_{l_i}  \overline{\nu}_{l_j}).
\end{equation*}
Therefore the BRs for various radiative decays can be written as:
\begin{equation*}
\begin{array}{lr}
\begin{array}{rcl}
\text {BR}(\mu\rightarrow e \gamma)&=&\frac{\alpha_{e m}}{192\pi}\frac{|({\mathcal{Y}}^\dagger \mathcal{Y})_{e\mu}|^2}{G_F^2}
	\Big(\frac{1}{M_{H^{\pm}}^2}+\frac{8}{M_{H^{\pm\pm}}^2}\Big)^2 \;\text{BR}(\mu\rightarrow e \bar{\nu}_e \nu_\mu),  \\
\text {BR}(\tau\rightarrow e \gamma)&=&\frac{\alpha_{e m}}{192\pi}\frac{|({\mathcal{Y}}^\dagger \mathcal{Y})_{e\tau}|^2}{G_F^2}
	\Big(\frac{1}{M_{H^\pm}^2}+\frac{8}{M_{H^{\pm\pm}}^2}\Big)^2 \;\text{BR}(\tau\rightarrow e \bar{\nu}_e \nu_\tau),  \\
\text {BR}(\tau\rightarrow \mu \gamma)&=&\frac{\alpha_{e m}}{192\pi}\frac{|({\mathcal{Y}}^\dagger \mathcal{Y})_{\mu\tau}|^2}{G_F^2}
	\Big(\frac{1}{M_{H^\pm}^2}+\frac{8}{M_{H^{\pm\pm}}^2}\Big)^2 \;\text{BR}(\tau\rightarrow \mu \bar{\nu}_\mu \nu_\tau),  \\
\end{array}
&
\begin{array}{rcl}
 	\text{BR}(\mu\rightarrow e \bar{\nu}_e \nu_e) 	&=& 100 \%, \\
 	\text{BR}(\tau\rightarrow e \bar{\nu}_e \nu_\tau)	&=& 17.83\%, \\
	\text{BR}(\tau\rightarrow \mu \bar{\nu}_\mu \nu_\tau) 	&=& 17.41 \%.\\
\end{array}
\end{array}
\end{equation*}

The contribution of $H^{\pm\pm}$ to the branching ratios is eight times larger than by $H^{\pm}$ because of the difference in a magnitude of couplings  between 
$\mathcal{V}^\pm$ and $\mathcal{V}^{\pm\pm}$ in Eq.~(\ref{Vpm_and_Vpmpm}), in addition the amplitude is proportional to the particles charge (which gives an additional  factor of 4).

\subsubsection*{Three body decays $\boldsymbol{l \to l_i l_j l_k}$:}

\begin{equation*}
\begin{array}{lcrrr}
 \text{BR}(\mu \rightarrow eee) & = & 
	\frac{1}{4G_F^2}\frac{|(\mathcal{Y}^\dagger)_{ee}(\mathcal{Y})_{\mu e}|^2}{M_{H^{\pm\pm}}^4} \;\text{BR}(\mu\rightarrow e \bar{\nu} \nu), & \\

 \text{BR}(\tau \rightarrow l_i l_j l_k)&=&\frac{S}{4G_F^2}\frac{|(\mathcal{Y}^\dagger)_{\tau i}(\mathcal{Y})_{j k}|^2}{M_{H^{\pm\pm}}^4} \;\text{BR}(\tau\rightarrow \mu \bar{\nu} \nu), & \qquad
  S = \left\{ \begin{array}{lll} 1 & \mbox{if} & j = k \\ 2 & \mbox{if} & j \neq k \\ \end{array} \right. .\\
\end{array}
\end{equation*}

\subsubsection*{$\boldsymbol{\mu}-$to$-\boldsymbol{e}$ conversion}
In the computation of conversion rate $\mu$ to $e$, both form factors contribute and the analytic form can be written as
\begin{eqnarray}
\text{CR}(\mu \mathcal{N}\rightarrow e\mathcal{N}^*) &=& \frac{\Gamma_{conv}}{\Gamma_{capt}} \cong\frac{2\alpha_{em}^5 G_F^2 m_\mu^5 Z_{eff}^4 Z |F(q^2)|^2}{\Gamma_{capt}}\Big|8A_R+\frac{2}{3}A_L\Big|^2.
\end{eqnarray}
 Therefore, the $\mu$-to-$e$ conversion ratio in the nuclei field can be given as~\cite{Dinh:2012bp,Chakrabortty:2015zpm}:
\begin{eqnarray}
\text{CR}(\mu \mathcal{N}\rightarrow e\mathcal{N}^*) &=& \frac{\Gamma_{conv}}{\Gamma_{capt}} \cong \frac{\alpha_{em}^5}{36\pi^4} \frac{m_\mu^5}{\Gamma_{capt}}Z_{eff}^4 Z |F(q^2=-m_\mu^2)|^2 \times \\
& & \Bigg |\frac{(M_\nu^\dagger M_\nu)_{e\mu}}{2v_\Delta^2} \Big[\frac{5}{24 m_{H^{\pm}}^2}+\frac{1}{m_{H^{\pm\pm}}^2}\Big]+\frac{1}{2v_\Delta^2 m_{H^{\pm\pm}}^2} \sum_{l=e,\mu,\tau} (M_\nu)^\dagger_{el} f(r,s_i) (M_\nu)_{l\mu}\Bigg |^2, \nonumber
\end{eqnarray}

\begin{tabular}{ll}
where: & ${\Gamma_{capt}} \quad - \; \mbox{total muon capture rate (see Tab.~\ref{Gcapt})}$, \\
\\
& $Z_{eff} \quad - \; \mbox{effective charge for the muon in the 1s state}$,\\
\end{tabular}

\begin{table}
\centering
\begin{tabular}{c|c|c|c}
 & $\Gamma_{capt} \; [\mbox{s}^{-1}]$ & $\Gamma_{capt} \; [\mbox{eV}]$ & $Z_{eff}$\\
\hline
 & & & \\
${}^{197}_{\phantom{1}79} \mbox{Au}$ & $13.07\times 10^6$ & $8.60\times 10^{-9}$ & 33.5 \\
 & & & \\
${}^{48}_{22} \mbox{Ti}$  & $2.59\times 10^6$ & $1.71\times 10^{-9}$ & 17.5\\
 & & & \\
${}^{27}_{13} \mbox{Al}$  & $0.7054\times 10^6$ & $0.4643\times 10^{-9}$ & 11.5 \\
\end{tabular}
\caption{Total muon capture rate and effective charge for ${}^{197} \mbox{Au}$, ${}^{48} \mbox{Ti}$ and ${}^{27} \mbox{Al}$ \label{Gcapt}\cite{Kitano:2002mt}.}.
\end{table}

\subsubsection*{Muon $({g-2})_\mu$}

In case of doubly charged scalars, cumulative effects of muon $(g-2)_\mu$ and lepton flavor violation have been discussed in details in \cite{Leveille:1977rc,Moore:1984eg,Chakrabortty:2015zpm} and for a triplet scalar it has been discussed in \cite{Fukuyama:2009xk}.

\begin{center}
\begin{figure}[h]
\begin{center}
\includegraphics[angle=0,width=0.8\textwidth]{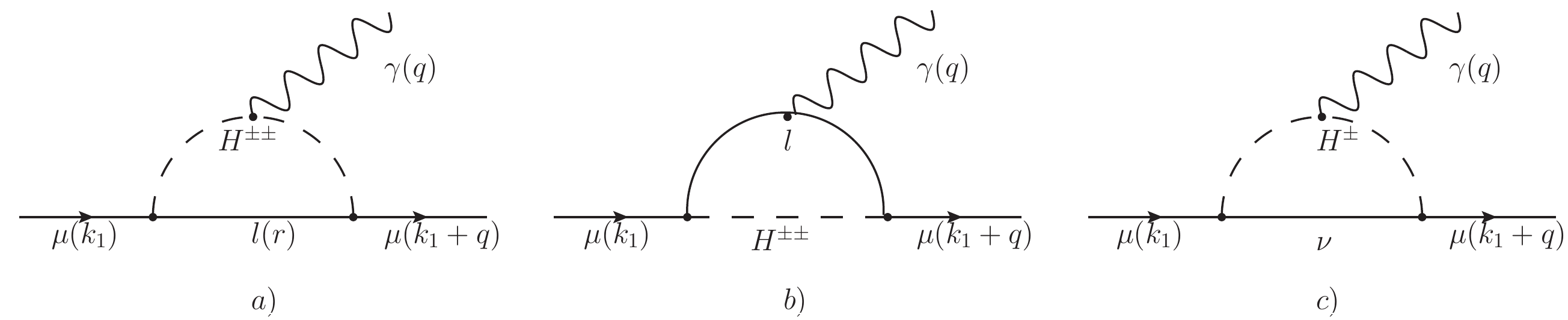}
\end{center}
\caption{ Feynman diagrams represent the contribution to $(g-2)_\mu$ within HTM\label{g-2_fig}}
\end{figure}
\end{center}

Contribution to muon $(g-2)$ from doubly and singly charged scalar are shown in Fig.~\ref{g-2_fig}. 
 Final formulas for $H^{\pm \pm}$ and $H^{\pm}$ reads

\begin{eqnarray}
\label{g-2_contribution_1}
\left[\Delta a_\mu\right]_{H^{\pm \pm}} &=& - \sum_l f^l \times\Bigg\{\frac{2  m_\mu^2 |\mathcal{V}_{\mu l}^{{\pm\pm}}|^2}{8\pi^2} 
	\underbrace{\int_0^1 dx \Bigg[ \frac{\Big\{ (x^3-x^2)+\frac{m_l}{m_\mu}(x^2-x)\Big\}}{\Big(m_\mu^2 x^2+(M_{H^{\pm\pm}}^2-m_\mu^2)x+(1-x)m_{l}^2\Big)} \Bigg]}_{C_1}\\
& & -\frac{  m_\mu^2 |\mathcal{V}_{\mu l}^{{\pm\pm}}|^2}{8\pi^2}
	\underbrace{\int_0^1 dx \Bigg[\frac{\Big\{x^2-x^3+\frac{m_l}{m_\mu} x^2\Big\}}{\Big(m_\mu^2 x^2 +(m_{l}^2-m_\mu^2) x +M_{H^{\pm\pm}}^2 (1-x)\Big)}\Bigg]\Bigg\}}_{C_2},\\
\label{g-2_contribution_3}
\left[\Delta a_\mu\right]_{H^{\pm}} &=& -\frac{1}{2} \sum_\nu \frac{ m_\mu^2 |\mathcal{V}_{\mu \nu}^{\pm}|^2}{8\pi^2} 
	\underbrace{\int_0^1 dx \Bigg[ \frac{ (x^3-x^2)}{\Big(m_\mu^2 x^2+(M_{H^{\pm}}^2-m_\mu^2)x\Big)} \Bigg]}_{C_3}.
 \end{eqnarray}

\begin{center}
\begin{figure}[h]
\begin{center}
\includegraphics[angle=270,width=0.6\textwidth]{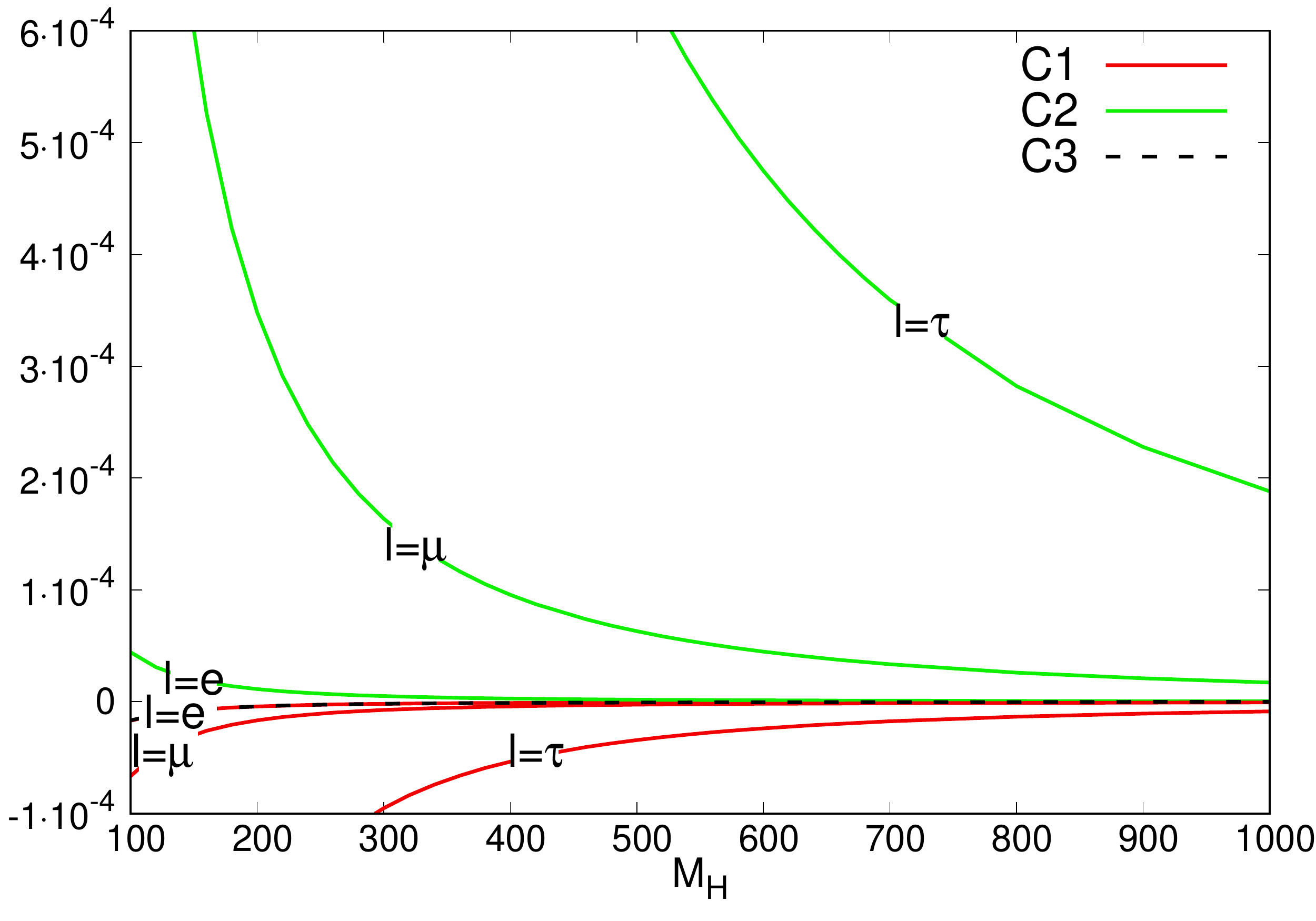}
\end{center}
\caption{$C_1$, $C_2$ and $C_3$ integrals  Eqs.~(\ref{g-2_contribution_1})-(\ref{g-2_contribution_3}) as a function of charged scalars mass 
$M_{H^\pm}=M_{H^{\pm\pm}}\equiv M_H$. \label{integrals_plot}}
\end{figure}
\end{center}

$f^l$ is a symmetric factor equals to 4 for $l=\mu$ and 1 otherwise.  
The term proportional to the $C_1$ integral is connected with the diagram a) in Fig.~\ref{g-2_fig}, the term with the $C_2$ integral corresponds to Fig.~\ref{g-2_fig}~b). Equation (\ref{g-2_contribution_3}) presents a contribution from a singly charged particle $H^\pm$ (Fig.~\ref{g-2_fig}~c)~) to $(g-2)_\mu$. Since both $\mathcal{V}^\pm$ and $\mathcal{V}^{\pm\pm}$ vertices are comparable, see Eq.~(\ref{Vpm_and_Vpmpm}), the contributions from different diagrams depend mostly on $C_1$, $C_2$ and $C_3$ integrals.  Fig.~\ref{integrals_plot} shows that the strongest contribution  $(g-2)_\mu$ comes from the doubly charged scalar $H^{\pm\pm}$.

\begin{thebibliography}{100}

\bibitem{Aad:2012tfa}
{\bf ATLAS} Collaboration, G.~Aad et~al., {\it {Observation of a new particle
  in the search for the Standard Model Higgs boson with the ATLAS detector at
  the LHC}},  {\em Phys. Lett.} {\bf B716} (2012) 1--29,
  [\href{http://arxiv.org/abs/1207.7214}{{\tt arXiv:1207.7214}}].

\bibitem{Chatrchyan:2012xdj}
{\bf CMS} Collaboration, S.~Chatrchyan et~al., {\it {Observation of a new boson
  at a mass of 125 GeV with the CMS experiment at the LHC}},  {\em Phys. Lett.}
  {\bf B716} (2012) 30--61, [\href{http://arxiv.org/abs/1207.7235}{{\tt
  arXiv:1207.7235}}].

\bibitem{1576}
G.~Aad et~al., {\it {A Particle Consistent with the Higgs Boson Observed with
  the ATLAS Detector at the Large Hadron Collider}},  {\em Science} {\bf 338}
  (2012), no.~6114 1576--1582.

\bibitem{Chatrchyan:2012jja}
{\bf CMS} Collaboration, S.~Chatrchyan et~al., {\it {Study of the Mass and
  Spin-Parity of the Higgs Boson Candidate Via Its Decays to Z Boson Pairs}},
  {\em Phys. Rev. Lett.} {\bf 110} (2013), no.~8 081803,
  [\href{http://arxiv.org/abs/1212.6639}{{\tt arXiv:1212.6639}}].

\bibitem{Aad:2013xqa}
{\bf ATLAS} Collaboration, G.~Aad et~al., {\it {Evidence for the spin-0 nature
  of the Higgs boson using ATLAS data}},  {\em Phys. Lett.} {\bf B726} (2013)
  120--144, [\href{http://arxiv.org/abs/1307.1432}{{\tt arXiv:1307.1432}}].

\bibitem{Chatrchyan:2011tz}
{\bf CMS} Collaboration, S.~Chatrchyan et~al., {\it {Measurement of $W^+ W^-$
  production and search for the Higgs boson in pp collisions at $\sqrt s=7$
  TeV}},  {\em Phys. Lett.} {\bf B699} (2011) 25--47,
  [\href{http://arxiv.org/abs/1102.5429}{{\tt arXiv:1102.5429}}].

\bibitem{Khachatryan:2014qaa}
{\bf CMS} Collaboration, V.~Khachatryan et~al., {\it {Search for the associated
  production of the Higgs boson with a top-quark pair}},  {\em JHEP} {\bf 09}
  (2014) 087, [\href{http://arxiv.org/abs/1408.1682}{{\tt arXiv:1408.1682}}].
  [Erratum: JHEP 10, 106 (2014)].

\bibitem{Aad:2014lma}
{\bf ATLAS} Collaboration, G.~Aad et~al., {\it {Search for $H \to \gamma\gamma$
  produced in association with top quarks and constraints on the Yukawa
  coupling between the top quark and the Higgs boson using data taken at 7 TeV
  and 8 TeV with the ATLAS detector}},  {\em Phys. Lett.} {\bf B740} (2015)
  222--242, [\href{http://arxiv.org/abs/1409.3122}{{\tt arXiv:1409.3122}}].

\bibitem{Khachatryan:2016vau}
{\bf ATLAS, CMS} Collaboration, G.~Aad et~al., {\it {Measurements of the Higgs
  boson production and decay rates and constraints on its couplings from a
  combined ATLAS and CMS analysis of the LHC pp collision data at $ \sqrt{s}=7
  $ and 8 TeV}},  {\em JHEP} {\bf 08} (2016) 045,
  [\href{http://arxiv.org/abs/1606.02266}{{\tt arXiv:1606.02266}}].

\bibitem{Aad:2015iha}
{\bf ATLAS} Collaboration, G.~Aad et~al., {\it {Search for the associated
  production of the Higgs boson with a top quark pair in multilepton final
  states with the ATLAS detector}},  {\em Phys. Lett.} {\bf B749} (2015)
  519--541, [\href{http://arxiv.org/abs/1506.05988}{{\tt arXiv:1506.05988}}].

\bibitem{Djouadi:1993ji}
A.~Djouadi, M.~Spira, and P.~M. Zerwas, {\it {Two photon decay widths of Higgs
  particles}},  {\em Phys. Lett.} {\bf B311} (1993) 255--260,
  [\href{http://arxiv.org/abs/hep-ph/9305335}{{\tt hep-ph/9305335}}].

\bibitem{Aad:2015gba}
{\bf ATLAS} Collaboration, G.~Aad et~al., {\it {Measurements of the Higgs boson
  production and decay rates and coupling strengths using pp collision data at
  $\sqrt{s}=7$ and 8 TeV in the ATLAS experiment}},  {\em Eur. Phys. J.} {\bf
  C76} (2016), no.~1 6, [\href{http://arxiv.org/abs/1507.04548}{{\tt
  arXiv:1507.04548}}].

\bibitem{Aad:2015mxa}
{\bf ATLAS} Collaboration, G.~Aad et~al., {\it {Study of the spin and parity of
  the Higgs boson in diboson decays with the ATLAS detector}},  {\em Eur. Phys.
  J. C} {\bf 75} (2015), no.~10 476,
  [\href{http://arxiv.org/abs/1506.05669}{{\tt arXiv:1506.05669}}]. [Erratum:
  Eur.Phys.J.C 76, 152 (2016)].

\bibitem{Aad:2015rwa}
{\bf ATLAS} Collaboration, G.~Aad et~al., {\it {Determination of spin and
  parity of the Higgs boson in the $WW^*\rightarrow e \nu \mu \nu $ decay
  channel with the ATLAS detector}},  {\em Eur. Phys. J.} {\bf C75} (2015),
  no.~5 231, [\href{http://arxiv.org/abs/1503.03643}{{\tt arXiv:1503.03643}}].

\bibitem{DiMarco:2016uwi}
{\bf CMS} Collaboration, E.~Di~Marco, {\it {Studies of the Higgs boson spin and
  parity using the $\gamma$$\gamma$ , ZZ, and WW decay channels with the CMS
  detector}},  {\em Nucl. Part. Phys. Proc.} {\bf 273-275} (2016) 746--752.

\bibitem{Sirunyan:2017khh}
{\bf CMS} Collaboration, A.~M. Sirunyan et~al., {\it {Observation of the Higgs
  boson decay to a pair of $\tau$ leptons with the CMS detector}},  {\em Phys.
  Lett.} {\bf B779} (2018) 283--316,
  [\href{http://arxiv.org/abs/1708.00373}{{\tt arXiv:1708.00373}}].

\bibitem{plb201859}
M.~Aaboud et~al., {\it {Observation of $H\to b \bar{b}$ decays and $VH$
  production with the ATLAS detector}},  {\em Physics Letters B} {\bf 786}
  (2018) 59 -- 86.

\bibitem{CMS-PAS-HIG-19-006}
{\bf CMS Collaboration} Collaboration, {\it {Measurement of Higgs boson decay
  to a pair of muons in proton-proton collisions at
  $\sqrt{s}=13\,\mathrm{TeV}$}},  Tech. Rep. CMS-PAS-HIG-19-006, CERN, Geneva,
  2020.

\bibitem{Aad:2020xfq}
{\bf ATLAS} Collaboration, G.~Aad et~al., {\it {A search for the dimuon decay
  of the Standard Model Higgs boson with the ATLAS detector, }},
  \href{http://arxiv.org/abs/2007.07830}{{\tt arXiv:2007.07830}}.

\bibitem{Blondel:2018aan}
A.~Blondel and P.~Janot, {\it {Future strategies for the discovery and the
  precise measurement of the Higgs self coupling, }},
  \href{http://arxiv.org/abs/1809.10041}{{\tt arXiv:1809.10041}}.

\bibitem{Branco:2011iw}
G.~C. Branco, P.~M. Ferreira, L.~Lavoura, M.~N. Rebelo, M.~Sher, and J.~P.
  Silva, {\it {Theory and phenomenology of two-Higgs-doublet models}},  {\em
  Phys. Rept.} {\bf 516} (2012) 1--102,
  [\href{http://arxiv.org/abs/1106.0034}{{\tt arXiv:1106.0034}}].

\bibitem{Gunion:1984yn}
J.~Gunion and H.~E. Haber, {\it {Higgs Bosons in Supersymmetric Models. 1.}},
  {\em Nucl. Phys. B} {\bf 272} (1986) 1. [Erratum: Nucl.Phys.B 402, 567--569
  (1993)].

\bibitem{Gunion:1989we}
J.~F. Gunion, H.~E. Haber, G.~L. Kane, and S.~Dawson, {\it {The Higgs Hunter's
  Guide}},  {\em Front. Phys.} {\bf 80} (2000) 1--404.

\bibitem{ESPINOSA1992113}
J.~Espinosa and M.~Quirós, {\it Higgs triplets in the supersymmetric standard
  model},  {\em Nuclear Physics B} {\bf 384} (1992), no.~1 113 -- 146.

\bibitem{DiChiara:2008rg}
S.~Di~Chiara and K.~Hsieh, {\it {Triplet Extended Supersymmetric Standard
  Model}},  {\em Phys. Rev.} {\bf D78} (2008) 055016,
  [\href{http://arxiv.org/abs/0805.2623}{{\tt arXiv:0805.2623}}].

\bibitem{KONETSCHNY1977433}
W.~Konetschny and W.~Kummer, {\it Nonconservation of total lepton number with
  scalar bosons},  {\em Physics Letters B} {\bf 70} (1977), no.~4 433 -- 435.

\bibitem{GELMINI1981411}
G.~Gelmini and M.~Roncadelli, {\it Left-handed neutrino mass scale and
  spontaneously broken lepton number},  {\em Physics Letters B} {\bf 99}
  (1981), no.~5 411 -- 415.

\bibitem{PhysRevLett.44.912}
R.~N. Mohapatra and G.~Senjanovi\ifmmode~\acute{c}\else \'{c}\fi{}, {\it
  Neutrino mass and spontaneous parity nonconservation},  {\em Phys. Rev.
  Lett.} {\bf 44} (Apr, 1980) 912--915.

\bibitem{PhysRevD.40.1546}
J.~F. Gunion, J.~Grifols, A.~Mendez, B.~Kayser, and F.~Olness, {\it Higgs
  bosons in left-right-symmetric models},  {\em Phys. Rev. D} {\bf 40} (Sep,
  1989) 1546--1561.

\bibitem{VEGA1990533}
R.~Vega and D.~A. Dicus, {\it Doubly charged higgs and $w^+w^+$ production},
  {\em Nuclear Physics B} {\bf 329} (1990), no.~3 533 -- 546.

\bibitem{PhysRevD.42.1673}
J.~F. Gunion, R.~Vega, and J.~Wudka, {\it Higgs triplets in the standard
  model},  {\em Phys. Rev. D} {\bf 42} (Sep, 1990) 1673--1691.

\bibitem{Huitu:1996su}
K.~Huitu, J.~Maalampi, A.~Pietila, and M.~Raidal, {\it {Doubly charged Higgs at
  LHC}},  {\em Nucl. Phys.} {\bf B487} (1997) 27--42,
  [\href{http://arxiv.org/abs/hep-ph/9606311}{{\tt hep-ph/9606311}}].

\bibitem{Chun:2003ej}
E.~J. Chun, K.~Y. Lee, and S.~C. Park, {\it {Testing Higgs triplet model and
  neutrino mass patterns}},  {\em Phys. Lett.} {\bf B566} (2003) 142--151,
  [\href{http://arxiv.org/abs/hep-ph/0304069}{{\tt hep-ph/0304069}}].

\bibitem{Georgi:1985nv}
H.~Georgi and M.~Machacek, {\it {Doubly charged Higgs bosons}},  {\em Nucl.
  Phys.} {\bf B262} (1985) 463--477.

\bibitem{Chiang:2012cn}
C.-W. Chiang and K.~Yagyu, {\it {Testing the custodial symmetry in the Higgs
  sector of the Georgi-Machacek model}},  {\em JHEP} {\bf 01} (2013) 026,
  [\href{http://arxiv.org/abs/1211.2658}{{\tt arXiv:1211.2658}}].

\bibitem{Mohapatra:1974gc}
R.~N. Mohapatra and J.~C. Pati, {\it {A Natural Left-Right Symmetry}},  {\em
  Phys. Rev.} {\bf D11} (1975) 2558.

\bibitem{Senjanovic:1975rk}
G.~Senjanovic and R.~N. Mohapatra, {\it {Exact Left-Right Symmetry and
  Spontaneous Violation of Parity}},  {\em Phys. Rev.} {\bf D12} (1975) 1502.

\bibitem{PhysRevD.44.837}
N.~G. Deshpande, J.~F. Gunion, B.~Kayser, and F.~Olness, {\it
  {Left-right-symmetric electroweak models with triplet Higgs field}},  {\em
  Phys. Rev. D} {\bf 44} (Aug, 1991) 837--858.

\bibitem{Duka:1999uc}
P.~Duka, J.~Gluza, and M.~Zralek, {\it {Quantization and renormalization of the
  manifest left-right symmetric model of electroweak interactions}},  {\em
  Annals Phys.} {\bf 280} (2000) 336--408,
  [\href{http://arxiv.org/abs/hep-ph/9910279}{{\tt hep-ph/9910279}}].

\bibitem{Barenboim:2001vu}
G.~Barenboim, M.~Gorbahn, U.~Nierste, and M.~Raidal, {\it {Higgs sector of the
  minimal left-right symmetric model}},  {\em Phys. Rev.} {\bf D65} (2002)
  095003, [\href{http://arxiv.org/abs/hep-ph/0107121}{{\tt hep-ph/0107121}}].

\bibitem{PhysRevD.84.095005}
A.~Arhrib, R.~Benbrik, M.~Chabab, G.~Moultaka, M.~C. Peyran\`ere, L.~Rahili,
  and J.~Ramadan, {\it {Higgs potential in the type II seesaw model}},  {\em
  Phys. Rev. D} {\bf 84} (Nov, 2011) 095005.

\bibitem{PhysRevD.85.055018}
A.~Melfo, M.~Nemev\ifmmode~\check{s}\else \v{s}\fi{}ek, F.~Nesti,
  G.~Senjanovi\ifmmode~\acute{c}\else \'{c}\fi{}, and Y.~Zhang, {\it {Type II
  neutrino seesaw mechanism at the LHC: The roadmap}},  {\em Phys. Rev. D} {\bf
  85} (Mar, 2012) 055018.

\bibitem{PhysRevD.85.055007}
M.~Aoki, S.~Kanemura, and K.~Yagyu, {\it {Testing the Higgs triplet model with
  the mass difference at the LHC}},  {\em Phys. Rev. D} {\bf 85} (Mar, 2012)
  055007.

\bibitem{PhysRevD.85.115009}
S.~Kanemura and K.~Yagyu, {\it {Radiative corrections to electroweak parameters
  in the Higgs triplet model and implication with the recent Higgs boson
  searches}},  {\em Phys. Rev. D} {\bf 85} (Jun, 2012) 115009.

\bibitem{PhysRevD.86.035015}
A.~G. Akeroyd and S.~Moretti, {\it {Enhancement of
  $H\ensuremath{\rightarrow}\ensuremath{\gamma}\ensuremath{\gamma}$ from doubly
  charged scalars in the Higgs triplet model}},  {\em Phys. Rev. D} {\bf 86}
  (Aug, 2012) 035015.

\bibitem{Blum2015}
K.~Blum, R.~T. D'Agnolo, and J.~Fan, {\it Vacuum stability bounds on higgs
  coupling deviations in the absence of new bosons},  {\em Journal of High
  Energy Physics} {\bf 2015} (Mar, 2015) 166.

\bibitem{Blunier:2016peh}
S.~Blunier, G.~Cottin, M.~A. Díaz, and B.~Koch, {\it {Phenomenology of a Higgs
  triplet model at future $e^{+}e^{-}$ colliders}},  {\em Phys. Rev.} {\bf D95}
  (2017), no.~7 075038, [\href{http://arxiv.org/abs/1611.07896}{{\tt
  arXiv:1611.07896}}].

\bibitem{Dev:2017ouk}
P.~S.~B. Dev, C.~M. Vila, and W.~Rodejohann, {\it {Naturalness in testable type
  II seesaw scenarios}},  {\em Nucl. Phys.} {\bf B921} (2017) 436--453,
  [\href{http://arxiv.org/abs/1703.00828}{{\tt arXiv:1703.00828}}].

\bibitem{Biswas:2017tnw}
A.~Biswas, {\it {All about $H^{\pm\pm}$ in Higgs Triplet Model. }},
  \href{http://arxiv.org/abs/1702.03847}{{\tt arXiv:1702.03847}}.

\bibitem{Du:2018eaw}
Y.~Du, A.~Dunbrack, M.~J. Ramsey-Musolf, and J.-H. Yu, {\it {Type-II Seesaw
  Scalar Triplet Model at a 100 TeV $pp$ Collider: Discovery and Higgs Portal
  Coupling Determination}},  {\em JHEP} {\bf 01} (2019) 101,
  [\href{http://arxiv.org/abs/1810.09450}{{\tt arXiv:1810.09450}}].

\bibitem{deMelo:2019asm}
T.~B. de~Melo, F.~S. Queiroz, and Y.~Villamizar, {\it {Doubly Charged Scalar at
  the High-Luminosity and High-Energy LHC}},  {\em Int. J. Mod. Phys. A} {\bf
  34} (2019), no.~27 1950157, [\href{http://arxiv.org/abs/1909.07429}{{\tt
  arXiv:1909.07429}}].

\bibitem{Primulando:2019evb}
R.~Primulando, J.~Julio, and P.~Uttayarat, {\it {Scalar phenomenology in
  type-II seesaw model}},  {\em JHEP} {\bf 08} (2019) 024,
  [\href{http://arxiv.org/abs/1903.02493}{{\tt arXiv:1903.02493}}].

\bibitem{Dev:2019hev}
P.~B. Dev, S.~Khan, M.~Mitra, and S.~K. Rai, {\it {Doubly-charged Higgs boson
  at a future electron-proton collider}},  {\em Phys. Rev. D} {\bf 99} (2019),
  no.~11 115015, [\href{http://arxiv.org/abs/1903.01431}{{\tt
  arXiv:1903.01431}}].

\bibitem{Fuks:2019clu}
B.~Fuks, M.~Nemevsek, and R.~Ruiz, {\it {Doubly Charged Higgs Boson Production
  at Hadron Colliders}},  {\em Phys. Rev. D} {\bf 101} (2020), no.~7 075022,
  [\href{http://arxiv.org/abs/1912.08975}{{\tt arXiv:1912.08975}}].

\bibitem{Aaboud:2017qph}
{\bf ATLAS} Collaboration, M.~Aaboud et~al., {\it {Search for doubly charged
  Higgs boson production in multi-lepton final states with the ATLAS detector
  using proton--proton collisions at $\sqrt{s}=13\,\text {TeV}$}},  {\em Eur.
  Phys. J. C} {\bf 78} (2018), no.~3 199,
  [\href{http://arxiv.org/abs/1710.09748}{{\tt arXiv:1710.09748}}].

\bibitem{Chakrabortty:2016wkl}
J.~Chakrabortty, J.~Gluza, T.~Jelinski, and T.~Srivastava, {\it {Theoretical
  constraints on masses of heavy particles in Left-Right Symmetric Models}},
  {\em Phys. Lett.} {\bf B759} (2016) 361--368,
  [\href{http://arxiv.org/abs/1604.06987}{{\tt arXiv:1604.06987}}].

\bibitem{Lindner:2016bgg}
M.~Lindner, M.~Platscher, and F.~S. Queiroz, {\it {A Call for New Physics : The
  Muon Anomalous Magnetic Moment and Lepton Flavor Violation}},  {\em Phys.
  Rept.} {\bf 731} (2018) 1--82, [\href{http://arxiv.org/abs/1610.06587}{{\tt
  arXiv:1610.06587}}].

\bibitem{Calibbi:2017uvl}
L.~Calibbi and G.~Signorelli, {\it {Charged Lepton Flavour Violation: An
  Experimental and Theoretical Introduction}},  {\em Riv. Nuovo Cim.} {\bf 41}
  (2018), no.~2 1, [\href{http://arxiv.org/abs/1709.00294}{{\tt
  arXiv:1709.00294}}].

\bibitem{Kuno:2013mha}
{\bf COMET} Collaboration, Y.~Kuno, {\it {A search for muon-to-electron
  conversion at J-PARC: The COMET experiment}},  {\em PTEP} {\bf 2013} (2013)
  022C01.

\bibitem{Brown:2015cka}
{\bf Mu2e} Collaboration, D.~Brown, {\it {The Mu2e Experiment: Searching for
  Muon to Electron Conversion}},  {\em Nucl. Part. Phys. Proc.} {\bf 260}
  (2015) 151--154.

\bibitem{Abgrall:2013rze}
{\bf Majorana} Collaboration, N.~Abgrall et~al., {\it {The Majorana
  Demonstrator Neutrinoless Double-Beta Decay Experiment}},  {\em Adv. High
  Energy Phys.} {\bf 2014} (2014) 365432,
  [\href{http://arxiv.org/abs/1308.1633}{{\tt arXiv:1308.1633}}].

\bibitem{Bolton:2019pcu}
P.~D. Bolton, F.~F. Deppisch, and P.~Bhupal~Dev, {\it {Neutrinoless double beta
  decay versus other probes of heavy sterile neutrinos}},  {\em JHEP} {\bf 03}
  (2020) 170, [\href{http://arxiv.org/abs/1912.03058}{{\tt arXiv:1912.03058}}].

\bibitem{Strategy:2019vxc}
R.~K. Ellis et~al., {\it {Physics Briefing Book}: {Input for the European
  Strategy for Particle Physics Update 2020. }},
  \href{http://arxiv.org/abs/1910.11775}{{\tt arXiv:1910.11775}}.

\bibitem{Abada:2019zxq}
{\bf FCC} Collaboration, A.~Abada et~al., {\it {FCC-ee: The Lepton Collider}:
  {Future Circular Collider Conceptual Design Report Volume 2}},  {\em Eur.
  Phys. J. ST} {\bf 228} (2019), no.~2 261--623.

\bibitem{fcccdrweb}
FCC -- Future Circular Collider, Conceptual Design Report
  \url{https://fcc-cdr.web.cern.ch/}.

\bibitem{Linssen:2012hp}
L.~Linssen, A.~Miyamoto, M.~Stanitzki, and H.~Weerts, {\it {Physics and
  Detectors at CLIC: CLIC Conceptual Design Report, }},
  \href{http://arxiv.org/abs/1202.5940}{{\tt arXiv:1202.5940}}.

\bibitem{clic}
CLIC -- Compact Linear International Collider Project, CERN,
  \url{http://clicdp.web.cern.ch/}.

\bibitem{ilcwww}
The Linear Collider Collaboration, \url{http://www.linearcollider.org/}.

\bibitem{Djouadi:2007ik}
G.~Aarons et~al., {\it {International Linear Collider Reference Design Report
  Volume 2: Physics at the ILC, }},  \href{http://arxiv.org/abs/0709.1893}{{\tt
  arXiv:0709.1893}}.

\bibitem{Moortgat-Picka:2015yla}
A.~Arbey et~al., {\it {Physics at the $e^+ e^-$ Linear Collider}},  {\em Eur.
  Phys. J.} {\bf C75} (2015), no.~8 371,
  [\href{http://arxiv.org/abs/1504.01726}{{\tt arXiv:1504.01726}}].

\bibitem{CEPCStudyGroup:2018rmc}
{\it {CEPC Conceptual Design Report: Volume 1 - Accelerator. }},
  \href{http://arxiv.org/abs/1809.00285}{{\tt arXiv:1809.00285}}.

\bibitem{cepc}
CEPC -- Circular Electron Positron Collider Project, China,
  \url{http://cepc.ihep.ac.cn/}.

\bibitem{Adli:2019hnt}
E.~Adli, {\it {Plasma Wakefield Linear Colliders - Opportunities and
  Challenges. }},  \href{http://arxiv.org/abs/1905.01879}{{\tt
  arXiv:1905.01879}}.

\bibitem{Blondel:2018mad}
A.~Blondel et~al., {\it {Standard model theory for the FCC-ee Tera-Z stage}},
  {\em CERN Yellow Rep. Monogr.} {\bf 3} (2019)
  [\href{http://arxiv.org/abs/1809.01830}{{\tt arXiv:1809.01830}}].

\bibitem{Blondel:2019qlh}
A.~Blondel, A.~Freitas, J.~Gluza, and T.~e.~a. Riemann, {\it {Theory
  Requirements and Possibilities for the FCC-ee and other Future High Energy
  and Precision Frontier Lepton Colliders. }},
  \href{http://arxiv.org/abs/1901.02648}{{\tt arXiv:1901.02648}}.

\bibitem{hllhc}
High-Luminosity LHC, CERN,
  \url{https://home.cern/science/accelerators/high-luminosity-lhc}.

\bibitem{Contino:2016spe}
R.~Contino et~al., {\it {Physics at a 100 TeV pp collider: Higgs and EW
  symmetry breaking studies}},  {\em CERN Yellow Report} (2017), no.~3
  255--440, [\href{http://arxiv.org/abs/1606.09408}{{\tt arXiv:1606.09408}}].

\bibitem{Golling:2016gvc}
T.~Golling et~al., {\it {Physics at a 100 TeV pp collider: beyond the Standard
  Model phenomena}},  {\em CERN Yellow Report} (2017), no.~3 441--634,
  [\href{http://arxiv.org/abs/1606.00947}{{\tt arXiv:1606.00947}}].

\bibitem{MadGraph:2014}
{Alwall, J., Frederix, R., Frixione, S. et al.}, {\it {The automated
  computation of tree-level and next-to-leading order differential cross
  section, and their maching to parton shower simulations}},  {\em JHEP} {\bf
  07} (2014) 079, [\href{http://arxiv.org/abs/1405.0301}{{\tt
  arXiv:1405.0301}}].

\bibitem{Sjostrand:2007gs}
T.~Sjostrand, S.~Mrenna, and P.~Z. Skands, {\it {A Brief Introduction to PYTHIA
  8.1}},  {\em Comput. Phys. Commun.} {\bf 178} (2008) 852--867,
  [\href{http://arxiv.org/abs/0710.3820}{{\tt arXiv:0710.3820}}].

\bibitem{Sjostrand:2006za}
T.~Sjostrand, S.~Mrenna, and P.~Z. Skands, {\it {PYTHIA 6.4 Physics and
  Manual}},  {\em JHEP} {\bf 05} (2006) 026,
  [\href{http://arxiv.org/abs/hep-ph/0603175}{{\tt hep-ph/0603175}}].

\bibitem{FeynRules:2013}
A.~Alloul, N.~D. Christensen, C.~Degrange, C.~Duhr, and B.~Fuks, {\it Feynrules
  2.0 - a complete toolbox for tree-level phenomenology},  {\em Comput. Phys.
  Commun.} {\bf 185} (2014) 2250--2300,
  [\href{http://arxiv.org/abs/1310.1921}{{\tt arXiv:1310.1921}}].

\bibitem{Chun:2012jw}
E.~J. Chun, H.~M. Lee, and P.~Sharma, {\it {Vacuum Stability, Perturbativity,
  EWPD and Higgs-to-diphoton rate in Type II Seesaw Models}},  {\em JHEP} {\bf
  11} (2012) 106, [\href{http://arxiv.org/abs/1209.1303}{{\tt
  arXiv:1209.1303}}].

\bibitem{Akeroyd:2012ms}
A.~G. Akeroyd and S.~Moretti, {\it {Enhancement of H to gamma gamma from doubly
  charged scalars in the Higgs Triplet Model}},  {\em Phys. Rev.} {\bf D86}
  (2012) 035015, [\href{http://arxiv.org/abs/1206.0535}{{\tt
  arXiv:1206.0535}}].

\bibitem{Shen_2015_EPL_2}
J.-F. Shen, Y.-P. Bi, and Z.-X. Li, {\it {Pair production of scalars at the ILC
  in the Higgs triplet model under the non-degenerate case}},  {\em EPL} {\bf
  112} (2015), no.~3 31002.

\bibitem{Das:2016}
D.~Das and A.~Santamaria, {\it Updated scalar sector constraints in higgs
  triplet model},  {\em Phys. Rev.} {\bf D94} (2016) 015015,
  [\href{http://arxiv.org/abs/1604.08099}{{\tt arXiv:1604.08099}}].

\bibitem{Gluza:2020icp}
J.~Gluza, M.~Kordiaczy\'nska, and T.~Srivastava, {\it {Doubly Charged Higgs
  Bosons and Spontaneous Symmetry Breaking at eV and TeV Scales}},  {\em
  Symmetry} {\bf 12} (2020), no.~1 153.

\bibitem{Bambhaniya:2013wza}
G.~Bambhaniya, J.~Chakrabortty, J.~Gluza, M.~Kordiaczyńska, and R.~Szafron,
  {\it {Left-Right Symmetry and the Charged Higgs Bosons at the LHC}},  {\em
  JHEP} {\bf 05} (2014) 033, [\href{http://arxiv.org/abs/1311.4144}{{\tt
  arXiv:1311.4144}}].

\bibitem{Xing:2010}
Z.-z. Xing and Y.-L. Zhou, {\it {A Generic Diagonalization of the 3x3 Neutrino
  Mass Matrix and Its Implications on the mu-tau Flavor Symmetry and Maximal CP
  Violation}},  {\em Phys. Lett.} {\bf B696} (2010) 584--590,
  [\href{http://arxiv.org/abs/1008.4906}{{\tt arXiv:1008.4906}}].

\bibitem{Ma:2000xh}
E.~Ma, M.~Raidal, and U.~Sarkar, {\it {Phenomenology of the neutrino mass
  giving Higgs triplet and the low-energy seesaw violation of lepton number}},
  {\em Nucl. Phys.} {\bf B615} (2001) 313--330,
  [\href{http://arxiv.org/abs/hep-ph/0012101}{{\tt hep-ph/0012101}}].

\bibitem{Fukuyama:2009xk}
T.~Fukuyama, H.~Sugiyama, and K.~Tsumura, {\it {Constraints from LFV processes
  in the Higgs triplet model}},  {\em JHEP} {\bf 03} (2010) 044,
  [\href{http://arxiv.org/abs/0909.4943}{{\tt arXiv:0909.4943}}].

\bibitem{Akeroyd:2009nu}
A.~G. Akeroyd, M.~Aoki, and H.~Sugiyama, {\it {Lepton Flavour Violating Decays
  $\tau \to \bar{l}ll$ and $\mu \to e \gamma$ in the Higgs Triplet Model}},
  {\em Phys. Rev.} {\bf D79} (2009) 113010,
  [\href{http://arxiv.org/abs/0904.3640}{{\tt arXiv:0904.3640}}].

\bibitem{Dinh:2012bp}
D.~Dinh, A.~Ibarra, E.~Molinaro, and S.~Petcov, {\it {The $\mu - e$ Conversion
  in Nuclei, $\mu \to e \gamma, \mu \to 3e$ Decays and TeV Scale See-Saw
  Scenarios of Neutrino Mass Generation}},  {\em JHEP} {\bf 08} (2012) 125,
  [\href{http://arxiv.org/abs/1205.4671}{{\tt arXiv:1205.4671}}]. [Erratum:
  JHEP 09, 023 (2013)].

\bibitem{Chakrabortty:2012vp}
J.~Chakrabortty, P.~Ghosh, and W.~Rodejohann, {\it {Lower Limits on $\mu \to e
  \gamma$ from New Measurements on $U_{e3}$}},  {\em Phys. Rev.} {\bf D86}
  (2012) 075020, [\href{http://arxiv.org/abs/1204.1000}{{\tt
  arXiv:1204.1000}}].

\bibitem{Crivellin:2018ahj}
A.~Crivellin, M.~Ghezzi, L.~Panizzi, G.~M. Pruna, and A.~Signer, {\it {Low- and
  high-energy phenomenology of a doubly charged scalar}},  {\em Phys. Rev. D}
  {\bf 99} (2019), no.~3 035004, [\href{http://arxiv.org/abs/1807.10224}{{\tt
  arXiv:1807.10224}}].

\bibitem{Chakrabarty:2018qtt}
N.~Chakrabarty, C.-W. Chiang, T.~Ohata, and K.~Tsumura, {\it {Charged scalars
  confronting neutrino mass and muon $g-2$ anomaly}},  {\em JHEP} {\bf 12}
  (2018) 104, [\href{http://arxiv.org/abs/1807.08167}{{\tt arXiv:1807.08167}}].

\bibitem{Dinh:2013tvc}
N.~D. Dinh, {\em {Probing the Possible TeV Scale See-saw Origin of Neutrino
  Masses with Charged Lepton Flavour Violation Processes and Neutrino Mass
  Spectroscopy Using Atoms}}.
\newblock PhD thesis, SISSA, Trieste (2013).
\newblock
  \url{https://s3.cern.ch/inspire-prod-files-3/33d85e96ff45ced9c9fcc19bd8195233}.

\bibitem{Schechter:1981bd}
J.~Schechter and J.~Valle, {\it {Neutrinoless Double beta Decay in SU(2) x U(1)
  Theories}},  {\em Phys. Rev. D} {\bf 25} (1982) 2951.

\bibitem{Wolfenstein:1982bf}
L.~Wolfenstein, {\it {Triplet Scalar Bosons and Double Beta Decay}},  {\em
  Phys. Rev.} {\bf D26} (1982) 2507.

\bibitem{Petcov:2009zr}
S.~T. Petcov, H.~Sugiyama, and Y.~Takanishi, {\it {Neutrinoless Double Beta
  Decay and $H^{\pm\pm} \to l'^\pm l^\pm$ Decays in the Higgs Triplet Model}},
  {\em Phys. Rev.} {\bf D80} (2009) 015005,
  [\href{http://arxiv.org/abs/0904.0759}{{\tt arXiv:0904.0759}}].

\bibitem{Actis:2008br}
S.~Actis, M.~Czakon, J.~Gluza, and T.~Riemann, {\it {Virtual hadronic and
  heavy-fermion $\mathcal{O}({\ensuremath{\alpha}}^{2})$ corrections to Bhabha
  scattering}},  {\em Phys. Rev.} {\bf D78} (2008) 085019,
  [\href{http://arxiv.org/abs/0807.4691}{{\tt arXiv:0807.4691}}].

\bibitem{Achard:2003mv}
{\bf L3} Collaboration, P.~Achard et~al., {\it {Search for doubly charged Higgs
  bosons at LEP}},  {\em Phys. Lett.} {\bf B576} (2003) 18--28,
  [\href{http://arxiv.org/abs/hep-ex/0309076}{{\tt hep-ex/0309076}}].

\bibitem{Campanario:2019mjh}
F.~Campanario, H.~Czy\.z, J.~Gluza, T.~Jeli\'nski, G.~Rodrigo, S.~Tracz, and
  D.~Zhuridov, {\it {Standard model radiative corrections in the pion form
  factor measurements do not explain the $a_\mu$ anomaly}},  {\em Phys. Rev. D}
  {\bf 100} (2019), no.~7 076004, [\href{http://arxiv.org/abs/1903.10197}{{\tt
  arXiv:1903.10197}}].

\bibitem{PhysRevD.98.030001}
{\bf Particle Data Group} Collaboration, M.~Tanabashi et~al., {\it Review of
  particle physics},  {\em Phys. Rev. D} {\bf 98} (Aug, 2018) 030001.

\bibitem{Chakrabortty:2015zpm}
J.~Chakrabortty, P.~Ghosh, S.~Mondal, and T.~Srivastava, {\it {Reconciling
  $(g-2)_\mu$ and charged lepton flavor violating processes through a doubly
  charged scalar}},  {\em Phys. Rev.} {\bf D93} (2016), no.~11 115004,
  [\href{http://arxiv.org/abs/1512.03581}{{\tt arXiv:1512.03581}}].

\bibitem{TheMEG:2016wtm}
{\bf MEG} Collaboration, A.~M. Baldini et~al., {\it {Search for the lepton
  flavour violating decay $\mu ^+ \rightarrow \mathrm {e}^+ \gamma $ with the
  full dataset of the MEG experiment}},  {\em Eur. Phys. J.} {\bf C76} (2016),
  no.~8 434, [\href{http://arxiv.org/abs/1605.05081}{{\tt arXiv:1605.05081}}].

\bibitem{Baldini:2013ke}
A.~M. Baldini et~al., {\it {MEG Upgrade Proposal. }},
  \href{http://arxiv.org/abs/1301.7225}{{\tt arXiv:1301.7225}}.

\bibitem{Aubert:2009ag}
{\bf BaBar} Collaboration, B.~Aubert et~al., {\it {Searches for Lepton Flavor
  Violation in the Decays $\tau^\pm \to e^\pm \gamma$ and $\tau^\pm \to \mu^\pm
  \gamma$}},  {\em Phys. Rev. Lett.} {\bf 104} (2010) 021802,
  [\href{http://arxiv.org/abs/0908.2381}{{\tt arXiv:0908.2381}}].

\bibitem{Wang:2015hdf}
B.~Wang, {\it {Searches for New Physics at the Belle II Experiment}},  in {\em
  {Meeting of the APS Division of Particles and Fields}}, 11, 2015.
\newblock \href{http://arxiv.org/abs/1511.00373}{{\tt arXiv:1511.00373}}.

\bibitem{Aushev:2010bq}
T.~Aushev et~al., {\it {Physics at Super B Factory. }},
  \href{http://arxiv.org/abs/1002.5012}{{\tt arXiv:1002.5012}}.

\bibitem{Bellgardt:1987du}
{\bf SINDRUM} Collaboration, U.~Bellgardt et~al., {\it {Search for the Decay
  $\mu^+ \to e^+ e^+ e^-$}},  {\em Nucl. Phys.} {\bf B299} (1988) 1--6.

\bibitem{Blondel:2013ia}
A.~Blondel et~al., {\it {Research Proposal for an Experiment to Search for the
  Decay $\mu \to eee$. }},  \href{http://arxiv.org/abs/1301.6113}{{\tt
  arXiv:1301.6113}}.

\bibitem{Hayasaka:2010np}
K.~Hayasaka et~al., {\it {Search for Lepton Flavor Violating Tau Decays into
  Three Leptons with 719 Million Produced Tau+Tau- Pairs}},  {\em Phys. Lett.}
  {\bf B687} (2010) 139--143, [\href{http://arxiv.org/abs/1001.3221}{{\tt
  arXiv:1001.3221}}].

\bibitem{SINDRUMII:2006}
W.~H. Bertl et~al., {\it {A Search for muon to electron conversion in muonic
  gold}},  {\em Eur.Phys.J.} {\bf C47} (2006) 337--346.

\bibitem{Bartoszek:2014mya}
{\bf Mu2e} Collaboration, L.~Bartoszek et~al., {\it {Mu2e Technical Design
  Report. }},  \href{http://arxiv.org/abs/1501.05241}{{\tt arXiv:1501.05241}}.

\bibitem{Maalampi:2002vx}
J.~Maalampi and N.~Romanenko, {\it {Single production of doubly charged Higgs
  bosons at hadron colliders}},  {\em Phys. Lett.} {\bf B532} (2002) 202--208,
  [\href{http://arxiv.org/abs/hep-ph/0201196}{{\tt hep-ph/0201196}}].

\bibitem{Nomura:2017abh}
T.~Nomura, H.~Okada, and H.~Yokoya, {\it {Discriminating leptonic Yukawa
  interactions with doubly charged scalar at the ILC}},  {\em Nucl. Phys. B}
  {\bf 929} (2018) 193--206, [\href{http://arxiv.org/abs/1702.03396}{{\tt
  arXiv:1702.03396}}].

\bibitem{Dev:2018sel}
P.~B. Dev, M.~J. Ramsey-Musolf, and Y.~Zhang, {\it {Doubly-Charged Scalars in
  the Type-II Seesaw Mechanism: Fundamental Symmetry Tests and High-Energy
  Searches}},  {\em Phys. Rev. D} {\bf 98} (2018), no.~5 055013,
  [\href{http://arxiv.org/abs/1806.08499}{{\tt arXiv:1806.08499}}].

\bibitem{NYFFELER:2014pta}
A.~Nyffeler, {\it {Status of hadronic light-by-light scattering in the muon
  $g-2$}},  {\em Nuovo Cim.} {\bf C037} (2014), no.~02 173--178,
  [\href{http://arxiv.org/abs/1312.4804}{{\tt arXiv:1312.4804}}]. [Int. J. Mod.
  Phys. Conf. Ser.35,1460456(2014)].

\bibitem{Cai:2017mow}
Y.~Cai, T.~Han, T.~Li, and R.~Ruiz, {\it {Lepton Number Violation: Seesaw
  Models and Their Collider Tests}},  {\em Front. in Phys.} {\bf 6} (2018) 40,
  [\href{http://arxiv.org/abs/1711.02180}{{\tt arXiv:1711.02180}}].

\bibitem{NuFIT}
``Nufit 4.1.'' \url{http://www.nu-fit.org/}, 2019.

\bibitem{Abe:2019vii}
{\bf T2K} Collaboration, K.~Abe et~al., {\it {Constraint on the
  matter--antimatter symmetry-violating phase in neutrino oscillations}},  {\em
  Nature} {\bf 580} (2020), no.~7803 339--344,
  [\href{http://arxiv.org/abs/1910.03887}{{\tt arXiv:1910.03887}}].

\bibitem{Girardi:2016zwz}
I.~Girardi, S.~T. Petcov, and A.~V. Titov, {\it {Predictions for the Majorana
  CP Violation Phases in the Neutrino Mixing Matrix and Neutrinoless Double
  Beta Decay}},  {\em Nucl. Phys.} {\bf B911} (2016) 754--804,
  [\href{http://arxiv.org/abs/1605.04172}{{\tt arXiv:1605.04172}}].

\bibitem{Olive:2016xmw}
{\bf Particle Data Group} Collaboration, C.~Patrignani et~al., {\it {Review of
  Particle Physics}},  {\em Chin. Phys.} {\bf C40} (2016), no.~10 100001.

\bibitem{Rodejohann:2011mu}
W.~Rodejohann, {\it {Neutrino-less Double Beta Decay and Particle Physics}},
  {\em Int. J. Mod. Phys.} {\bf E20} (2011) 1833--1930,
  [\href{http://arxiv.org/abs/1106.1334}{{\tt arXiv:1106.1334}}].

\bibitem{Ade:2013zuv}
{\bf Planck} Collaboration, P.~A.~R. Ade et~al., {\it {Planck 2013 results.
  XVI. Cosmological parameters}},  {\em Astron. Astrophys.} {\bf 571} (2014)
  A16, [\href{http://arxiv.org/abs/1303.5076}{{\tt arXiv:1303.5076}}].

\bibitem{Barger:1999na}
V.~D. Barger and K.~Whisnant, {\it {Majorana neutrino masses from neutrinoless
  double beta decay and cosmology}},  {\em Phys. Lett. B} {\bf 456} (1999)
  194--200, [\href{http://arxiv.org/abs/hep-ph/9904281}{{\tt hep-ph/9904281}}].

\bibitem{Czakon:2001uh}
M.~Czakon, J.~Gluza, J.~Studnik, and M.~Zralek, {\it {In quest of neutrino
  masses at O (eV) scale}},  {\em Phys. Rev. D} {\bf 65} (2002) 053008,
  [\href{http://arxiv.org/abs/hep-ph/0110166}{{\tt hep-ph/0110166}}].

\bibitem{Czakon:1999ha}
M.~Czakon, J.~Gluza, F.~Jegerlehner, and M.~Zralek, {\it {Confronting
  electroweak precision measurements with new physics models}},  {\em Eur.
  Phys. J.} {\bf C13} (2000) 275--281,
  [\href{http://arxiv.org/abs/hep-ph/9909242}{{\tt hep-ph/9909242}}].

\bibitem{PhysRevD.21.1404}
T.~G. Rizzo, {\it Tests of the fermion and higgs multiplet structure of the
  su(2)\ifmmode\times\else\texttimes\fi{}u(1) model},  {\em Phys. Rev. D} {\bf
  21} (Mar, 1980) 1404--1409.

\bibitem{PhysRevD.77.095009}
M.~Aoki and S.~Kanemura, {\it {Unitarity bounds in the Higgs model including
  triplet fields with custodial symmetry}},  {\em Phys. Rev. D} {\bf 77} (May,
  2008) 095009.

\bibitem{Barger:2003rs}
V.~Barger, T.~Han, P.~Langacker, B.~McElrath, and P.~Zerwas, {\it {Effects of
  genuine dimension-six Higgs operators}},  {\em Phys. Rev.} {\bf D67} (2003)
  115001, [\href{http://arxiv.org/abs/hep-ph/0301097}{{\tt hep-ph/0301097}}].

\bibitem{Kanemura:2004mg}
S.~Kanemura, Y.~Okada, E.~Senaha, and C.~P. Yuan, {\it {Higgs coupling
  constants as a probe of new physics}},  {\em Phys. Rev.} {\bf D70} (2004)
  115002, [\href{http://arxiv.org/abs/hep-ph/0408364}{{\tt hep-ph/0408364}}].

\bibitem{Tanabashi:2018oca}
{\bf Particle Data Group} Collaboration, M.~Tanabashi et~al., {\it {Review of
  Particle Physics}},  {\em Phys. Rev.} {\bf D98} (2018), no.~3 030001.

\bibitem{Aaboud:2017yvp}
{\bf ATLAS} Collaboration, M.~Aaboud et~al., {\it {Search for new phenomena in
  dijet events using 37 fb$^{-1}$ of $pp$ collision data collected at
  $\sqrt{s}=$13 TeV with the ATLAS detector}},  {\em Phys. Rev.} {\bf D96}
  (2017), no.~5 052004, [\href{http://arxiv.org/abs/1703.09127}{{\tt
  arXiv:1703.09127}}].

\bibitem{Sirunyan:2018pom}
{\bf CMS} Collaboration, A.~M. Sirunyan et~al., {\it {Search for a heavy
  right-handed W boson and a heavy neutrino in events with two same-flavor
  leptons and two jets at $\sqrt{s}=$ 13 TeV}},  {\em JHEP} {\bf 05} (2018),
  no.~05 148, [\href{http://arxiv.org/abs/1803.11116}{{\tt arXiv:1803.11116}}].

\bibitem{Sirunyan:2017ukk}
{\bf CMS} Collaboration, A.~M. Sirunyan et~al., {\it {Searches for $W^{'}$
  bosons decaying to a top quark and a bottom quark in proton-proton collisions
  at 13 TeV}},  {\em JHEP} {\bf 08} (2017) 029,
  [\href{http://arxiv.org/abs/1706.04260}{{\tt arXiv:1706.04260}}].

\bibitem{Aaboud:2019wfg}
{\bf ATLAS} Collaboration, M.~Aaboud et~al., {\it {Search for a right-handed
  gauge boson decaying into a high-momentum heavy neutrino and a charged lepton
  in $pp$ collisions with the ATLAS detector at $\sqrt{s}=13$ TeV}},  {\em
  Phys. Lett. B} {\bf 798} (2019) 134942,
  [\href{http://arxiv.org/abs/1904.12679}{{\tt arXiv:1904.12679}}].

\bibitem{Gluza:2015goa}
J.~Gluza and T.~Jeli\'nski, {\it {Heavy neutrinos and the pp$\rightarrow$lljj
  CMS data}},  {\em Phys. Lett. B} {\bf 748} (2015) 125--131,
  [\href{http://arxiv.org/abs/1504.05568}{{\tt arXiv:1504.05568}}].

\bibitem{Dev:2015pga}
P.~Bhupal~Dev and R.~Mohapatra, {\it {Unified explanation of the $eejj$,
  diboson and dijet resonances at the LHC}},  {\em Phys. Rev. Lett.} {\bf 115}
  (2015), no.~18 181803, [\href{http://arxiv.org/abs/1508.02277}{{\tt
  arXiv:1508.02277}}].

\bibitem{Das:2017hmg}
A.~Das, P.~S.~B. Dev, and R.~N. Mohapatra, {\it {Same Sign versus Opposite Sign
  Dileptons as a Probe of Low Scale Seesaw Mechanisms}},  {\em Phys. Rev. D}
  {\bf 97} (2018), no.~1 015018, [\href{http://arxiv.org/abs/1709.06553}{{\tt
  arXiv:1709.06553}}].

\bibitem{Frank:2018ifw}
{Frank, Mariana and \"Ozdal, \"Ozer and Poulose, Poulose}, {\it {Relaxing LHC
  constraints on the $W_R$ mass}},  {\em Phys. Rev.} {\bf D99} (2019), no.~3
  035001, [\href{http://arxiv.org/abs/1812.05681}{{\tt arXiv:1812.05681}}].

\bibitem{Sirunyan:2018vhk}
{\bf CMS} Collaboration, A.~M. Sirunyan et~al., {\it {Search for heavy
  neutrinos and third-generation leptoquarks in hadronic states of two $\tau$
  leptons and two jets in proton-proton collisions at $\sqrt{s} =$ 13 TeV}},
  {\em JHEP} {\bf 03} (2019) 170, [\href{http://arxiv.org/abs/1811.00806}{{\tt
  arXiv:1811.00806}}].

\bibitem{Aaboud:2018spl}
{\bf ATLAS} Collaboration, M.~Aaboud et~al., {\it {Search for heavy Majorana or
  Dirac neutrinos and right-handed $W$ gauge bosons in final states with two
  charged leptons and two jets at $ \sqrt{s}=13 $ TeV with the ATLAS
  detector}},  {\em JHEP} {\bf 01} (2019) 016,
  [\href{http://arxiv.org/abs/1809.11105}{{\tt arXiv:1809.11105}}].

\bibitem{Gluza:1993gf}
J.~Gluza and M.~Zralek, {\it {Neutrino production in e+ e- collisions in a
  left-right symmetric model}},  {\em Phys. Rev.} {\bf D48} (1993) 5093--5105.

\bibitem{Czakon:1999ga}
M.~Czakon, J.~Gluza, and M.~Zralek, {\it {Low-energy physics and left-right
  symmetry: Bounds on the model parameters}},  {\em Phys. Lett.} {\bf B458}
  (1999) 355--360, [\href{http://arxiv.org/abs/hep-ph/9904216}{{\tt
  hep-ph/9904216}}].

\bibitem{Gluza:2016qqv}
J.~Gluza, T.~Jelinski, and R.~Szafron, {\it {Lepton number violation and
  ‘Diracness’ of massive neutrinos composed of Majorana states}},  {\em
  Phys. Rev.} {\bf D93} (2016), no.~11 113017,
  [\href{http://arxiv.org/abs/1604.01388}{{\tt arXiv:1604.01388}}].

\bibitem{Dev:2019rxh}
P.~Bhupal~Dev, R.~N. Mohapatra, and Y.~Zhang, {\it {CP Violating Effects in
  Heavy Neutrino Oscillations: Implications for Colliders and Leptogenesis}},
  {\em JHEP} {\bf 11} (2019) 137, [\href{http://arxiv.org/abs/1904.04787}{{\tt
  arXiv:1904.04787}}].

\bibitem{Czakon:2002wm}
M.~Czakon, J.~Gluza, and J.~Hejczyk, {\it {Muon decay to one loop order in the
  left-right symmetric model}},  {\em Nucl. Phys. B} {\bf 642} (2002) 157--172,
  [\href{http://arxiv.org/abs/hep-ph/0205303}{{\tt hep-ph/0205303}}].

\bibitem{Beall:1981ze}
G.~Beall, M.~Bander, and A.~Soni, {\it {Constraint on the Mass Scale of a
  Left-Right Symmetric Electroweak Theory from the K(L) K(S) Mass Difference}},
   {\em Phys. Rev. Lett.} {\bf 48} (1982) 848.

\bibitem{GagyiPalffy:1997hh}
Z.~Gagyi-Palffy, A.~Pilaftsis, and K.~Schilcher, {\it {Gauge independent
  analysis of $K_L \to e \mu$ in left-right models}},  {\em Nucl. Phys.} {\bf
  B513} (1998) 517--554, [\href{http://arxiv.org/abs/hep-ph/9707517}{{\tt
  hep-ph/9707517}}].

\bibitem{Pilaftsis:1995tf}
A.~Pilaftsis, {\it {Confronting left-right symmetric models with electroweak
  precision data at the Z peak}},  {\em Phys. Rev.} {\bf D52} (1995) 459--471,
  [\href{http://arxiv.org/abs/hep-ph/9502330}{{\tt hep-ph/9502330}}].

\bibitem{Ball:1999mb}
P.~Ball, J.~M. Frere, and J.~Matias, {\it {Anatomy of mixing induced CP
  asymmetries in left-right symmetric models with spontaneous CP violation}},
  {\em Nucl. Phys.} {\bf B572} (2000) 3--35,
  [\href{http://arxiv.org/abs/hep-ph/9910211}{{\tt hep-ph/9910211}}].

\bibitem{Pospelov:1996fq}
M.~E. Pospelov, {\it {FCNC in left-right symmetric theories and constraints on
  the right-handed scale}},  {\em Phys. Rev.} {\bf D56} (1997) 259--264,
  [\href{http://arxiv.org/abs/hep-ph/9611422}{{\tt hep-ph/9611422}}].

\bibitem{Kiers:2002cz}
K.~Kiers, J.~Kolb, J.~Lee, A.~Soni, and G.-H. Wu, {\it {Ubiquitous CP violation
  in a top inspired left-right model}},  {\em Phys. Rev.} {\bf D66} (2002)
  095002, [\href{http://arxiv.org/abs/hep-ph/0205082}{{\tt hep-ph/0205082}}].

\bibitem{Rizzo:1994aj}
T.~G. Rizzo, {\it {Constraints from
  b\ensuremath{\rightarrow}s\ensuremath{\gamma} on the left-right symmetric
  model}},  {\em Phys. Rev.} {\bf D50} (1994) 3303--3309,
  [\href{http://arxiv.org/abs/hep-ph/9401319}{{\tt hep-ph/9401319}}].

\bibitem{Cho:1993zb}
P.~L. Cho and M.~Misiak, {\it {$b\ensuremath{\rightarrow}s\ensuremath{\gamma}$
  decay in
  ${\mathrm{SU}(2)}_{L}\ifmmode\times\else\texttimes\fi{}{\mathrm{SU}(2)}_{R}\ifmmode\times\else\texttimes\fi{}\mathrm{U}(1)$
  extensions of the standard model}},  {\em Phys. Rev.} {\bf D49} (1994)
  5894--5903, [\href{http://arxiv.org/abs/hep-ph/9310332}{{\tt
  hep-ph/9310332}}].

\bibitem{Senjanovic:1978av}
G.~Senjanovic and A.~Sokorac, {\it {Left-right Symmetric Gauge Theory and Its
  Prediction for Parity Violation in Atoms}},  {\em Phys. Lett.} {\bf 76B}
  (1978) 610--614.

\bibitem{Senjanovic:1978ee}
G.~Senjanovic and A.~Sokorac, {\it {Effects of Heavy Higgs Scalars at
  Low-energies}},  {\em Phys. Rev.} {\bf D18} (1978) 2708.

\bibitem{Chakrabortty:2012pp}
J.~Chakrabortty, J.~Gluza, R.~Sevillano, and R.~Szafron, {\it {Left-Right
  Symmetry at LHC and Precise 1-Loop Low Energy Data}},  {\em JHEP} {\bf 07}
  (2012) 038, [\href{http://arxiv.org/abs/1204.0736}{{\tt arXiv:1204.0736}}].

\bibitem{Deppisch:2014zta}
F.~F. Deppisch, T.~E. Gonzalo, S.~Patra, N.~Sahu, and U.~Sarkar, {\it {Double
  beta decay, lepton flavor violation, and collider signatures of left-right
  symmetric models with spontaneous $D$-parity breaking}},  {\em Phys. Rev.}
  {\bf D91} (2015), no.~1 015018, [\href{http://arxiv.org/abs/1410.6427}{{\tt
  arXiv:1410.6427}}].

\bibitem{Borah:2017ldt}
D.~Borah, A.~Dasgupta, and S.~Patra, {\it {Neutrinoless double beta decay in
  minimal left--right symmetric model with universal seesaw}},  {\em Int. J.
  Mod. Phys. A} {\bf 33} (2018), no.~35 1850198,
  [\href{http://arxiv.org/abs/1706.02456}{{\tt arXiv:1706.02456}}].

\bibitem{FileviezPerez:2017zwm}
P.~Fileviez~Perez and C.~Murgui, {\it {Lepton Flavour Violation in Left-Right
  Theory}},  {\em Phys. Rev.} {\bf D95} (2017), no.~7 075010,
  [\href{http://arxiv.org/abs/1701.06801}{{\tt arXiv:1701.06801}}].

\bibitem{Borah:2016iqd}
D.~Borah and A.~Dasgupta, {\it {Charged lepton flavour violcxmation and
  neutrinoless double beta decay in left-right symmetric models with type I+II
  seesaw}},  {\em JHEP} {\bf 07} (2016) 022,
  [\href{http://arxiv.org/abs/1606.00378}{{\tt arXiv:1606.00378}}].

\bibitem{Guadagnoli:2010sd}
D.~Guadagnoli and R.~N. Mohapatra, {\it {TeV Scale Left Right Symmetry and
  Flavor Changing Neutral Higgs Effects}},  {\em Phys. Lett.} {\bf B694} (2011)
  386--392, [\href{http://arxiv.org/abs/1008.1074}{{\tt arXiv:1008.1074}}].

\bibitem{Shaban:1992he}
N.~T. Shaban and W.~J. Stirling, {\it {Minimal left-right symmetry and SO(10)
  grand unification using LEP coupling constant measurements}},  {\em Phys.
  Lett.} {\bf B291} (1992) 281--287.

\bibitem{Lindner:1996tf}
M.~Lindner and M.~Weiser, {\it {Gauge coupling unification in left-right
  symmetric models}},  {\em Phys. Lett.} {\bf B383} (1996) 405--414,
  [\href{http://arxiv.org/abs/hep-ph/9605353}{{\tt hep-ph/9605353}}].

\bibitem{Das:2016bir}
D.~Das and A.~Santamaria, {\it {Updated scalar sector constraints in the Higgs
  triplet model}},  {\em Phys. Rev.} {\bf D94} (2016), no.~1 015015,
  [\href{http://arxiv.org/abs/1604.08099}{{\tt arXiv:1604.08099}}].

\bibitem{Bambhaniya:2014cia}
G.~Bambhaniya, J.~Chakrabortty, J.~Gluza, T.~Jeliński, and M.~Kordiaczynska,
  {\it {Lowest limits on the doubly charged Higgs boson masses in the minimal
  left-right symmetric model}},  {\em Phys. Rev.} {\bf D90} (2014), no.~9
  095003, [\href{http://arxiv.org/abs/1408.0774}{{\tt arXiv:1408.0774}}].

\bibitem{Mitra:2016kov}
M.~Mitra, R.~Ruiz, D.~J. Scott, and M.~Spannowsky, {\it {Neutrino Jets from
  High-Mass $W_R$ Gauge Bosons in TeV-Scale Left-Right Symmetric Models}},
  {\em Phys. Rev. D} {\bf 94} (2016), no.~9 095016,
  [\href{http://arxiv.org/abs/1607.03504}{{\tt arXiv:1607.03504}}].

\bibitem{Ruiz:2017nip}
R.~Ruiz, {\it {Lepton Number Violation at Colliders from Kinematically
  Inaccessible Gauge Bosons}},  {\em Eur. Phys. J. C} {\bf 77} (2017), no.~6
  375, [\href{http://arxiv.org/abs/1703.04669}{{\tt arXiv:1703.04669}}].

\bibitem{Nemevsek:2018bbt}
M.~Nemev\v{s}ek, F.~Nesti, and G.~Popara, {\it {Keung-Senjanovi\'c process at
  the LHC: From lepton number violation to displaced vertices to invisible
  decays}},  {\em Phys. Rev. D} {\bf 97} (2018), no.~11 115018,
  [\href{http://arxiv.org/abs/1801.05813}{{\tt arXiv:1801.05813}}].

\bibitem{Das:2017pvt}
A.~Das, {\it {Pair production of heavy neutrinos in next-to-leading order QCD
  at the hadron colliders in the inverse seesaw framework, }},
  \href{http://arxiv.org/abs/1701.04946}{{\tt arXiv:1701.04946}}.

\bibitem{Ruiz:2015zca}
R.~Ruiz, {\it {QCD Corrections to Pair Production of Type III Seesaw Leptons at
  Hadron Colliders}},  {\em JHEP} {\bf 12} (2015) 165,
  [\href{http://arxiv.org/abs/1509.05416}{{\tt arXiv:1509.05416}}].

\bibitem{Padhan:2019jlc}
R.~Padhan, D.~Das, M.~Mitra, and A.~Kumar~Nayak, {\it {Probing doubly and
  singly charged Higgs bosons at the $pp$ collider HE-LHC}},  {\em Phys. Rev.
  D} {\bf 101} (2020), no.~7 075050,
  [\href{http://arxiv.org/abs/1909.10495}{{\tt arXiv:1909.10495}}].

\bibitem{Gallinaro:2020cte}
M.~Gallinaro et~al., {\it {Beyond the Standard Model in Vector Boson Scattering
  Signatures}},  5, 2020.
\newblock \href{http://arxiv.org/abs/2005.09889}{{\tt arXiv:2005.09889}}.

\bibitem{Agrawal:2018}
P.~Agrawal, M.~Mitra, S.~Niyogi, S.~Shil, and M.~Spannowsky, {\it {Probing the
  Type-II Seesaw Mechanism through the Production of Higgs Bosons at a Lepton
  Collider}},  {\em Phys. Rev. D} {\bf 98} (2018), no.~1 015024,
  [\href{http://arxiv.org/abs/1803.00677}{{\tt arXiv:1803.00677}}].

\bibitem{Perez:2008ha}
P.~Fileviez~Perez, T.~Han, G.-y. Huang, T.~Li, and K.~Wang, {\it {Neutrino
  Masses and the CERN LHC: Testing Type II Seesaw}},  {\em Phys. Rev.} {\bf
  D78} (2008) 015018, [\href{http://arxiv.org/abs/0805.3536}{{\tt
  arXiv:0805.3536}}].

\bibitem{Moultaka:2020dmb}
G.~Moultaka and M.~C. Peyran\`ere, {\it {Vacuum Stability Conditions for Higgs
  Potentials with $SU(2)_L$ Triplets, }},
  \href{http://arxiv.org/abs/2012.13947}{{\tt arXiv:2012.13947}}.

\bibitem{Garayoa:2007fw}
J.~Garayoa and T.~Schwetz, {\it {Neutrino mass hierarchy and Majorana CP phases
  within the Higgs triplet model at the LHC}},  {\em JHEP} {\bf 03} (2008) 009,
  [\href{http://arxiv.org/abs/0712.1453}{{\tt arXiv:0712.1453}}].

\bibitem{Dekens:2014ina}
W.~Dekens and D.~Boer, {\it {Viability of minimal left--right models with
  discrete symmetries}},  {\em Nucl. Phys. B} {\bf 889} (2014) 727--756,
  [\href{http://arxiv.org/abs/1409.4052}{{\tt arXiv:1409.4052}}].

\bibitem{Bambhaniya:2015wna}
G.~Bambhaniya, J.~Chakrabortty, J.~Gluza, T.~Jelinski, and R.~Szafron, {\it
  {Search for doubly charged Higgs bosons through vector boson fusion at the
  LHC and beyond}},  {\em Phys. Rev. D} {\bf 92} (2015), no.~1 015016,
  [\href{http://arxiv.org/abs/1504.03999}{{\tt arXiv:1504.03999}}].

\bibitem{Bambhaniya:2013yca}
G.~Bambhaniya, J.~Chakrabortty, S.~Goswami, and P.~Konar, {\it {Generation of
  neutrino mass from new physics at TeV scale and multilepton signatures at the
  LHC}},  {\em Phys. Rev.} {\bf D88} (2013), no.~7 075006,
  [\href{http://arxiv.org/abs/1305.2795}{{\tt arXiv:1305.2795}}].

\bibitem{Pumplin:2002vw}
J.~Pumplin, D.~Stump, J.~Huston, H.~Lai, P.~M. Nadolsky, and W.~Tung, {\it {New
  generation of parton distributions with uncertainties from global QCD
  analysis}},  {\em JHEP} {\bf 07} (2002) 012,
  [\href{http://arxiv.org/abs/hep-ph/0201195}{{\tt hep-ph/0201195}}].

\bibitem{Hou:2019efy}
T.-J. Hou et~al., {\it {New CTEQ global analysis of quantum chromodynamics with
  high-precision data from the LHC. }},
  \href{http://arxiv.org/abs/1912.10053}{{\tt arXiv:1912.10053}}.

\bibitem{Cacciari:2008zb}
M.~Cacciari, S.~Frixione, M.~L. Mangano, P.~Nason, and G.~Ridolfi, {\it
  {Updated predictions for the total production cross sections of top and of
  heavier quark pairs at the Tevatron and at the LHC}},  {\em JHEP} {\bf 09}
  (2008) 127, [\href{http://arxiv.org/abs/0804.2800}{{\tt arXiv:0804.2800}}].

\bibitem{Alwall:2014hca}
J.~Alwall, R.~Frederix, S.~Frixione, V.~Hirschi, F.~Maltoni, O.~Mattelaer,
  H.~S. Shao, T.~Stelzer, P.~Torrielli, and M.~Zaro, {\it {The automated
  computation of tree-level and next-to-leading order differential cross
  sections, and their matching to parton shower simulations}},  {\em JHEP} {\bf
  07} (2014) 079, [\href{http://arxiv.org/abs/1405.0301}{{\tt
  arXiv:1405.0301}}].

\bibitem{Chiesa:2020ttl}
M.~Chiesa, C.~Oleari, and E.~Re, {\it {NLO QCD+NLO EW corrections to diboson
  production matched to parton shower}},  {\em Eur. Phys. J. C} {\bf 80}
  (2020), no.~9 849, [\href{http://arxiv.org/abs/2005.12146}{{\tt
  arXiv:2005.12146}}].

\bibitem{Benedikt:2018csr}
{\bf FCC} Collaboration, A.~Abada et~al., {\it {FCC-hh: The Hadron Collider}:
  {Future Circular Collider Conceptual Design Report Volume 3}},  {\em Eur.
  Phys. J. ST} {\bf 228} (2019), no.~4 755--1107.

\bibitem{Arhrib:2011uy}
A.~Arhrib, R.~Benbrik, M.~Chabab, G.~Moultaka, M.~C. Peyranere, L.~Rahili, and
  J.~Ramadan, {\it {The Higgs Potential in the Type II Seesaw Model}},  {\em
  Phys. Rev.} {\bf D84} (2011) 095005,
  [\href{http://arxiv.org/abs/1105.1925}{{\tt arXiv:1105.1925}}].

\bibitem{Gunion:1989ci}
J.~F. Gunion, R.~Vega, and J.~Wudka, {\it {Higgs triplets in the standard
  model}},  {\em Phys. Rev.} {\bf D42} (1990) 1673--1691.

\bibitem{Dey:2008jm}
P.~Dey, A.~Kundu, and B.~Mukhopadhyaya, {\it {Some consequences of a Higgs
  triplet}},  {\em J. Phys.} {\bf G36} (2009) 025002,
  [\href{http://arxiv.org/abs/0802.2510}{{\tt arXiv:0802.2510}}].

\bibitem{Georgi:1981pg}
H.~M. Georgi, S.~L. Glashow, and S.~Nussinov, {\it {Unconventional Model of
  Neutrino Masses}},  {\em Nucl. Phys.} {\bf B193} (1981) 297--316.

\bibitem{Melfo:2011nx}
A.~Melfo, M.~Nemevsek, F.~Nesti, G.~Senjanovic, and Y.~Zhang, {\it {Type II
  Seesaw at LHC: The Roadmap}},  {\em Phys. Rev.} {\bf D85} (2012) 055018,
  [\href{http://arxiv.org/abs/1108.4416}{{\tt arXiv:1108.4416}}].

\bibitem{Kanemura:2012rs}
S.~Kanemura and K.~Yagyu, {\it {Radiative corrections to electroweak parameters
  in the Higgs triplet model and implication with the recent Higgs boson
  searches}},  {\em Phys. Rev.} {\bf D85} (2012) 115009,
  [\href{http://arxiv.org/abs/1201.6287}{{\tt arXiv:1201.6287}}].

\bibitem{Babu:2016rcr}
K.~S. Babu and S.~Jana, {\it {Probing Doubly Charged Higgs Bosons at the LHC
  through Photon Initiated Processes}},  {\em Phys. Rev.} {\bf D95} (2017),
  no.~5 055020, [\href{http://arxiv.org/abs/1612.09224}{{\tt
  arXiv:1612.09224}}].

\bibitem{Mohapatra:1980yp}
R.~N. Mohapatra and G.~Senjanovic, {\it {Neutrino Masses and Mixings in Gauge
  Models with Spontaneous Parity Violation}},  {\em Phys. Rev. D} {\bf 23}
  (1981) 165.

\bibitem{Gunion:1989in}
J.~Gunion, J.~Grifols, A.~Mendez, B.~Kayser, and F.~I. Olness, {\it {Higgs
  Bosons in Left-Right Symmetric Models}},  {\em Phys. Rev. D} {\bf 40} (1989)
  1546.

\bibitem{Deshpande:1990ip}
N.~Deshpande, J.~Gunion, B.~Kayser, and F.~I. Olness, {\it {Left-right
  symmetric electroweak models with triplet Higgs}},  {\em Phys. Rev. D} {\bf
  44} (1991) 837--858.

\bibitem{Gluza:1991wj}
J.~Gluza and M.~Zralek, {\it {Feynman rules for Majorana neutrino
  interactions}},  {\em Phys. Rev. D} {\bf 45} (1992) 1693--1700.

\bibitem{Gluza:1995ky}
J.~Gluza and M.~Zralek, {\it {Inverse neutrinoless double beta decay in gauge
  theories with CP violation}},  {\em Phys. Rev.} {\bf D52} (1995) 6238--6248,
  [\href{http://arxiv.org/abs/hep-ph/9502284}{{\tt hep-ph/9502284}}].

\bibitem{Gluza:2002vs}
J.~Gluza, {\it {On teraelectronvolt Majorana neutrinos}},  {\em Acta Phys.
  Polon.} {\bf B33} (2002) 1735--1746,
  [\href{http://arxiv.org/abs/hep-ph/0201002}{{\tt hep-ph/0201002}}].

\bibitem{Kakizaki:2003jk}
M.~Kakizaki, Y.~Ogura, and F.~Shima, {\it {Lepton flavor violation in the
  triplet Higgs model}},  {\em Phys. Lett.} {\bf B566} (2003) 210--216,
  [\href{http://arxiv.org/abs/hep-ph/0304254}{{\tt hep-ph/0304254}}].

\bibitem{Kitano:2002mt}
R.~Kitano, M.~Koike, and Y.~Okada, {\it {Detailed calculation of lepton flavor
  violating muon electron conversion rate for various nuclei}},  {\em Phys.
  Rev. D} {\bf 66} (2002) 096002,
  [\href{http://arxiv.org/abs/hep-ph/0203110}{{\tt hep-ph/0203110}}]. [Erratum:
  Phys.Rev.D 76, 059902 (2007)].

\bibitem{Leveille:1977rc}
J.~P. Leveille, {\it {The Second Order Weak Correction to $(g-2)$ of the Muon
  in Arbitrary Gauge Models}},  {\em Nucl. Phys.} {\bf B137} (1978) 63--76.

\bibitem{Moore:1984eg}
S.~R. Moore, K.~Whisnant, and B.-L. Young, {\it {Second Order Corrections to
  the Muon Anomalous Magnetic Moment in Alternative Electroweak Models}},  {\em
  Phys. Rev.} {\bf D31} (1985) 105.

\end{thebibliography}
\providecommand{\href}[2]{#2}\begingroup\raggedright\endgroup

\end{document}